\documentclass[preprint2]{aa} 
\usepackage{graphicx}
\usepackage{url}
\usepackage{txfonts}
\usepackage{natbib}
\bibpunct{(}{)}{;}{a}{}{,} 
\newcommand{\1}{{~\sc i}}
\newcommand{\2}{{~\sc ii}}
\newcommand{\3}{{~\sc iii}}
\newcommand{\4}{{~\sc iv}}
\newcommand{\5}{{~\sc v}}

\newcommand{\wm}{{\,W\,m$^{-2}$}}

\newcommand{\kms}{{\,km\,s$^{-1}$}}
\newcommand{\cc}{{\,cm$^{-3}$}}
\newcommand{\mic}{{\,$\mu$m}}

\newcommand{\izw}{I\,Zw\,18}

\usepackage{pdflscape}

\begin{document}

 \title{Neutral gas heating by X-rays in primitive galaxies: Infrared observations of the blue compact dwarf \object{I Zw 18} with \textit{Herschel}\thanks{{\it Herschel} is an ESA space observatory with science instruments provided by European-led Principal Investigator consortia and with important participation from NASA.} }

   
	   \titlerunning{Neutral gas heating by X-rays in primitive galaxies: Infrared observations of \izw\ with \textit{Herschel} }
   	\authorrunning{Lebouteiller et al.}

   \author{V.\ Lebouteiller\inst{1}, D.\ P\'equignot\inst{2}, D.\ Cormier\inst{1,3}, S.\ Madden\inst{1}, M.\,W.\ Pakull\inst{4}, D.\ Kunth\inst{5}, F.\ Galliano\inst{1}, M.\ Chevance\inst{1,6}, S.\,R.\ Heap\inst{7}, M.-Y.\ Lee\inst{1}, F.\ L.\ Polles\inst{1}}

   \institute{$^1$ Laboratoire AIM Paris-Saclay, CEA/IRFU - CNRS/INSU - Universit\'e Paris Diderot, Service d'Astrophysique, B\^at. 709, CEA-Saclay, 91191, Gif-sur-Yvette Cedex, France \email{vianney.lebouteiller@cea.fr} \\
   $^2$ LUTH, Observatoire de Paris, CNRS, Universit\'e Paris Diderot, 5 place Jules Janssen, 92190 Meudon, France \\
   $^3$ Zentrum f\"ur Astronomie der Universit\"at Heidelberg, Institut f\"ur Theoretische Astrophysik, Albert-Ueberle-Str. 2, 69120 Heidelberg, Germany \\
   $^4$ Observatoire astronomique de Strasbourg, Universit\'e de Strasbourg, CNRS, UMR 7550, 11 rue de l'Universit\'e, F-67000 Strasbourg, France \\
   $^5$ Institut d'Astrophysique de Paris, 98 bis Boulevard Arago, F-75014 Paris, France \\
	$^6$ Astronomisches Rechen-Institut, Zentrum f\"ur Astronomie der Universit\"at Heidelberg, M\"onchhofstra{\ss}e 12-14, 	D-69120 Heidelberg, Germany \\
   $^7$ NASA Goddard Space Flight Center, Greenbelt, MD 20771, USA 
               }

   \date{Received 9 September 2016; accepted 23 February 2017}

 
  \abstract
   {The neutral interstellar medium of galaxies acts as a reservoir to fuel star formation. The dominant heating and cooling mechanisms in this phase are uncertain in extremely metal-poor star-forming galaxies. The low dust-to-gas mass ratio and low polycyclic aromatic hydrocarbon abundance in such objects suggest that the traditional photoelectric effect heating may not be effective. }
   {Our objective is to identify the dominant thermal mechanisms in one such galaxy, \izw\ ($1/30$\,Z$_\odot$), assess the diagnostic value of fine-structure cooling lines, and estimate the molecular gas content. Even though molecular gas is an important catalyst and tracer of star formation, constraints on the molecular gas mass remain elusive in the most metal-poor galaxies.  }
   {Building on a previous photoionization model describing the giant H\2\ region of \izw-NW within a multi-sector topology, we provide additional constraints using, in particular, the [C\2] $157$\mic\ and [O\1] $63$\mic\ lines and the dust mass recently measured with the \textit{Herschel} Space Telescope. }
   {The heating of the H\1\ region appears to be mainly due to photoionization by radiation from a bright X-ray binary source, while the photoelectric effect is negligible. Significant cosmic ray heating is not excluded. Inasmuch as X-ray heating dominates in the H\1\ gas, the infrared fine-structure lines provide an average X-ray luminosity of order $4\times10^{40}$\,erg\,s$^{-1}$ over the last few $10^4$ years in the galaxy.  The upper limits to the [Ne\5] lines provide strong constraints on the soft X-ray flux arising from the binary. A negligible mass of H$_2$ is predicted. Nonetheless, up to $\sim10^7$\,M$_\odot$ of H$_2$ may be hidden in a few sufficiently dense clouds of order $\lesssim10$\,pc ($\lesssim0.1$\arcsec) in size. Regardless of the presence of significant amounts of H$_2$ gas, [C\2] and [O\1] do not trace the so-called ``CO-dark gas'', but they trace the almost purely atomic medium. Although the [C\2]+[O\1] to total infrared ratio in \izw\ is similar to values in more metal-rich sources ($\sim1$\%), it cannot be safely used as a photoelectric heating efficiency proxy. This ratio seems to be kept stable owing to a correlation between the X-ray luminosity and the star formation rate.   } 
   {X-ray heating could be an important process in extremely metal-poor sources. The lack of photoelectric heating due to the low dust-to-gas ratio tends to be compensated for by the larger occurrence and power of X-ray binaries in low-metallicity galaxies. We speculate that X-ray heating may quench star formation.  }


   \maketitle

\section{Introduction}

\defcitealias{Pequignot08}{P08}
\defcitealias{Wu07}{W07}

Star formation in primordial (or quasi-primordial) gas is a fundamental process taking place in the first galaxies that are not yet enriched with elements produced by stellar nucleosynthesis. Star formation proceeds when a cloud is gravitationally bound, dense, and cold enough to be subject to the Jeans instability (e.g., \citealt{Krumholz12a}). Thermal pressure is removed 
by lowering the heating from UV photons through H$_2$ self-shielding or absorption by dust particles (and conversion to infrared radiation; IR). Furthermore, the presence of metals, even in small amounts, significantly cools down the gas through radiative transitions such as [C\2] $157$\mic, [O\1] $63$\mic, and [Si\2] $34$\mic\ in the neutral atomic medium, or CO in the molecular medium. In the diffuse interstellar medium (ISM), metal cooling is expected to become dominant over H$_2$ cooling when the metallicity is $\gtrsim1/10$\,Z$_\odot$ \citep{Glover13}. The cooling rate from metals in the neutral phase and the abundance of H$_2$ are therefore two critical parameters to understand the prerequisites for star formation in low-metallicity environments. At the same time, it is essential to identify the main heating mechanisms at work, especially in the neutral ISM, in order to establish the relationship between the thermal tracers and the star formation process. Infrared cooling lines are, for instance, widely used tracers to probe star formation at potentially all redshifts (e.g., \citealt{deLooze14a}), despite the lack of precise knowledge concerning the heating mechanisms.

The class of blue compact dwarf (BCD) galaxies contains some of the most metal-poor star-forming galaxies known. Apart from a subcomponent of \object{SBS 0335-052} and a low star formation rate (SFR) BCD recently discovered through a blind H\1\ survey (AGC\,198691; \citealt{Hirschauer16}), \izw\ is the nearby star-forming galaxy with the lowest metallicity known, i.e., $12+\log({\rm O/H})=7.22$ or $1/30$\,Z$_\odot$\footnote{We use the oxygen abundance obtained by \cite{Pequignot08} divided by the solar value from \cite{Asplund09}. }, as measured by optical emission lines in the H\2\ regions \citep{Searle72,Skillman93,Kunth94,Garnett97,Izotov98}. Observations of the neutral atomic medium probed by far-ultraviolet (FUV) absorption lines toward the massive stars suggest that the H\1\ region might be even more metal poor   \citep{Kunth94,Aloisi03,Lecavelier04,Lebouteiller13a}. According to \cite{Lebouteiller13a}, it is possible that as much as $50$\%\ of the H$^0$ mass in \izw\ is pristine. Blue compact dwarfs and in particular the well-studied galaxy \izw, thus represent important probes of the thermal balance of the ISM in primitive environments. 

The main heating mechanism in the ionized gas of H\2\ regions of star-forming galaxies is photoionization\footnote{Outside the H\2\ regions, in the so-called warm diffuse ionized medium of disk galaxies, some extra heating exists in extremely low-density regions (e.g., \citealt{Reynolds99}) and may be due, e.g., to photoelectric effect on dust or dissipation of interstellar turbulence.} of H, He, and sometimes He$^+$. 
In \izw, \cite{Stasinska99} proposed that the H\2\ regions may be heated by other energy sources as well (shocks, conductive heating at the interface of an X-ray plasma), mainly owing to the supposedly too large electron temperature observed (see also \citealt{Kehrig2016}). However, \citet[hereafter P08]{Pequignot08} later performed a detailed modeling of the \izw-NW region using the code Nebu \citep{Pequignot01}; these authors concluded that photoionization by hot stars could satisfactorily explain the entire optical line spectrum, provided that the H\2\ region topology is equivalent to an incomplete radiation-bounded shell embedded in a diffuse low-density matter-bounded medium of filling factor unity. In essence, the lower density of the diffuse ionized gas leads to a smaller fraction of H$^0$, which is a dominant cooling agent in low-metallicity H\2\ regions, and therefore to a higher electron temperature.

The heating of H\1\ regions is comparatively much less understood. While the main heating mechanism in the neutral ISM of our Galaxy is due to the photoelectric effect on polycyclic aromatic hydrocarbons (PAHs) and dust grains (e.g., \citealt{Weingartner01a}), both the low dust-to-gas mass ratio (D/G) and the low PAH abundance observed in BCDs (e.g., \citealt{Wu06,Remy13b}) may lead to important differences as compared to more metal-rich objects. 
\citetalias{Pequignot08} introduced in his model of \izw-NW the heating of the H\1\ region by the soft X-ray source observed in this galaxy and was able to account to order of magnitude for the low-ionization fine-structure lines then recently detected by \textit{Spitzer} (in particular [Si\2] $34.8$\mic\ and [Fe\2] $26.0$\mic). \citetalias{Pequignot08} also made tentative predictions for the far-infrared (FIR) lines [C\2] $157$\mic\ and [O\1] $63$\mic, which are the most important coolants in the neutral atomic medium. According to the models, these lines are mainly produced in an H\1\ region of moderate ionization and temperature, that is, in an X-ray dominated region (XDR), using the terminology introduced in the framework of the physics of active galactic nuclei (e.g., \citealt{Tine97}). The study of \citetalias{Pequignot08} implied that an X-ray source could provide an effective heating mechanism in the neutral ISM of low-metallicity BCD galaxies, and therefore a possible alternative to the traditional photoelectric effect heating. Thanks to the \textit{Herschel} Space Observatory \citep{Pilbratt10} and, in particular, the Photodetector Array Camera and Spectrometer (PACS; \citealt{Poglitsch10}), it is now possible to compare observations and models for [C\2] and [O\1].

More recently, using \textit{Hubble}/COS, \cite{Lebouteiller13a} observed \izw-NW in the FUV absorption-lines C\2\ $\lambda1334.5$ and C\2* $\lambda1335.7$, arising from the ground level and fine-structure level of C$^+$, respectively, and observed against the FUV continuum provided by the UV-bright stars in the stellar cluster. The authors roughly estimated an electron fraction $n_e/n_{\rm H}\sim0.1$\% and attempted to substantiate the assumption of photoelectric effect on dust and PAHs, but with mitigate success, which may be viewed retroactively as evidence in favor of X-ray heating in the H\1\ region, as proposed by \citetalias{Pequignot08}.

Here, building on the model of \citetalias{Pequignot08}, the heating by photoelectric effect, X-rays, and by other processes is examined. Consequences may pertain to other metal-poor star-forming galaxies, as there is growing evidence that ultraluminous X-ray sources (ULXs; $L_{\rm X}\gtrsim10^{39}$\,erg\,s$^{-1}$) are more numerous and more luminous in low-metallicity galaxies (e.g., \citealt{Kaaret11,Kaaret13,Brorby14,Brorby15,Basu16} and references therein). These ULXs are thought to be associated with high-mass X-ray binaries (HMXBs), involving either a stellar-mass or intermediate-mass black hole or, rather unexpectedly pulsating neutron stars \citep{Bachetti14,Furst16,IsraelX16,IsraelX17}. The effects of X-rays are numerous, since they can photoevaporate small molecules and PAHs (while heating larger grains), and at the same time penetrate deep inside the H\1\ region where they can ionize atomic and molecular hydrogen. The most obvious hallmark of X-ray photoionization of the ISM by luminous X-ray sources is the presence of highly ionized species such as He\2\ $\lambda4686$ recombination radiation. This effect has first been observed by \cite{Pakull86} for the luminous black hole candidate \object{LMC X-1}. In the case of the ULX \object{Holmberg II X-1} the detection of a X-ray ionized nebula has furthermore allowed an independent estimate of the total luminosity of the X-ray source from the He\2\ $\lambda4686$ emission \citep{Pakull02}. By modeling the observed ionization structure with Cloudy \citep{Ferland98} these observations imply a largely isotropic X-ray emission and largely exclude any significant beaming of the ULX into our line of sight.

Understanding the origin of [C\2] or [O\1] in metal-poor galaxies and the impact of X-rays is an important challenge not only to constrain the gas heating mechanism but also to evaluate the possible reservoir of molecular gas. The apparent lack of molecular gas in \izw\ \citep{Vidal00,Wu06,Leroy07b} is at variance with the present vigorous starburst episode. While it is possible that the earliest stages of star formation occur in the cold atomic gas, with molecular gas forming only at the onset of the star-forming cloud collapse (e.g., \citealt{Glover12a,Krumholz12a}), a significant reservoir of molecular gas that is not traced by CO may still exist, i.e., the so-called CO-dark gas (e.g., \citealt{Tielens85,Maloney88,vanDishoeck88,Grenier05,Wolfire10}). The low dust abundance in metal-poor galaxies results in a smaller photodissociated CO core while H$_2$ is self-shielded, resulting in a CO-free molecular gas layer with abundant C$^+$ and leading to an enhanced [C\2]/CO ratio for the global cloud emission (e.g., \citealt{Poglitsch96,Madden97}). \izw\ provides an opportunity to examine the origin of [C\2] and its hypothetical association with molecular gas.

A summary of relevant properties of \izw\ is provided in Section\,\ref{sec:general}.  Observations are described in Section\,\ref{sec:obs}. A topologically significant model of the NW region is then obtained using the photoionization and photodissociation code Cloudy (Sects.\,\ref{sec:modelprep}, \ref{sec:obsconst}). Various models, which are shown to be relevant to the full observed IR emitting region, are explored in Section\,\ref{sec:cloudy}. 
The presence of molecular gas and physical conditions in the diffuse gas are investigated in Section\,\ref{sec:discussion}. Implications of X-ray heating of the H\1\ gas are examined in Section\,\ref{sec:firdiag}. Conclusions are found in Section\,\ref{sec:conclusion}. Details about \textit{Herschel}, \textit{Spitzer}, and X-ray data treatments are provided in Appendices\,\ref{secapp:herschel}, \ref{secapp:spitzer}, and \ref{secapp:xspec} respectively.

\section{Characteristics of \izw}\label{sec:general}

Some of the main characteristics of \izw\ are listed in Table\,\ref{tab:obs_log}. The most important properties are described in the following. Gas and dust masses are discussed separately in Section\,\ref{sec:obsconst}. 

\begin{table}
\caption{Main \izw\ properties used in this study.\label{tab:obs_log}}
\centering
\begin{tabular}{lll}
\hline\hline
Parameter & Value & Reference \\
\hline
Coordinates (J2000) & $09$h$34$m$02.2$s, $+55$d$14$m$28.0$s \\
Distance & $18.2$\,Mpc & (1) \\
 M(H$^0$) (main body) &  $10^8$\,M$_\odot$& (2,3)   \\
 M(H$^0$) (NW) &  $\approx2\times10^7$\,M$_\odot$& (this study)   \\
 M(H$^+$) (NW) &  $3.4\times10^6$\,M$_\odot$& (4)   \\
M$_*$          &  $1-20\times10^6$\,M$_\odot$ &(5,6)   \\ 
M$_*$ (NW)     &  $\sim3\times10^5$\,M$_\odot$& (7) \\
M$_{\rm dust}^{\rm a}$ & $562^{+586}_{-287}$\,M$_\odot$, $724^{+535}_{-308}$\,M$_\odot$ & (6)    \\ 
D/G$^{\rm b}$ &  $5^{+9.3}_{-1.7}\times10^{-4}$ &(6)   \\ 
L$_{\rm FIR}$  & $2.88\pm0.13\times10^{7}$\,L$_\odot$ & (6)     \\
L$_{\rm UV}$  & $1.06\times10^{8}$\,L$_\odot$  &(8)     \\
SFR & $\approx0.1$\,M$_\odot$\,yr$^{-1}$ &(9) \\
\hline
Abundances  &  $12+\log({\rm C/H})=6.59$ & (4)\\
   &        $12+\log({\rm N/H})=5.57$ &  (4)\\
   &        $12+\log({\rm O/H})=7.22$ &  (4)\\
   &        $12+\log({\rm Ne/H})=6.41$ &  (4)\\
     &  $12+\log({\rm Si/H})=6.00$&  (4)  \\
     &  $12+\log({\rm S/H})=5.63$ &  (4) \\
     &   $12+\log({\rm Fe/H})=5.79$ &  (4)\\
     &  $12+\log({\rm Ar/H})=5.79$ &   (4)\\
 \hline
 \end{tabular}\\
\tablebib{(1) \cite{Aloisi07}; see also Section\,\ref{sec:distance}, (2) \cite{Lelli12}, (3) \cite{vanZee98}, (4) \cite{Pequignot08}; see also Section\,\ref{sec:abundances}, (5) \cite{Schneider16}, (6) \cite{Remy15}, (7) \cite{Stasinska99},  (8) \cite{Heckman98}, (9) \cite{Cannon05b}; see also Section\,\ref{sec:sfrlit}.} 
\tablefoottext{a}{The dust mass is given for two dust compositions (\citealt{Galliano11}, ``standard'' and ``AC'') with different emissivities \citep{Remy15}.}
\tablefoottext{b}{Dust-to-gas mass ratio. }
\end{table}

\subsection{Distance}\label{sec:distance}

The distance to \izw\ has been the subject of much debate. Early determinations fell in the range $10-13$\,Mpc \citep{Ostlin00,Izotov04b}. The distance was then revised to $18.2$\,Mpc when using the red giant branch tip \citep{Aloisi07} and $19.0$\,Mpc using Cepheids \citep{Fiorentino10,Marconi10}. The distance used by \citetalias{Pequignot08} for modeling the NW region was $13$\,Mpc although he briefly considered one model at $18.3$\,Mpc. A distance of $18.2$\,Mpc is adopted in our models. An update to the \citetalias{Pequignot08} model using this distance is presented in Section\,\ref{sec:p08} for consistency and for comparison with the present study.

\subsection{Constituents}\label{sec:evcon}

Although \izw\ has been often considered to be a young galaxy, possibly showing its first episode of star formation, several studies have identified an old ($>1$\,Gyr) stellar population (e.g., \citealt{Aloisi07,Annibali13}). 
The current onset of star formation could be due to the merging of dwarfs or sub-damped Lyman $\alpha$ systems, as suggested by the somewhat disrupted H\1\ morphology observed,  for instance, by \cite{Lelli12}.

\begin{figure}
\centering
\includegraphics[angle=0,width=9cm,clip,trim=0 0 0 0]{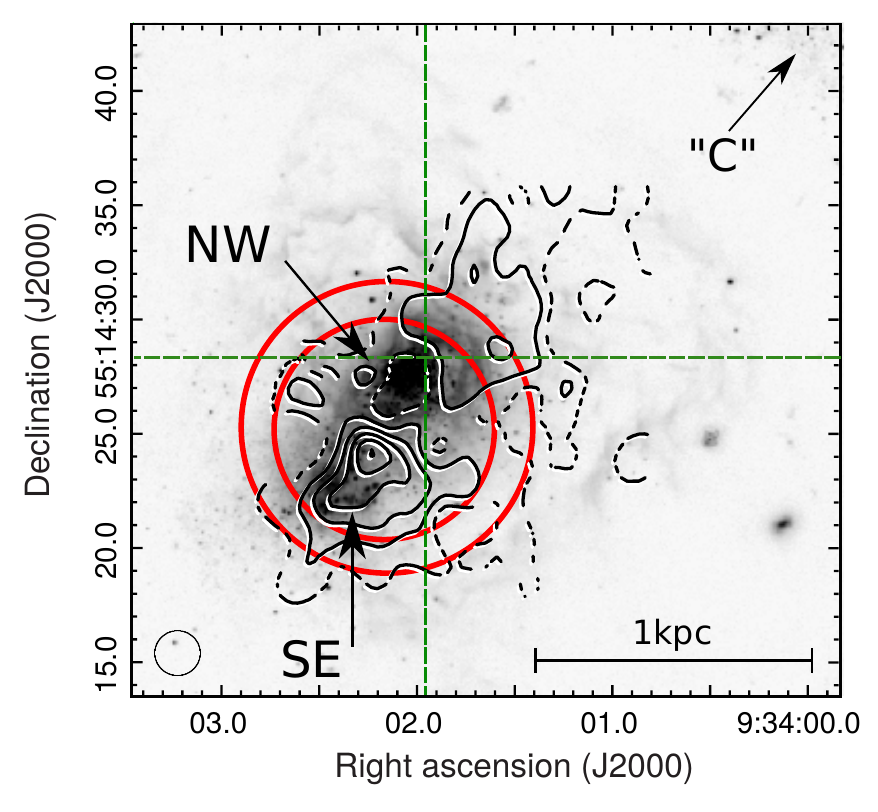}
\caption{H\1\ column density contours from \cite{Lelli12}, with $2$\arcsec\ resolution (beam size in the bottom left). Contours are drawn for $3$ (dashed), $6$, $9$, $12$, and $15\times10^{21}$\,cm$^{-2}$. The largest red circle shows the PACS beam at $157$\mic\ ([C\2]) and the smallest red circle shows the beam at $63$\mic\ ([O\1]). Both beams are centered at the emission centroid derived by the PACS optimal extraction method (Sect.\,\ref{sec:pacsobs}). The green cross shows the location of the X-ray point source (Sect.\,\ref{sec:xray}). The background image is HST/ACS F555W. The NW region coincides with an H\1\ hole. The H\1\ column density peak lies between NW and SE. }
\label{fig:hipacs}
\end{figure}

\izw\ contains a main body and a secondary body. The main body contains two massive stellar clusters, NW and SE (Fig.\,\ref{fig:hipacs}), associated with giant H\2\ regions and surrounded by an irregular and filamentary halo of diffuse ionized gas (e.g., \citealt{Izotov01a}). Although the secondary body is gravitationally bound to the galaxy  \citep{Petrosian97,vanZee98}, it is disconnected from the main body and contains stars that are older on average \citep{Contreras11}. 
Both NW and SE contain a young stellar population but NW has been the more active recently \citep{Contreras11}.

For comparison, the diameter of \izw, as observed at 21\,cm (e.g., \citealt{Lelli12}), is close to that of the \object{Large Magellanic Cloud}, i.e., $\approx7$\,kpc (e.g., \citealt{Kim98,StaveleySmith03}), and the diameter of the NW H\2\ region, ($\approx400$\,pc), is about twice that of the \object{LMC-30 Dor} region.

\subsection{Star formation rate}\label{sec:sfrlit}

\cite{Legrand00} found that a low SFR ($\sim10^{-4}$\,M$_{\odot}$\,yr$^{-1}$) over the Hubble time could explain the metal enrichment of \izw. The instantaneous SFR derived from H$\alpha$ is $0.1-0.2$\,M$_{\odot}$\,yr$^{-1}$ (e.g., \citealt{Dufour90,Petrosian97,Cannon02,Cannon05b}). \cite{deLooze14a} calculated a similar SFR using the combination of FUV and $24$\mic, providing $0.06$\,M$_\odot$\,yr$^{-1}$. \cite{deLooze14a} also investigated the applicability of several FIR lines for tracing SFR and obtained $\approx0.02$\,M$_\odot$\,yr$^{-1}$ using preliminary measurements of [C\2], [O\1], and [O\3] with \textit{Herschel}/PACS. The radio continuum emission, which consists of the combination of thermal free-free emission and synchrotron radiation, provides yet another independent SFR estimate. In \izw, the $1.4$\,GHz emission is dominated by synchrotron radiation \citep{Cannon05b}. Using $L_{\rm 1.4 GHz} = 8\times10^{19}$\,W\,Hz$^{-1}$ \citep{Hunt05} and the SFR calibration from \cite{Bell03}, we obtain $0.13$\,M$_\odot$\,yr$^{-1}$.

\cite{Annibali13} examined color-magnitude diagrams (CMD) from deep \textit{Hubble}/ACS images and found a larger value of $\approx1$\,M$_{\odot}$\,yr$^{-1}$ over the last $10$\,Myr in the most crowded regions (including NW). Part of the discrepancy could be explained by somewhat different timescales probed by each tracer ($\sim100$\,Myr for the FUV vs.\ $\sim10$\,Myr for H$\alpha$ or CMD). Some SFR determinations are sensitive to the escape of ionizing photons from the galaxy, but, considering the large amount of surrounding neutral gas, it is unlikely that this could explain the scatter of the different determinations. Various SFR values are considered to scale the cosmic ray (CR)  ionization rate in Section\,\ref{sec:cosmicrays}.

\subsection{Chemical abundances}\label{sec:abundances}

The H\2 region abundances are adopted (Table\,\ref{tab:obs_log}). A discontinuity seems to exist between these values measured from optical emission lines in the H\2\ regions and those determined from FUV absorption lines in the diffuse H\1\ region \citep{Aloisi03,Lecavelier04,Lebouteiller13a}. According to \cite{Lebouteiller13a}, the oxygen abundance may be slightly lower by $0.18\pm0.16$\,dex ($2\sigma$ error bar) in the H\1\ region. A greater discontinuity might exist for C and Si, but absorption-line saturation prohibits a reliable estimate. The origin of this discontinuity, which is in fact much larger in BCDs that are more metal rich than \izw\ (see summary in \citealt{Lebouteiller09}), is subject to debate. Local self-enrichment of the H\2\ regions was initially proposed by \cite{Kunth86}, but contamination by metal-poor gas along the lines of sight was the favored explanation in \cite{Lebouteiller13a}. We use hereafter the H\2\ region abundances while keeping in mind that all abundances in the H\1\ region may be slightly lower. 

The choice of using individual observed elemental abundances (as opposed to the solar abundance pattern scaled to the metallicity of \izw) has some impact on the IR line ratio interpretation. The C/O abundance ratio in \izw\ is $\approx2.5$ times lower than the solar ratio. In general, C/O tends to decrease with decreasing metallicity in BCDs (e.g., \citealt{Garnett95}), which is consistent with enrichment by massive stars at low metallicity. It can also be noted that the Si/O abundance ratio in \izw\ is about solar (in both the ionized gas and neutral gas, as discussed in \citealt{Lebouteiller13a}), indicating that both elements are produced in the same massive stars, that silicon is not significantly depleted on dust grains, and that [Si\2] may therefore be an important gas coolant (see Sect.\,\ref{sec:obsconst}). 
Iron is not depleted either \citepalias{Pequignot08}. Thus, there is no sign of depletion on dust grains in \izw, which is consistent with the low dust-to-metal ratio in this galaxy \citep{Remy15}.

\section{Observations}\label{sec:obs}

We present here observational data that were either not used or unavailable in \citetalias{Pequignot08}, namely the dust mass and spectral energy distribution (SED), the [C\2] $157$\mic, [O\1] $63$\mic, and [O\3] $88$\mic\ line fluxes from \textit{Herschel}, the suite of \textit{Spitzer} lines remeasured,  recent X-ray observations, and the H$^0$ mass. A summary of the observational constraints used in the models is provided in Section\,\ref{sec:obsconst}.

\subsection{Herschel/PACS}

\subsubsection{Datasets}\label{sec:pacsobs}

\izw\ was observed by \textit{Herschel} as part of the Dwarf Galaxy Survey Key Program (DGS; \citealt{Madden13}). Observations are described in detail in \cite{Cormier15}, we describe in the following specific information relevant to the observation of \izw. The PACS spectroscopy observations were performed in two steps. The [C\2] $157$\mic\ line was observed first in May 2011 (OBSID 1342220973) as part of the SHINING program (PI E.\ Sturm, KPGT\_esturm\_1) for $3.7$\,ks. The [O\1] $63$\mic\ (OBSID 1342253757) and [O\3] $88$\mic\ (OBSID 1342253758) lines were then observed in October 2012 as part of the DGS (PI S.\ Madden, OT2\_smadde01) for $13.8$\,ks and for $4.3$\,ks respectively. The input coordinates for the [C\2] observation were slightly different than the [O\1] and [O\3] observations.

As explained in \cite{Cormier15}, the projection of the PACS array on the sky is a footprint of $5\times5$ spatial pixels (``spaxels''), corresponding to a $\approx47\arcsec\times47\arcsec$ field of view. Each spaxel is $\approx9.4\arcsec$ in size. A single footprint observation was performed since \izw\ appears smaller than the footprint size. According to the PACS Observer's Manual\footnote{\url{http://herschel.esac.esa.int/Docs/PACS/html/pacs_om.html}}, the point spread function (PSF) full width at half maximum (FWHM) ranges from $\approx9.5\arcsec$ ($\approx0.8$\,kpc at the adopted distance of $18.2$\,Mpc) between $55$\mic\ and $100$\mic\ to about $\approx14\arcsec$ at $200$\mic\ ($\approx1.2$\,kpc). The spectral resolution is about $90$, $125$, and $240$\kms\ for [O\1], [O\3], and [C\2] respectively.

The data reduction was performed in HIPE \texttt{12.0} \citep{Ott10} using the default chop/nod pipeline script. The level 1 product (calibrated in flux and in wavelength, with bad pixels masks according to the HIPE reduction criteria) was then exported and processed by our in-house PACSman tool \citep{Lebouteiller12b} for empirical error estimates and line flux extraction.

\begin{figure}
\centering
\includegraphics[angle=0,width=7.5cm,clip,trim=10 10 0 10]{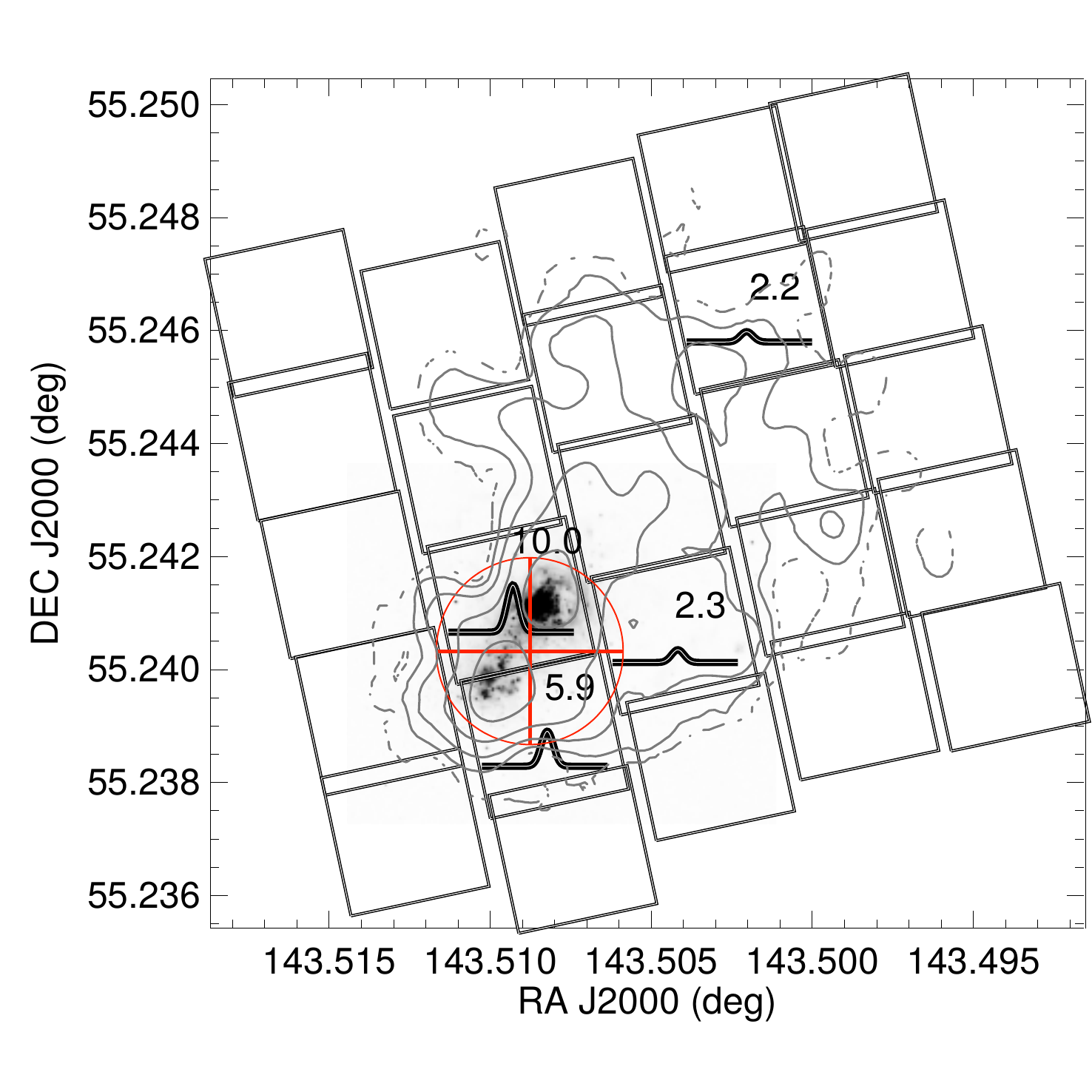}
\includegraphics[angle=0,width=7.5cm,clip,trim=10 10 0 10]{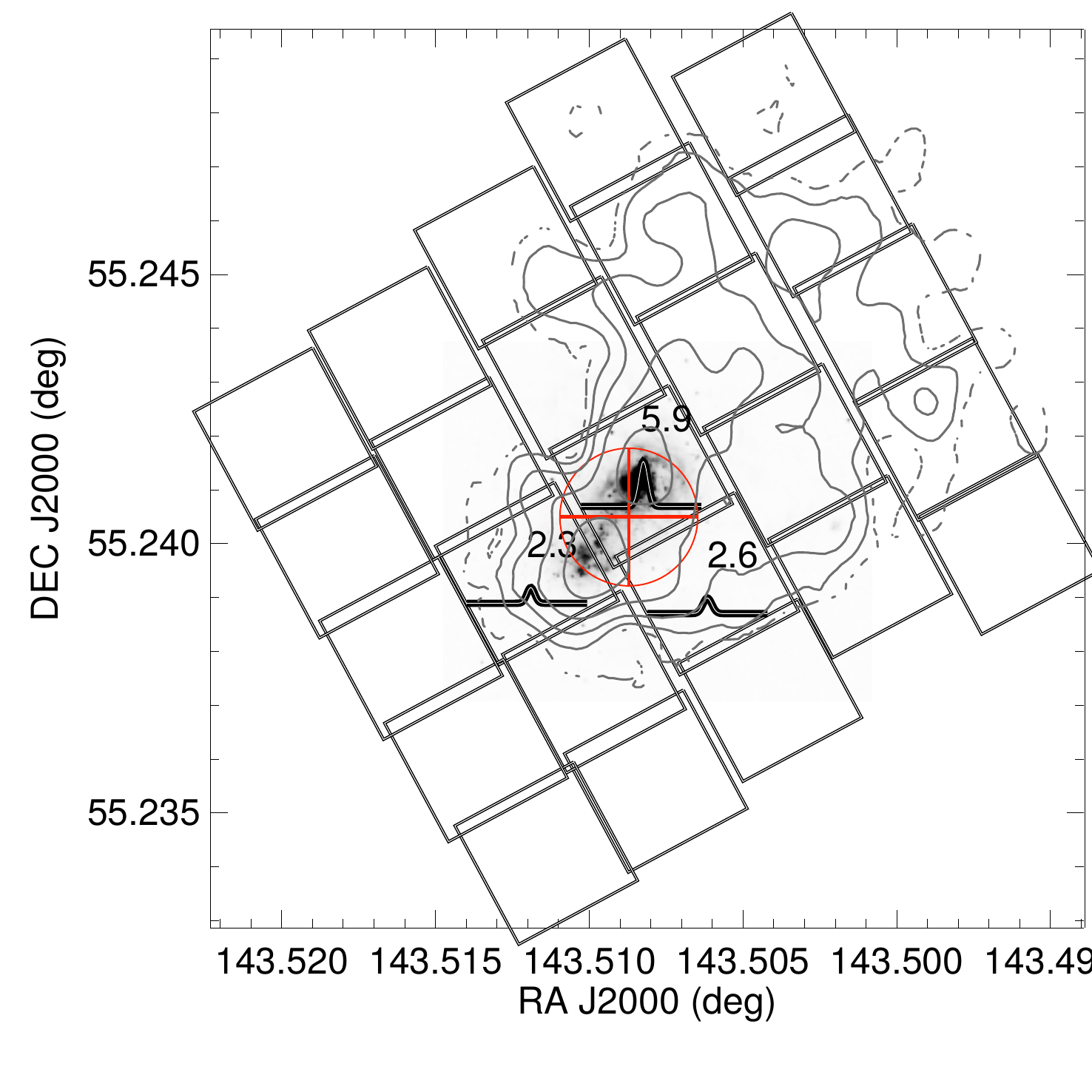}
\includegraphics[angle=0,width=7.5cm,clip,trim=10 10 0 10]{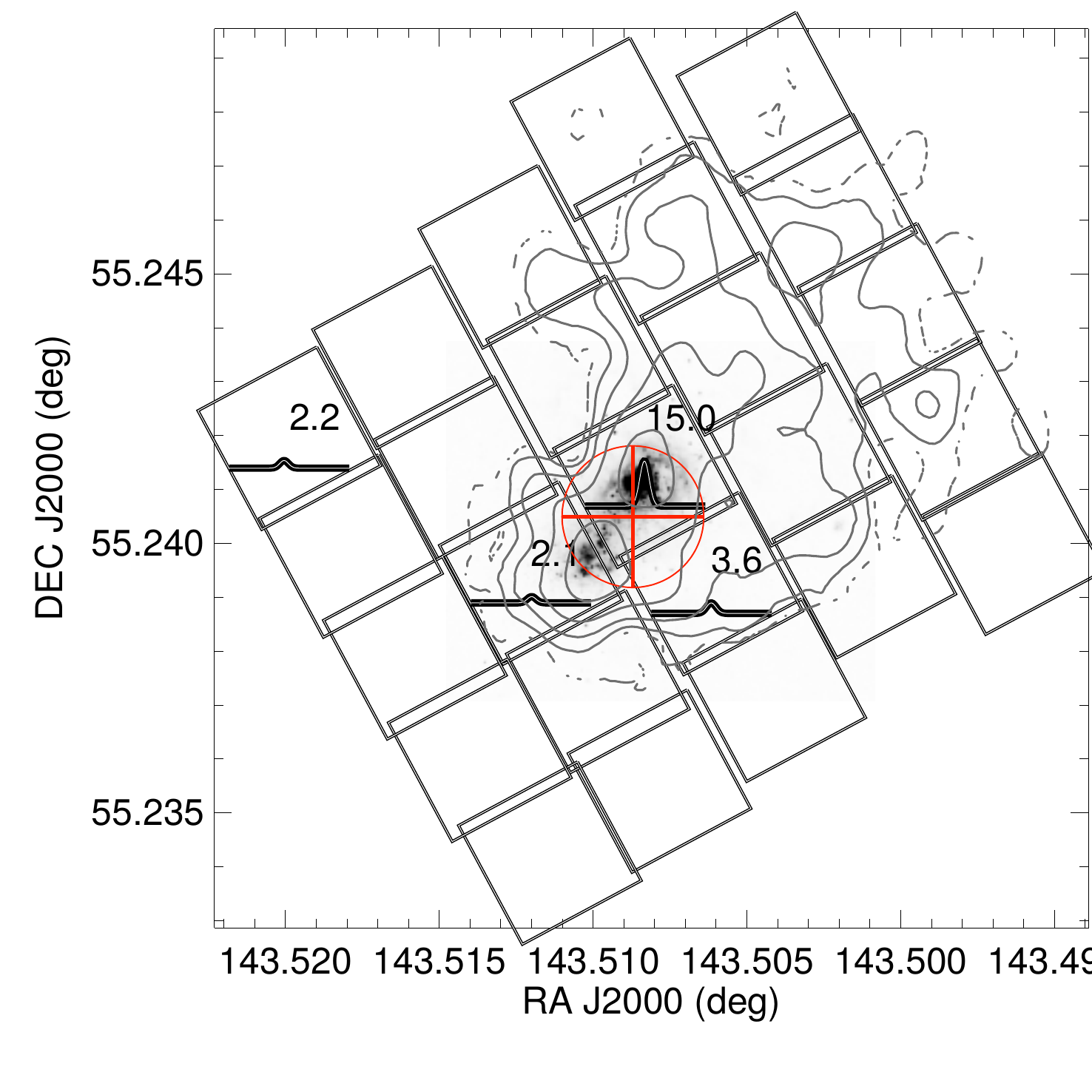}
\caption{\textit{Herschel}/PACS map of [C\2] (top), [O\1] (middle), and [O\3] (bottom) emission in \izw. The $25$ spaxels of the PACS footprint are overplotted on the F555W HST/ACS image. The contours show the H\1\ column density at $5\arcsec$ resolution \citep{Lelli12}. For display purposes, only the fits are shown for each spaxel with detection level $>2\sigma$, and the number indicates the detection level in $\sigma$. The red circle shows the beam size and the red cross shows the emission centroid as calculated by the optimal extraction (Section\,\ref{sec:optimal}). Individual spaxel spectra are presented in Appendix\,\ref{secapp:optimal}.  }
\label{fig:pacs}
\end{figure}

Figure\,\ref{fig:pacs} shows the footprint and line detections. Each line is well detected ($>5\sigma$) in at least one spaxel, and it is always unresolved in velocity. The observed spectra and the line fits are shown in Appendix\,\ref{secapp:optimal}.  
Another, independent, observation of the [O\3] line was performed as a small map, but with a lower integration time (see Appendix\,\ref{sec:oiiimap}), so we decided to use only the pointed observation described here.

\subsubsection{Line fluxes and spatial distribution}\label{sec:optimal}

The line profile in each spaxel is adjusted with a Gaussian component and a flat baseline (Appendix\,\ref{secapp:optimal}). The line width is fixed, constrained by the spectral resolution of the instrument, in particular for the spaxels for which the line is not detected.

Although each line is well detected in at least one spaxel, the low signal-to-noise ratio (S/N) in spaxels corresponding to the wings of the PSF together with the low spatial sampling of the PSF usually prevent an accurate determination of the emission spatial centroid. More specifically, for all lines, most of the spaxels around the brightest spaxel have a detection level $\lesssim3\sigma$ (Fig.\,\ref{fig:pacs}), so it is difficult to pinpoint the peak position or the source spatial shape with an accuracy smaller than the spaxel size ($\approx9.4$\arcsec). The optimal extraction algorithm of PACSman was used to obtain a more accurate estimate of the source centroid. The optimal extraction compares the spatial profile of the source with that of the instrument PSF, accounting for the uncertainties in the line flux measurements in all spaxels (see Appendix\,\ref{secapp:optimal}). We find that the emission is point-like for [C\2], [O\1], and [O\3] with an intrinsic extent $\lesssim6\arcsec$ ($530$\,pc at the adopted distance of $18.2$\,Mpc). We also find a remarkable agreement between the centroids, despite the different map position angle and pointing coordinates (Fig.\,\ref{fig:pacs}). The centroid location and the compact appearance both indicate that the line emission originates within the main body of \izw. The [O\1] and [O\3] observations, with a slightly higher spatial resolution than for [C\2], suggest that the centroid is closer to NW than SE (Fig.\,\ref{fig:pacs}). 
Overall, the low spatial resolution of PACS observations unfortunately prevents us from disentangling the emission of the NW and SE regions. The fluxes we can derive therefore correspond to the global emission (implications for models are discussed in Sect.\,\ref{sec:nwse}).

\cite{Cormier15} provide flux determinations for all DGS objects, including \izw, using various methods. We review these methods in the following and examine their applicability to the \izw\ observation in detail. 

Method $F_1$ scales the flux in the brightest spaxel
by applying a point-source correction (from $25\%$ to $67$\%\ between $50-220$\mic). This method is valid for a point-like source exactly centered in a spaxel. Any deviation from this hypothesis results in underestimating the flux determination. Method $F_1$ provides the best S/N since it uses only the brightest spaxel, but an additional systematic uncertainty exists because of the pointing issues and possible deviation from a point source. 

The second method uses the footprint subarray of $3\times3$ spaxels centered on the brightest spaxel and adds the line fluxes either from all spaxels ($F_{3\times3}$) or only from spaxels with $>3\sigma$ detections ($F'_{3\times3}$). A point-source correction factor is also required, although it is much smaller (from $4$\%\ to $17$\%\ for the range $50-220$\mic) than for $F_1$. The $3\times3$ methods are more reliable than $F_1$ when the source is not well centered in any spaxel, but it may increase the error bar on the flux determination by including spaxels with a low S/N.
The $F'_{3\times3}$ method uses an incomplete sampling of the PSF and therefore results in a lower limit on the flux determination\footnote{Alternative methods consist of combining the spaxel spectra (either $3\times3$ or $5\times5$) \textit{before} performing the line fitting. However, possible variations of the baseline between spaxels and co-addition of spurious features may introduce systematic errors. In any case, the fluxes measured this way, despite large systematic errors, are compatible with our final values. }.

Finally, method $F_{\rm opt}$ performs an optimal extraction by scaling the normalized instrument point-spread function. This is in principle the best method since it reaches a compromise between S/N and the ability to recover the total flux from a source that is not well centered in a spaxel. The flux calibration remains accurate as long as the source is point-like. Details on optimal extraction are given in Appendix\,\ref{secapp:optimal}, where it can be seen that the emission in \izw\ appears point-like. 

Table\,\ref{tab:fluxes} lists the various flux determinations. We consider $F_1$ to be a  lower limit because the emission can never be perfectly centered in any spaxel and we use $F_{\rm opt}$ for our final fluxes. \cite{Cormier15} used the $F_{3\times3}$ method for \izw\ as part of the global and systematic DGS analysis and our revised measurements agree within errors (Table\,\ref{tab:fluxes}). For all lines, we verified that the relatively large error bars of $F_{3\times3}$ encompass the $F_{\rm opt}$ determination. We discuss, in Section\,\ref{sec:scaling}, how PACS fluxes are normalized for comparison with the other tracers used in this study.

\begin{table}
\caption{\textit{Herschel}/PACS line flux determinations.\label{tab:fluxes}}
\centering
\begin{tabular}{llll}
\hline\hline
Method  & [C\2]\,157\mic  & [O\1]\,63\mic & [O\3]\,88\mic  \\
 \hline
\hline
$F_{\rm opt}$                                & $97\pm23$  & $76\pm20$  &  $248\pm31$  \\
$F_1$                                & $(>)\ 71\pm7$    & $(>)\ 69\pm12$  & $(>)\ 207\pm14$  \\
$F_{3\times3}$                              & $108\pm45$ & $143\pm82$  &  $267\pm102$ \\
$F'_{3\times3}$                              & $(>)\ 81\pm16$ & $(>)\ 80\pm21$  &  $(>)\ 217\pm38$ \\
\hline
\cite{Cormier15}$^{\rm a}$ &   $106\pm8$ & $115\pm21$  &  $284\pm34$ \\
\hline
Final                                           & $97\pm25$ & $76\pm25$   &  $248\pm35$  \\
\hline
\hline
 \end{tabular}\\
\tablefoot{Fluxes are given in $10^{-19}$\,\wm. 
}
\tablefoottext{a}{\cite{Cormier15} used the $F_{3\times3}$ method for all lines in \izw. }
\end{table}

\subsection{Spitzer/IRS}\label{sec:spitzerobs}

We used archival data from the Infrared Spectrograph (IRS; \citealt{Houck04a}) on board the \textit{Spitzer} Space Telescope \citep{Werner04} data to measure in a consistent way the suite of lines originating mostly in the ionized gas (e.g., [Ne\2] $12.8$\mic\ and [Ne\3] $15.5$\mic), but also [Si\2]\,$34.8$\mic\ and [Fe\2], which partly originate from the neutral gas. For the low-resolution spectrum (SL and LL modules; $R=\lambda/{\Delta\lambda}\sim57-126$ over $\approx5-14$\mic\ and $\approx14-36.5$\mic,  respectively), the deepest observation available was used (AORkey $16205568$). For the high-resolution spectrum (SH and LH; $R\sim600$ over $\approx10-19.5$\mic\ and $\approx20-36.5$\mic\ respectively), AORkey $16205568$ was used together with $9008640$ and $12622848$, which are shallower but less affected by bad pixels in some spectral regions. These observations were all performed in staring mode, in which the source is observed in two nod positions. The PSF FWHM ranges between $\approx2\arcsec$ at $5$\mic\ to $\approx11\arcsec$ at $38$\mic.

\begin{figure}
\centering
\includegraphics[angle=0,width=8cm,height=5.3cm,trim=15 40 0 0,clip=true]{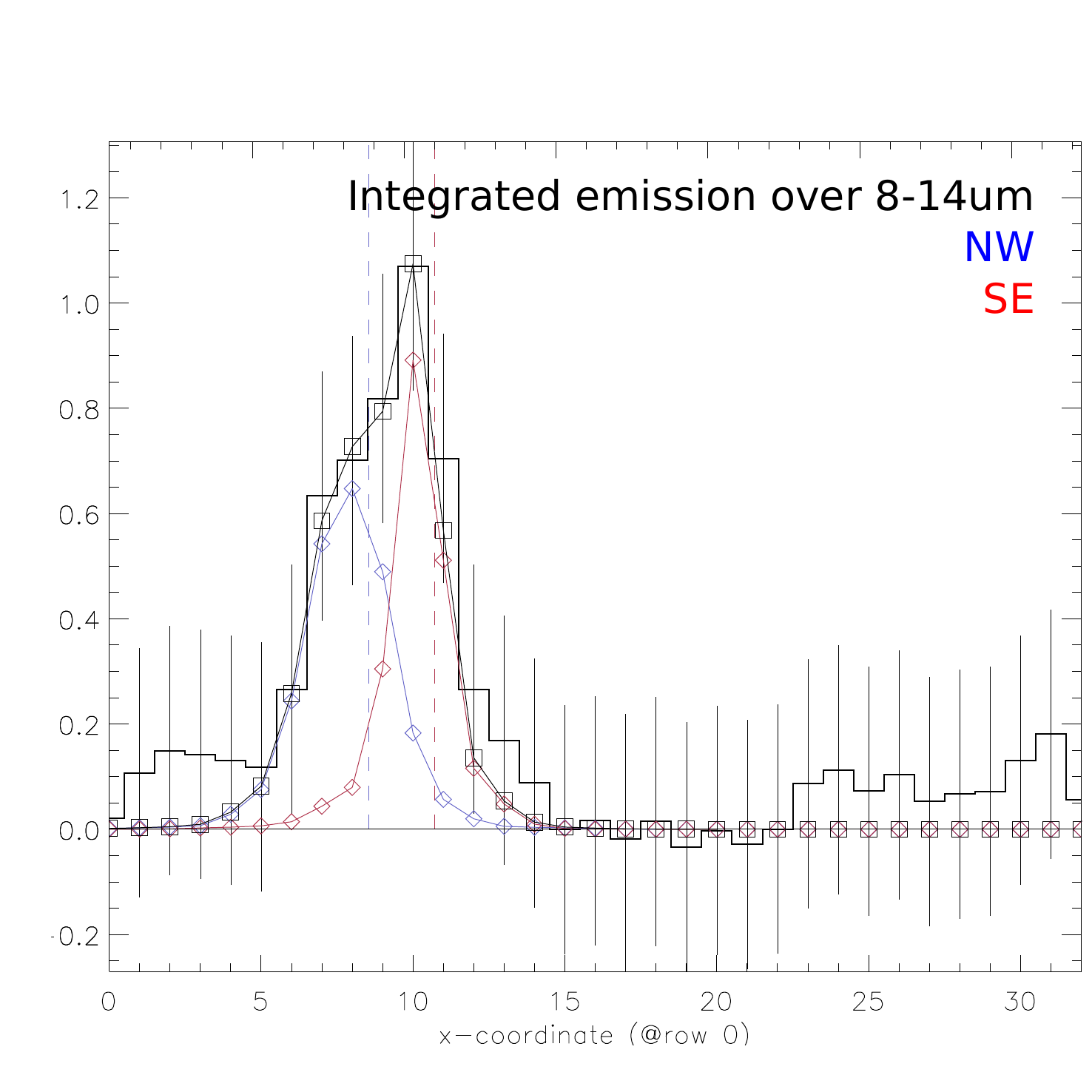}
\includegraphics[angle=0,width=8cm,height=5.3cm,trim=15 10 0 40,clip=true]{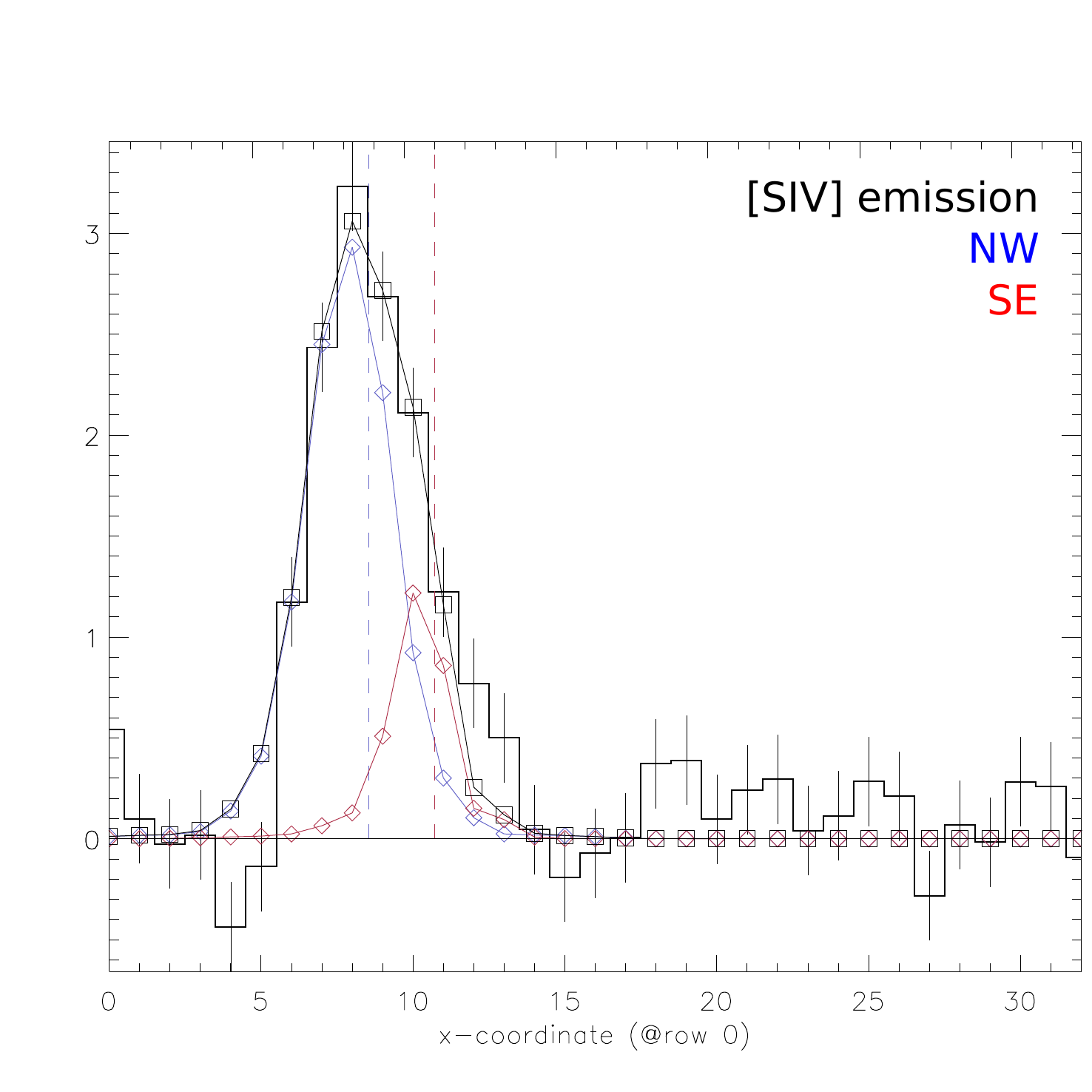}
\caption{Cross-dispersion profiles along the SL slit of the \textit{Spitzer}/IRS observation 16205568. The histogram shows the emission as a function of the spatial position along the slit, in pixel units ($1$\,px=$1.8$\arcsec). The spatial profile is modeled by two slightly extended sources, one of which corresponds to NW (blue, left) and the other to SE (red, right). The profile is shown for the entire integrated spectral order of module SL1 ($\sim8-14$\mic; \textit{top}) and for the [S\4]\,$10.5$\mic\ line only (\textit{bottom}). Left corresponds to north in Figure\,\ref{fig:irs_spatial}.   }
\label{fig:irs_spatial_prof}
\end{figure}

\begin{figure}
\centering
\includegraphics[angle=0,width=8.3cm,clip=true]{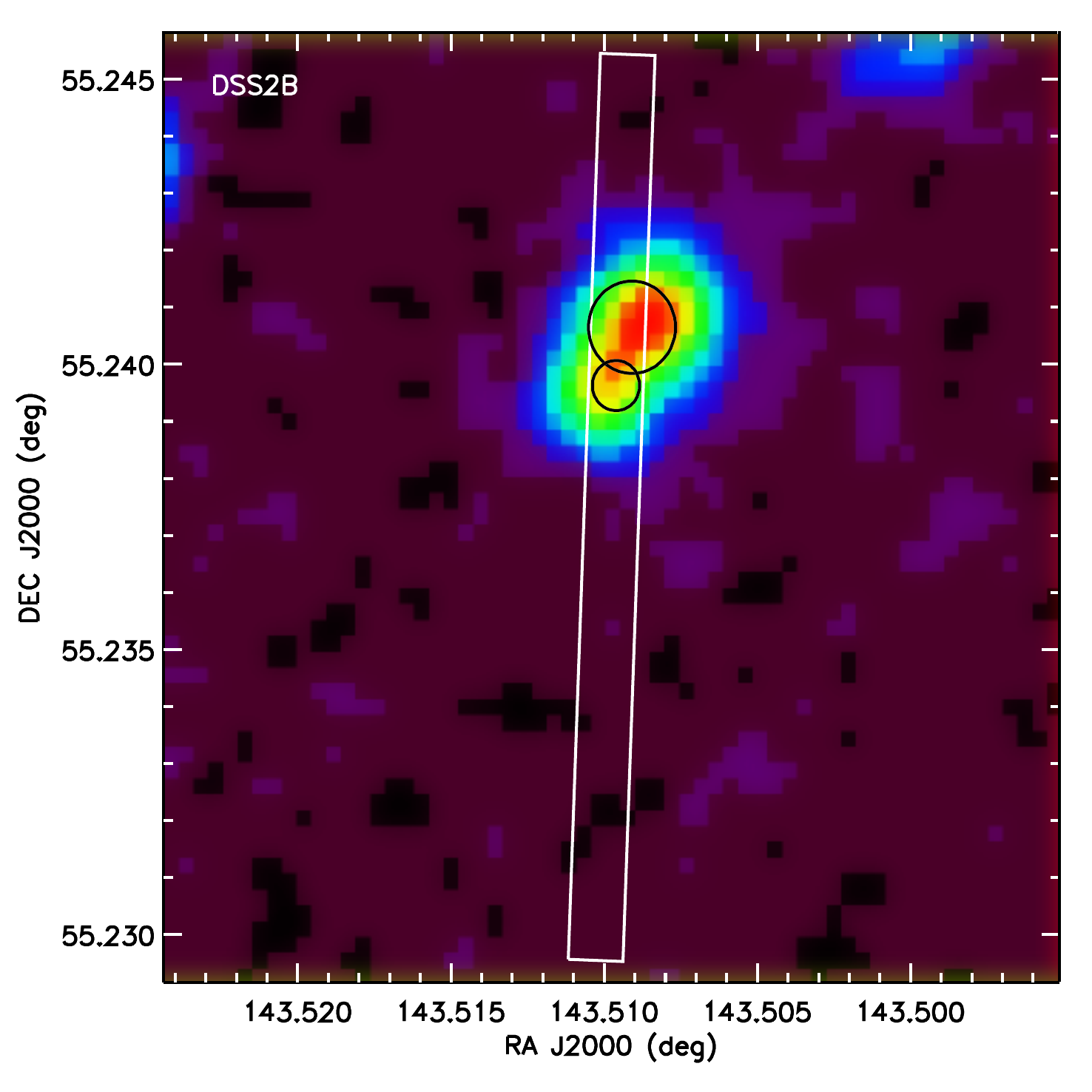}
\caption{Spatial positions of the two components seen in the IRS SL slit. The slit is shown with the white rectangle and the circles indicate the locations of the sources shown in Figure\,\ref{fig:irs_spatial_prof}. The size of the circles corresponds to the total FWHM of each source (i.e., including both the instrument PSF and the intrinsic broadening). 
The background image is the B band from the DSS2 survey.  }
\label{fig:irs_spatial}
\end{figure}

The investigation of the spatial profiles of AORkey $16205568$ in the cross-dispersion direction (Figs.\,\ref{fig:irs_spatial_prof} and \ref{fig:irs_spatial}) shows that the source appears somewhat extended in SL and SH (about $6\arcsec$ FWHM). One can distinguish two components in the SL profile. One component is located at 09h34m02.29s/+55$^\circ$14$\arcmin$27.52$\arcsec$, coinciding with NW, with a prominent [S\4] line and relatively shallow continuum, while the other component is located at 09h34m02.31s/55$^\circ$14$\arcmin$22.67$\arcsec$, coinciding with SE, with a much weaker [S\4] line and a relatively steeper continuum. The NW component is responsible for $\approx75$\%\ of the total [S\4] and $\approx68$\%\ of the total H\1\ recombination line Hu$\alpha$ 12.37\mic\ (see fluxes in Appendix\,\ref{secapp:spitzer}). These values are in good agreement with the H$\beta$ fraction originating from NW, $\approx78\%$ \citep{Skillman93}. 
The NW component appears extended in the SL module while the SE component appears quasi-point-like (intrinsic broadening of $\approx5\arcsec$ and $\approx1\arcsec$, respectively). The NW and SE components are not distinguishable in the LL and LH modules because of the relatively lower spatial resolution.

Line measurements are described in Appendix\,\ref{secapp:spitzer}, where a comparison is performed with \citet[hereafter W07]{Wu07}. \citetalias{Pequignot08} already compared photoionization models to the \citetalias{Wu07} fluxes and found an overall good agreement. However, \citetalias{Pequignot08} noted that the [S\3] 33.5\mic\ flux measured by \citetalias{Wu07} is more than a factor of $2$ larger than predictions; based on a similar discrepancy for the [Si\2] $34.8$\mic\ line arising in the same IRS module as [S\3] $33.5$\mic, \citetalias{Pequignot08} proposed that the fluxes of these two lines were overestimated.
We find new [S\3] $33.5$\mic\ and [Si\2] fluxes that are lower by factors of $2.3$ and $2.1$, respectively, as compared to \citetalias{Wu07}, thereby confirming the hypothesis of \citetalias{Pequignot08}. Our revised measurement of both [S\3] lines at $18.7$\mic\ and $33.5$\mic\ confirms that the gas density in the ionized gas is well below $1000$\cc. 
Furthermore, the [S\4] flux predicted by \citetalias{Pequignot08} is significantly larger than in \citetalias{Wu07}. Our measured value confirms the low value of \citetalias{Wu07}, hinting that the problem may be due to a doubtful S$^{2+}$ di-electronic recombination coefficient (see \citetalias{Pequignot08} for more details). 

The upper limit on the PAH emission was calculated using the deep low-resolution observation 16205568. We fitted the latter spectrum with the model of \cite{Galliano11}. The PAH optical properties are from \cite{Draine07a}. Rather than fitting individual PAH emission bands, we have adjusted a template with two PAH components (neutral and ionized) with fixed properties. The upper limit on the PAH emission in the range $6-15$\mic\ is $\leq1.9\times10^{-17}$\wm\ with little influence from the neutral/ionized PAH mixture. 
We review in Section\,\ref{sec:scaling} the extracted line fluxes from \textit{Spitzer} and \textit{Herschel} and how they are used as constraints for the models.

\subsection{X-ray observations}\label{sec:xray}

Following \citetalias{Pequignot08}, X-rays are to be considered a promising heating mechanism in the H\1\ region of \izw. We describe here the various X-ray measurements, and in particular the most recent observation by XMM-\textit{Newton}. 

\izw\ was observed with ROSAT with two instruments, first with PSPC (Position Sensitive Proportional Counters) in 1992 and then with the High Resolution Imager (HRI) in 1997. \cite{Fourniol96} reported the PSPC detection of an unresolved source and calculated an unabsorbed X-ray luminosity of $L_{\rm X}=6.2_{-4.9}^{+2.3}\times10^{39}$\,erg\,s$^{-1}$ between $0.1-2.4$\,keV. \cite{Martin96} independently examined the same observation, calculating a lower limit of $L_{\rm X}\gtrsim1\times10^{39}$\,erg\,s$^{-1}$ with only Galactic absorption considered. The higher spatial resolution enabled by the subsequent HRI observations led \cite{Bomans02} to conclude to the existence of both a point source located in NW and a diffuse component. 

\izw\ was later observed with \textit{Chandra} in 2000 and X-ray Multi-Mirror Mission (XMM-\textit{Newton}) in 2002. The \textit{Chandra} observation, first reported in \cite{Bomans02}, was analyzed in detail by \cite{Thuan04b}. As noted by \cite{Thuan04b}, the X-ray emission from \izw\ is dominated by a single point source associated with the NW region\footnote{\cite{Ott05a} reported a different position, located at the edge of the H$\alpha$ nebula, somewhat offset from the NW stellar cluster. While the coordinates of the X-ray source reported in their Table\,6 does in fact coincide with the NW cluster, their interpretation seems to rely on incorrect astrometry of HST images. } (see position in Fig.\,\ref{fig:hipacs}). The point-source luminosity is $3\times10^{39}$\,erg\,s$^{-1}$ in the $0.5-10$\,keV range (value renormalized to a distance of $18.2$\,Mpc). Faint diffuse emission was also detected but contributed to at most $4$\%\ of the point-source flux. 

\cite{Kaaret13} analyzed the XMM-\textit{Newton} observation and found a larger flux, $1.4\times10^{40}$\,erg\,s$^{-1}$, as compared to the \textit{Chandra} observation. Based on this increased flux, \cite{Kaaret13} suggested that the X-ray point source emission is likely dominated by a single X-ray binary, as already proposed by \cite{Thuan04b}. The X-ray source in \izw\ is located precisely within the NW cluster, according to the relative position with several quasars in the XMM-\textit{Newton} images, whose positions are known with high accuracy (Pakull et al.\ in prep.; Fig.\,\ref{fig:hipacs}). 

Diffuse X-ray emission might be present, possibly associated with a supernova (SN) cavity between the NW and SE regions \citep{Thuan04b}. The luminosity of the extended X-ray component measured by \cite{Thuan04b} is about $10^{38}$\,erg\,s$^{-1}$ (value renormalized to a distance of $18.2$\,Mpc). 
Although their studies are based on the same \textit{Chandra} data, \cite{Ott05a} and \cite{Kaaret11} both argue against the presence of detectable diffuse X-ray emission. 
In the following, only the X-ray point source is considered. The use of the X-ray observation as a constraint to our models is discussed in Section\,\ref{sec:const_xrays}.

\subsection{Photometry}\label{sec:photometry}

\begin{figure}
\centering
\includegraphics[width=9cm,height=12cm,clip,trim=0 0 0 0]{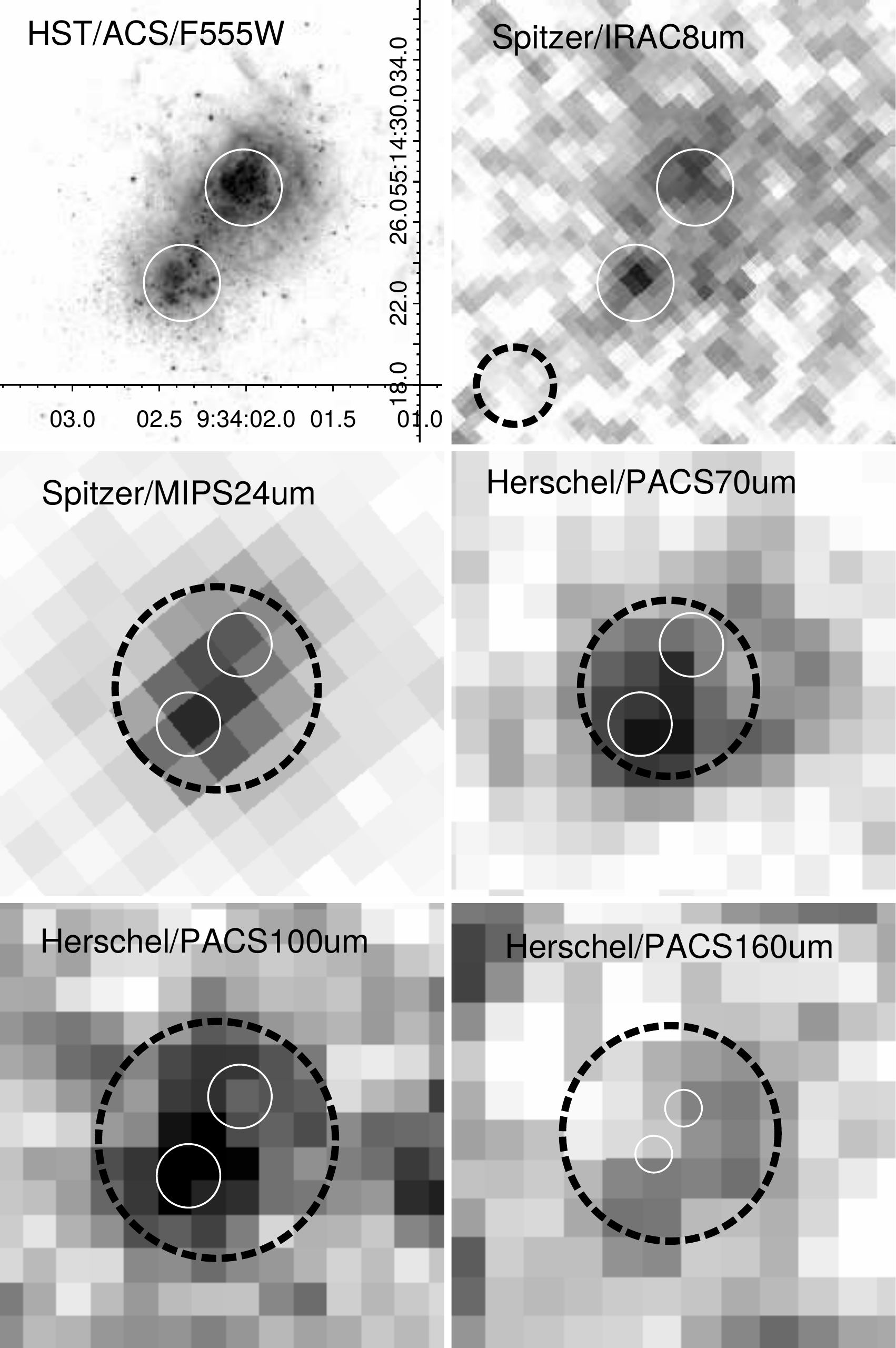}
\caption{Photometric maps of \izw\ with \textit{Spitzer} and \textit{Herschel} \citep{Remy15}. The first panel shows the HST/ACS F555W image for reference, with the NW and SE regions circled in white. The dashed black circle indicates the beam size (from $2$\arcsec\ for IRAC $8$\mic\ to $12$\arcsec\ for PACS $160$\mic). The 8\mic\ band is not dominated by PAH emission in \izw\ but by stochastically heated small grains and warm grains in thermal equilibrium with the interstellar radiation field. }
\label{fig:dustmaps}
\end{figure}

Photometry data is used in our study to examine the predicted SED from the models, in particular in the IR range. We use the photometry data measurements from radio to IR (with 2MASS, \textit{Spitzer}/IRAC, WISE, \textit{Herschel}/PACS, and  \textit{Herschel}/SPIRE) compiled in \cite{Remy15}. Figure\,\ref{fig:dustmaps} shows the \textit{Spitzer} and \textit{Herschel} dust emission maps. The measured photometry corresponds to the main body emission, and the NW and SE regions cannot be disentangled. 
Optical to FUV measurements (within a $20$\arcsec\ aperture), taken from the NED archival data, are only used for illustrative purposes.

\subsection{H\1\ observations}

While \citetalias{Pequignot08} explored the H\1\ region heating in \izw, the H$^0$ mass was not explicitly considered. The H$^0$ mass, notably responsible for the [C\2] emission, provides an important constraint to our models (Sect.\,\ref{sec:const_himass}). Here the interferometric H\1\ $21$\,cm observations of \cite{Lelli12} are adopted.

\section{Model description}\label{sec:modelprep}

Our objective is to build a photoionization model
for the incomplete H\2+H\1\ region shell of \izw-NW that takes into account new observational constraints presented in Section\,\ref{sec:obs}. In the following we describe the model components (Sect.\,\ref{sec:modeltopo}) and present the Nebu and Cloudy models (Sect.\,\ref{sec:p08}).

\subsection{Model topology}\label{sec:modeltopo}

\begin{figure}
\includegraphics[angle=0,width=8cm,clip=true,trim=10 300 340 10]{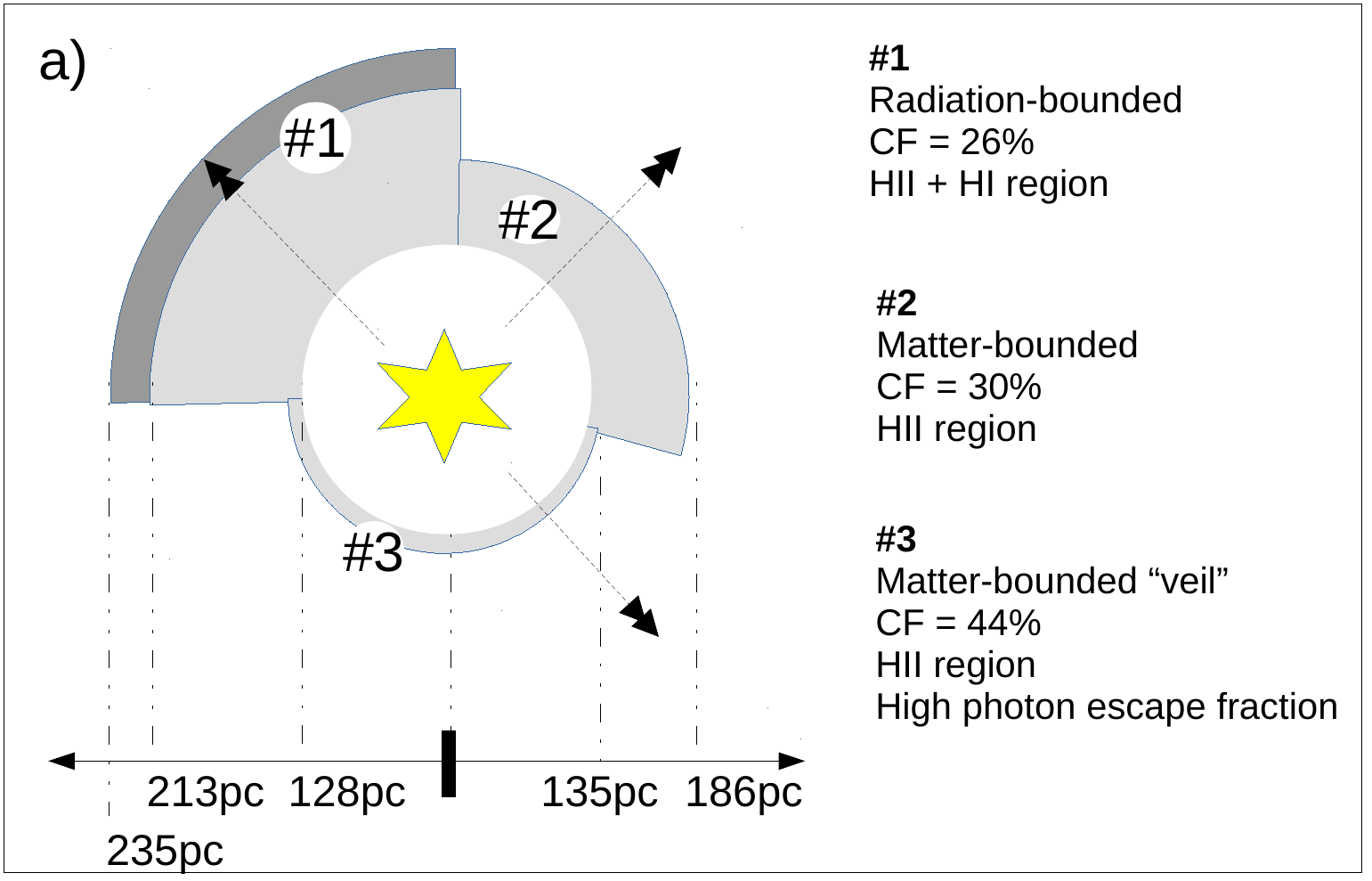}
\includegraphics[angle=0,width=8cm,clip=true,trim=10 350 340 10]{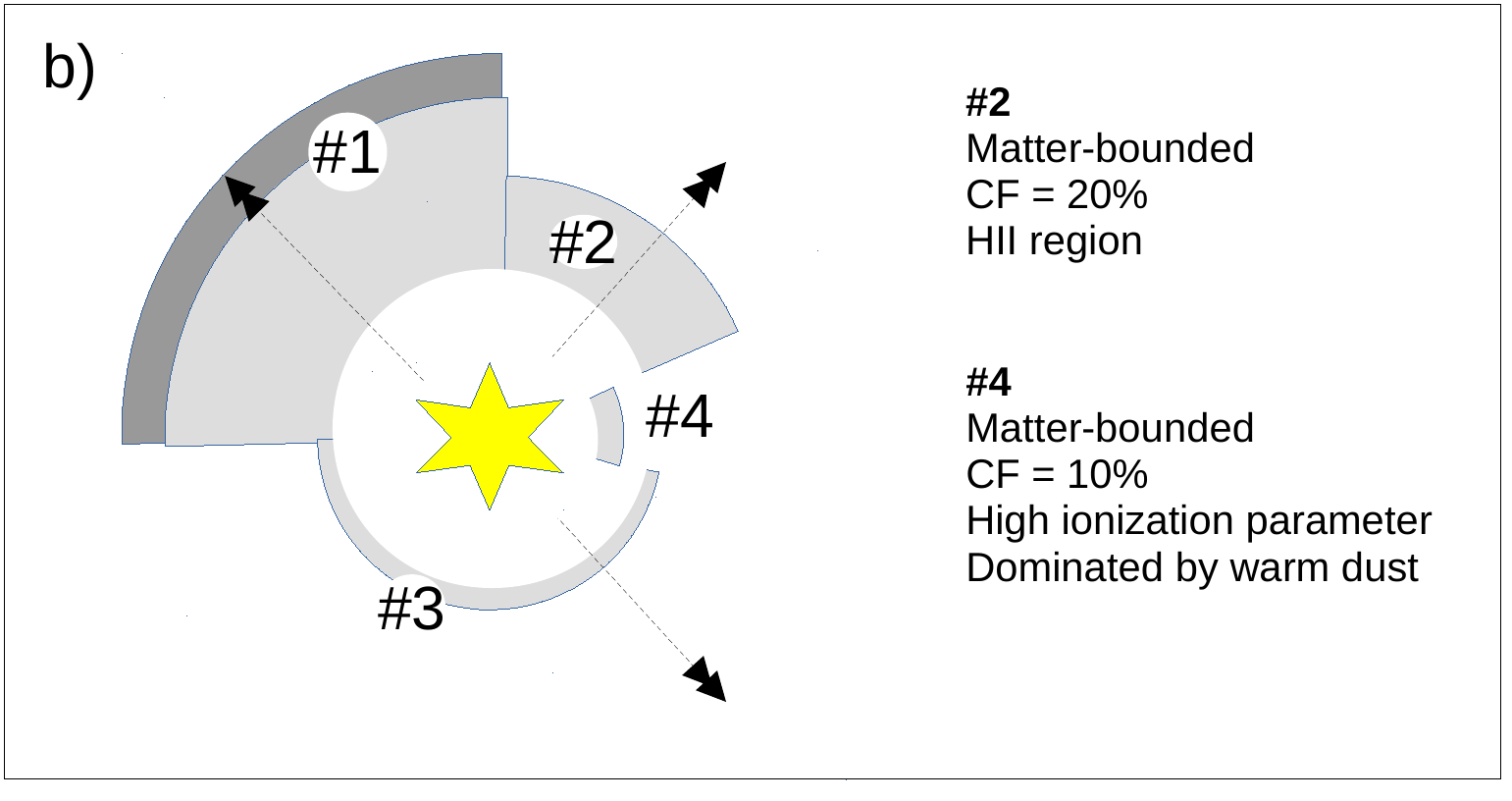}
\includegraphics[angle=0,width=8cm,clip=true,trim=10 350 340 10]{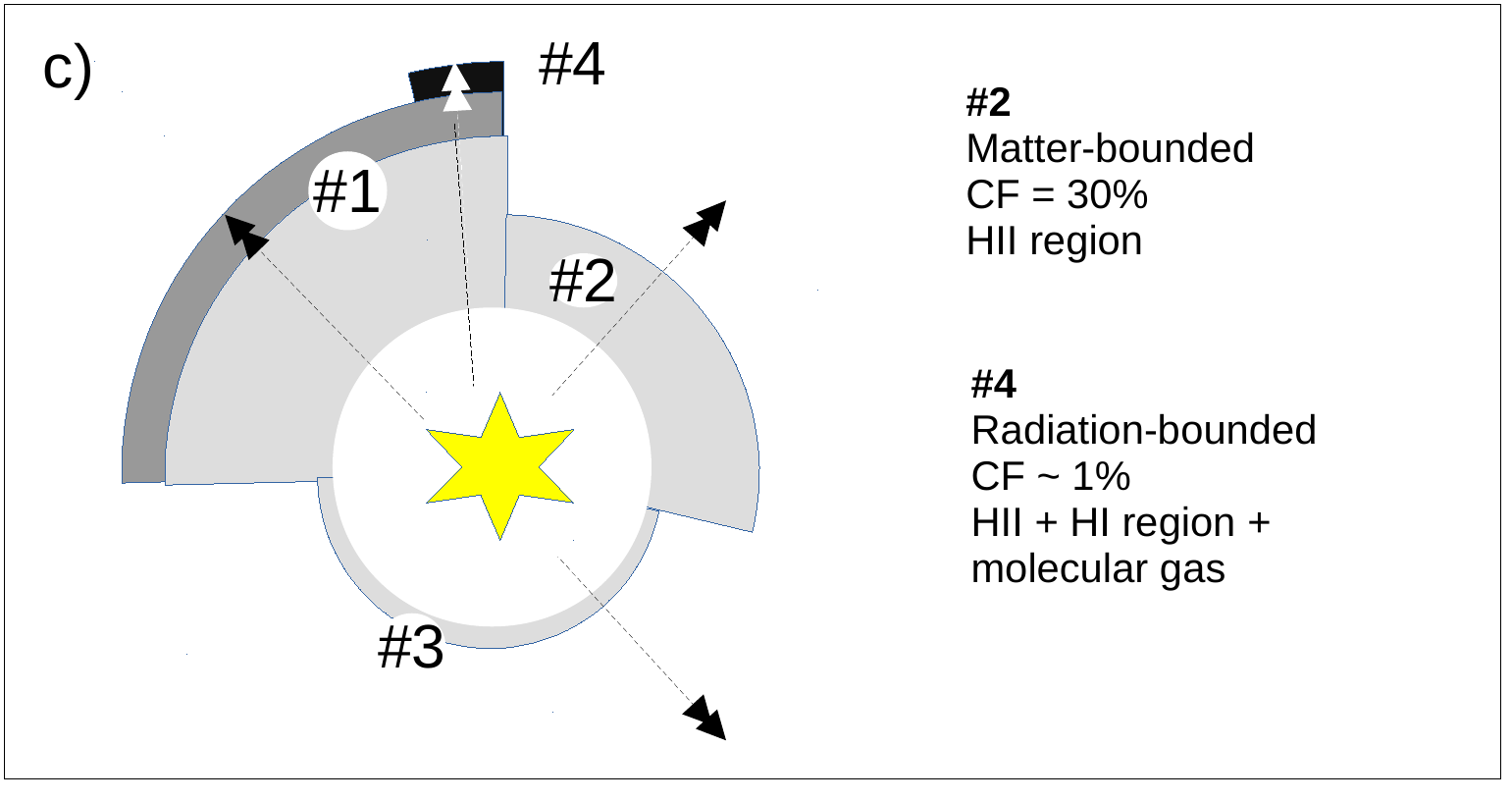}
\caption{Description of the sectors used in our models. The first panel shows the topology used in \citetalias{Pequignot08} and in most of our models (Table\,\ref{tab:models}). The arrows reaching out of sectors \#2 and \#3 illustrate the fact that the fraction of escaping photon is significant. Optical depths at $1$\,Ryd are $1.18$ and $0.05$ for sectors \#2 and \#3 respectively. ``CF'' stands for covering factor. The sectors are drawn according to their respective covering factors but in reality the lines of sight are intermixed.  }
\label{fig:sectors}
\end{figure}

In the description of \izw-NW by \citetalias{Pequignot08}, the primary radiation sources are (1) the central young star cluster (responsible for the H\2\ region shell) and (2) the point-like X-ray source (responsible for a partially ionized warm H\1\ region beyond the ionization front of the H\2\ region), resulting in a relatively simple geometry with a central UV+X source. The geometry is open, with $\approx65$\%\ of the ionizing photons actually escaping the NW region through two different (FUV-)matter-bounded sectors (deprived of ionization fronts), while all photons (except for the hardest ones) are absorbed in the radiation-bounded sector (Fig.\,\ref{fig:sectors}a). 
Both the Nebu and Cloudy calculations are performed in spherical symmetry (with no absorption by the far side) and the output parameters are post-processed using the covering factor of each sector. The model chemical composition is given in Table\,\ref{tab:obs_log}.

Based on H$\alpha$ morphology, and following \citetalias{Pequignot08}, the inner and outer radii of the H$^+$ shell are set to $1.5\arcsec$ and $2.5\arcsec$, respectively ($\approx130$\,pc and $\approx215$\,pc at $18.2$\,Mpc). Although the H\1\ 21\,cm emission extends over several kiloparsecs from the stellar cluster (e.g., \citealt{vanZee98}), the line and continuum IR emission seems to be concentrated within $\sim6\arcsec$ (Sects.\,\ref{sec:optimal}, \ref{sec:spitzerobs}), i.e., $530$\,pc at $18.2$\,Mpc. In the models, the H\1\ region extends from $215$ to $240$\,pc. The model diameter thus obtained is on the same order as the observed IR extent. In other words, the \citetalias{Pequignot08} model, focused on the NW H\2\ region, turns out to be suited for the IR emission as well since the latter is intimately linked to the H\2\ region (see also Sect.\,\ref{sec:nwse}). 

As in \citetalias{Pequignot08}, the radial density profile is obtained from a thermal pressure law as a function of radial optical depth at $1$\,Ryd. We refer to \citetalias{Pequignot08} for details about model convergence. The same density profile is used for all sectors, but the calculation is stopped in each matter-bounded sector at a given optical depth. The density then levels off in the H\1\ zone of the radiation-bounded sector, which ends at a given temperature.

\subsection{New Nebu and Cloudy models}\label{sec:p08}

The H\2\ and H\1\ region model selected in \citetalias{Pequignot08} ($M2_{\rm X}$) needs to be updated\footnote{The secondary ionization by X-ray photoelectrons was inadvertently skipped in the \citetalias{Pequignot08} computation. Secondary ionizations result in more energy going into ionization, at the expense of the thermal energy available for line excitation in the H\1\ region, so the predicted line fluxes become smaller than reported in \citetalias{Pequignot08} for a given primary X-ray luminosity.}  before it can be used to build an equivalent model using Cloudy. Some important input parameters and free parameters are summarized in Table\,\ref{tab:modelparams}. Several parameters are introduced here in order to manage the new heating processes and diagnostics, which are  studied by means of Cloudy (Sects.\,\ref{sec:cloudy}, \ref{sec:discussion}). We use Cloudy version c13.03\footnote{The default atomic data of Cloudy version c13.03 are used except in that \cite{Abrahamsson07}, also used in \citetalias{Pequignot08}, is preferred to \cite{Launay77b} for the critical collisional excitation ${\rm O}^0(^3P)+{\rm H}^0(^2S)$. The more recent rate is many times larger than the old rate.}  \citep{Ferland13}. 

The distance is now $18.2$\,Mpc instead of $13$\,Mpc with the primary luminosity and initial/final H\2\ region radii increased accordingly. The helium abundance by number is now He/H$=0.085$ (instead of $0.080$), as in one trial calculation of \citetalias{Pequignot08}. 

\begin{table}
\caption{Model summary and updates from \citetalias{Pequignot08}.\label{tab:modelparams}}
\centering
\begin{tabular}{lll}
\hline\hline
Parameter & Value/comment  \\
\hline
\textbf{Topology} \\
Number of sectors & Fixed (3 to 5) \\
Covering factors & Fixed, except for dense clump sector \\
Inner radius & Fixed ($130$\,pc) \\
Ionization front & Fixed ($220$\,pc) \\
\hline
\textbf{Physical conditions} \\
Density & Pressure law, maximum value free \\
Stopping temperature  & Minimum value free \\
(outer radius of H\1\ region) & (maximum value free) \\
\hline
\textbf{Radiation field} \\
Blackbody &  Fixed ($\log L=41.58$, $T=4\times10^4$\,K) \\
Blackbody$^\textrm{a}$ &  Fixed ($\log L=41.58$, $T=8\times10^4$\,K) \\
X-rays   & \texttt{diskbb} spectrum, luminosity free \\
Optical-UV & Added in present study (Fig.\,\ref{fig:radiationfield}) \\
CR & Several ionization rate values tested \\
\hline
\textbf{Chemical composition} \\
Metal abundances & Fixed (see Table\,\ref{tab:obs_log}) \\
             & He/H=$0.085$ (updated from \citetalias{Pequignot08}) \\
D/G & Several values tested \\
\hline
\end{tabular}\\
\tablefoottext{a}{Scaled down for energies $\geq4$\,Ryd to mimic the discontinuity observed in model stars (see \citetalias{Pequignot08} for details). }
\end{table}

Owing to the larger distance, the radial density profile, $n_{\rm H}(r)$, must be updated. After convergence, the new parameters in expression (1) of \citetalias{Pequignot08} are as follows: $P_{\rm in}=2.8$, $P_{\rm out}=24$ (in $/k/10^5$ CGS), and $\tau_{\rm c}=4.2$. The $n_{\rm H}(r)$ obtained from Nebu is numerically introduced step by step in the Cloudy computation. 

\begin{figure}
\includegraphics[angle=0,width=9cm,clip=true]{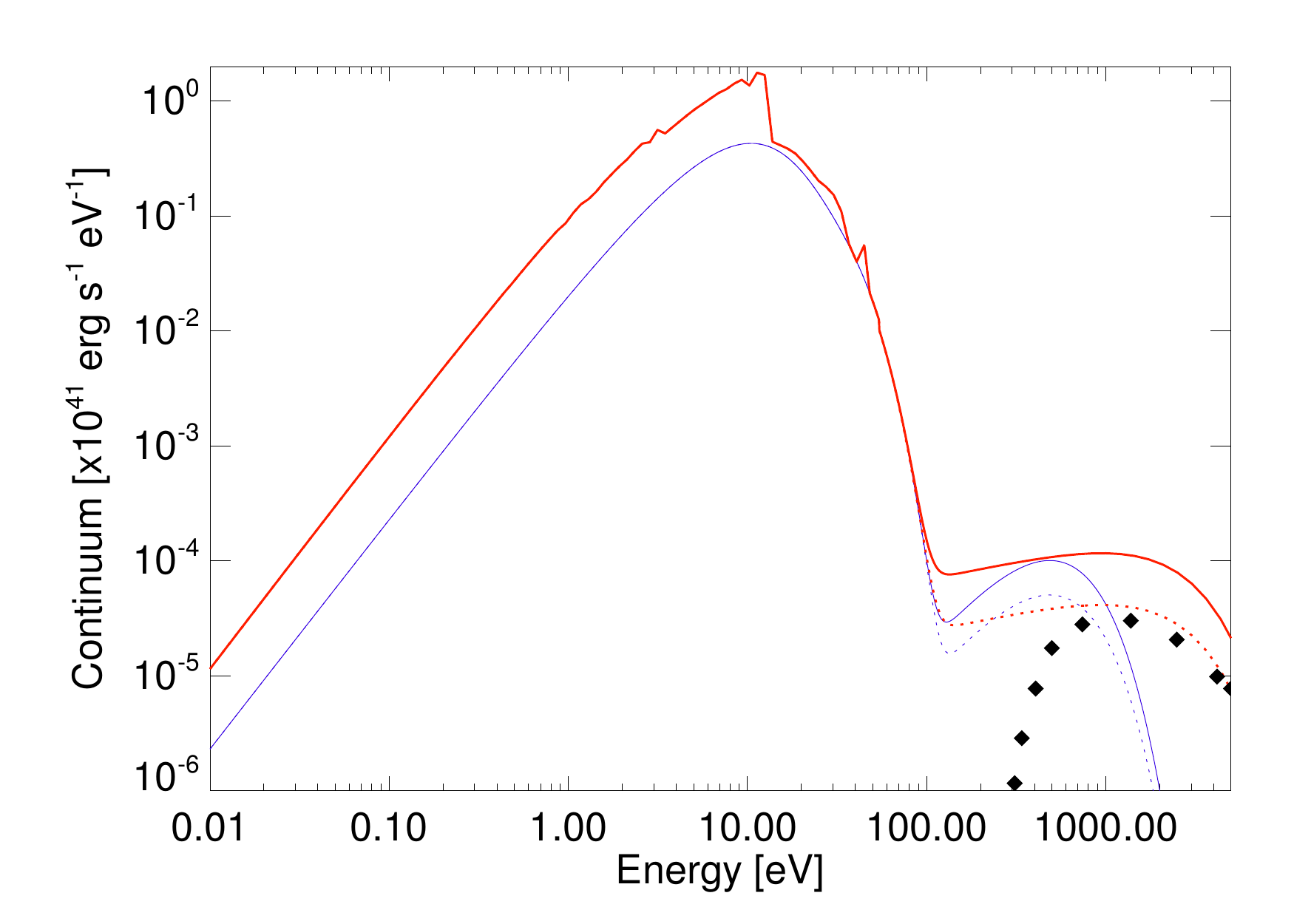}
\caption{Blue curves show the input radiation field used in \citetalias{Pequignot08}, comprised of three blackbodies with temperature $4\times10^4$, $8\times10^4$, and $2\times10^6$\,K. The dashed curve is for $L_{\rm X}=4\times10^{39}$\,erg\,s$^{-1}$ (\citetalias{Pequignot08} model M2$_{\rm X}$) and the solid curve is for $L_{\rm X}=8\times10^{39}$\,erg\,s$^{-1}$ (\citetalias{Pequignot08} model M2$_{\rm X2}$). The red curves show the radiation field used in this study, with additional UV-optical contribution (from \citealt{Lebouteiller13a}) and with an improved X-ray spectrum prescription as compared to \citetalias{Pequignot08}. The dashed curve is for $L_{\rm X}=1.4\times10^{40}$\,erg\,s$^{-1}$ (luminosity inferred from observations) and the solid curve is for $L_{\rm X}=4\times10^{40}$\,erg\,s$^{-1}$ (adopted standard). The black diamonds show the unfolded XMM-\textit{Newton} spectrum (see Section\,\ref{sec:const_xrays} for more details). }
\label{fig:radiationfield}
\end{figure}

As in \citetalias{Pequignot08}, the FUV radiation field is described as the sum of two blackbodies of similar power, with temperatures $4\times10^4$\,K and $8\times10^4$\,K,  respectively. 
Unlike in \citetalias{Pequignot08}, a realistic stellar continuum is implemented below $1$\,Ryd. Optical and UV photons contribute to the photoelectric heating on dust grains and to the photoionization of several neutral species (C$^0$, Si$^0$...), thereby influencing chemistry processes. The optical+UV continuum prescription used by \cite{Lebouteiller13a} is adopted. The radiation fields used in \citetalias{Pequignot08} and in the present study are compared in Figure\,\ref{fig:radiationfield}.

Although the usual signatures of Wolf-Rayet (WR) stars are relatively discrete in \izw-NW (e.g., \citealt{Brown02,Kehrig15}), \citetalias{Pequignot08} considered that the observed nebular He\2\ recombination emission was essentially due to hot stars.
It is now admitted that many low-metallicity massive stars could evolve into either very hot WR stars with weak winds (e.g., \citealt{Crowther06}) or else into chemically homogeneous transparent wind ultraviolet intense stars (``TWUIN stars'';  \citealt{Szecsi15}); these are both almost undetectable in the optical, but emit a plethora of radiation above $4$\,Ryd. 
Nonetheless, no star cluster synthetic model could pretend to predict the FUV continuum of \izw-NW around $4$\,Ryd with any certainty, in particular the amplitude of the discontinuity expected at $4$\,Ryd. 
In the \citetalias{Pequignot08} and present photoionization models, the low-energy tail of the assumed X-ray source spectrum (Appendix\,\ref{secapp:xspec}) contributes somewhat to He\2, and the discontinuity at $4$\,Ryd is empirically adjusted so that He\2\ is exactly fitted\footnote{Fitting He\2\ and a number of other nebular lines is a prerequisite to any model of the H\2\ region (see \citetalias{Pequignot08}). }. 
\citetalias{Pequignot08} provides an acceptable H\2\ region model that is useful to our H\1\ region models, but the adopted FUV continuum may not be unique. The important point to emphasize here is that the assumed $1-6$\,Ryd FUV continuum has strictly no impact on the properties of the present H\1\ region modeling.

Instead of the coarse representation of the intrinsic X-ray emission as a single blackbody at $2\times10^6$\,K \citepalias{Pequignot08}, we take advantage of satisfactory fits to the observed XMM-\textit{Newton} data, obtained by \cite{Kaaret13}, who assumed either Kerr black hole, cutoff power-law, or \texttt{diskbb} (distribution of blackbodies from an accretion disk) models. Here we adopt the \texttt{diskbb} model spectrum (see Fig.\,\ref{fig:radiationfield} and Sect.\,\ref{sec:const_xrays}).

In order to check the computations, we have compared the Nebu and Cloudy results in similar conditions, that is only with heating by UV and X-ray photoionization (no dust and no CR). Overall most significant line fluxes agree to better or much better than $15$\% (Table\,\ref{tab:cloudycompopt}).
This agreement between the Cloudy and Nebu model results is impressive, especially as the Cloudy computation was not performed in fully self-consistent conditions. Comparison between photoionization codes in standard conditions most often reveals larger discrepancies 
(e.g., \citealt{Pequignot01}). This success may be partly explained by the fact that, due to the very low metallicity and high ionization in the \izw\ NW H\2\ region shell, the energy balance, dominated by H and He, is much simplified and the physical conditions approach the theoretical limit allowed for H\2\ regions.

\begin{table}
\caption{Comparison between observed extinction-corrected optical and UV line fluxes and models.\label{tab:cloudycompopt}}
\centering
\begin{tabular}{l|l|l|ll}
\hline\hline
Line  & Obs. & Nebu$^\textrm{a}$ & $\mathcal{M}0a$ &  $\mathcal{M}4a$  \\
\hline
H\1\ 4861\AA   & 1000   & 1000 & 1000& 1000 \\
H\1\ 6563\AA    &   2860   & 2841 & 2858 & 2854 \\
He\1\ 4471\AA    &  21.4  & 36.6    & 36.9 &  35.2  \\
He\1\ 6678\AA    &  25.3  & 27.0     & 28.1 & 26.7    \\
He\1\ 7065\AA    &  24.4  & 23.9     & 22.4 & 21.8    \\
He\2\ 4686\AA   &  36.8  &  36.8   & 38.8 & 35.6 \\
$[$N\2$]$ 6548+84\AA   &  9.2 & 9.2&  7.8 & 7.3  \\
$[$O\1$]$ 6300+63\AA    &  8.5  &  8.6 & 9.9 & 9.0  \\
$[$O\2$]$ 3726+9\AA    & 238   & 236.8 & 207 & 203 \\
$[$O\2$]$ 7320+30\AA    &  6.3:   & 6.6 & 5.7 & 5.4   \\
$[$O\3$]$ 4363\AA    & 65.9   &  66.0 & 65.6 & 66.3 \\
$[$O\3$]$ 5007+4959\AA  &  2683   & 2679.4  & 2691 & 2653  \\
$[$Ne\3$]$ 3869\AA+968\AA   & 191   &  191.2  & 224 & 219.9  \\
$[$S\2$]$ 6716\AA  &  22.5   & 18.9 &  20.7 & 19.0  \\
$[$S\2$]$ 6731\AA  &  16.9   & 14.0 &  15.5 & 14.4  \\
$[$S\2$]$ 6716/6731   &  1.33     &  1.35   &  1.33 & 1.32   \\
$[$S\3$]$ 6312\AA   & 6.7   &  6.4  & 7.6 & 7.2 \\
$[$S\3$]$ 9069+532\AA   & 114   &  113  & 129.4 & 122   \\
$[$Ar\3$]$ 7136+751\AA    & 23.5   & 23.4  & 20 & 18.6   \\
$[$Ar\4$]$ 4740\AA    & 4.5   &  7.2 & 9.5 & 9.5  \\
$[$Fe\3$]$ 4658\AA    & 4.5 & 4.5 & 4.8 & 4.5  \\
$[$Fe\3$]$ 4986\AA    & 7.4  & 6.7 & ... & ...  \\
\hline
C\3$]$  1909+07\AA &    467 &   465 &  430   & 411 \\
Si\3$]$  1882\AA &       164: &  207 &  225  &  209 \\
C\4     1549\AA    &   512:  & 393. & 413  & ... \\
\hline
$\log$ $L$(H$\beta$) (erg\,s$^{-1}$)   &  39.43    &  39.43     & 39.36 & 39.43  \\
\hline
 \end{tabular}\\

\tablefoot{Fluxes are scaled with H$\beta$=1000. Observed values for the optical lines are from \cite{Izotov99a}, \cite{Skillman93}, and \citetalias{Pequignot08}. \cite{Kehrig15} provide a new integrated He\2\ flux, but the observed He\2\ intensity to be modeled, based on an average of the relatively stable He\2/H$\beta$ intensity ratio measured in different slit spectra of \izw-NW, is not changed.}
\tablefoottext{a}{See Section\,\ref{sec:p08}. }
\end{table}

\section{Summary of the observational constraints}\label{sec:obsconst}

The main observational constraints considered in this study are the optical and IR emission lines arising in the ionized gas, the IR lines [C\2], [O\1], and [Si\2] arising in the H\1\ region, and the H$^0$ mass. The dust mass, dust SED, and X-ray luminosity are also used, but they are as well input parameters of the models and are left some freedom for different reasons explained in this section. The suitability of these constraints to the NW model is also discussed.

\subsection{Homogenization of infrared and optical line fluxes}\label{sec:scaling}

Optical and IR line fluxes are used to constrain the physical conditions of both the H\2\ and H\1\ region in a consistent manner\footnote{We do not refer to the normalization of fluxes due to a possible contamination by regions outside NW (Sect.\,\ref{sec:nwse}) but to the normalization between observations with different apertures. }. The \textit{Herschel} and \textit{Spitzer} line fluxes (Sects.\,\ref{sec:optimal}, \ref{sec:spitzerobs}) first need to be normalized to be compared to the optical tracers. All fluxes are scaled to H$\beta=1000$. 

For the normalization of \textit{Herschel}/PACS measurements, we calculate the total deredenned H$\beta$ flux in the PACS footprint from the H$\alpha$ map \citep{depaz03}, assuming H$\alpha$/H$\beta=2.8$ and $E$(B-V)$=0.09$ \citep{Schlegel98,Schlafly11}. We find $F({\rm H}\beta)\approx11.5\times10^{-17}$\,W\,m$^{-2}$, 
implying a scaling factor of $\approx1.15\times10^{-19}$, which is the factor by which the PACS fluxes should be divided by to normalize to H$\beta=1000$. 

Since we consider the \citetalias{Pequignot08} predictions for the H\2\ region lines as robust, another, independent and informative, estimate of the PACS scaling factor is provided by the ratio between the observed and predicted ionized gas tracer [O\3] $88$\mic. We find $\approx1.2\times10^{-19}$, i.e., in good agreement with the geometrical factor derived directly from the H$\alpha$ observation. The same normalization is used in our study for all PACS tracers [O\3], [C\2], and [O\1], with a factor $1.2\times10^{-19}$ (Table\,\ref{tab:scaling}).

The \textit{Spitzer}/IRS fluxes measured in each module do not strictly correspond to the emission within the corresponding apertures, as a fraction of the source emission is lost outside the aperture owing to the PSF size. For this reason, we cannot simply use the H$\alpha$ fraction falling inside the IRS apertures. For point sources, the aperture correction is accounted for by the regular IRS flux calibration. In the case of \izw, the emission is somewhat extended, requiring a specific extraction method to recover the total flux (Sect.\,\ref{sec:spitzerobs}). With this method, we expect that our \textit{Spitzer}/IRS fluxes correspond to the total emission from the main body.

Similar to the scaling for PACS data, we can also test reference ionized gas tracers in the IRS range to estimate the required normalization by comparing the observed flux to the H\2\ region model predictions from \citetalias{Pequignot08}. Results are shown in Table\,\ref{tab:scaling} where it can be seen that the scaling factors for all modules agree well with each other, implying that the total flux was recovered in the small apertures of the IRS\footnote{The present absolute flux values are globally larger than those in \citetalias{Wu07}, and the scaling factor reduces the fluxes to levels corresponding approximately to the \citetalias{Pequignot08} normalization.
\citetalias{Pequignot08} did not want to consider all of the H$\beta$ emission, but that arising from the main NW shell. However, \citetalias{Pequignot08} calibrated the different instruments, using H$\alpha$ fluxes from comparable regions (not necessarily the region he wanted to model). In the same way, \citetalias{Pequignot08} derived $65$\% of ionizing photons escaping from the NW shell.}. The IRS scaling factors also agree well with those derived for PACS, implying that, as expected, the IRS line fluxes are well recovered by our extraction method.

The final normalized fluxes are provided in Table\,\ref{tab:finalfluxes}. The present study focuses on the H\1\ region and does not aim to improve the H\2\ region model. Our adopted fluxes for the IR ionized gas tracers are close to those used in \citetalias{Pequignot08} and we use the same optical line fluxes as in \citetalias{Pequignot08}. While our models are adapted to the NW region, we bear in mind that some of the tracers may be contaminated by the SE region (see discussion in Sect.\,\ref{sec:nwse}).

\begin{table}
\caption{Factors to normalize fluxes in W\,m$^{-2}$ to H$\beta=1000$.}
\centering
\begin{tabular}{lll}
\hline\hline
Observation & Reference tracer(s) & Factor  \\
\hline
\textit{Herschel} & [O\3]\,88\mic, aperture corrected H$\beta$          &  $\approx1.25\times10^{-19}$       \\
\hline
\textit{Spitzer} \\
...SL     & Hu$\alpha$\,12.37\mic, [Ne\2]\,12.8\mic               & $\sim1\times10^{-19}$      \\
...SH     & [Ne\3]\,15.6\mic                  & $\approx1.3\times10^{-19}$ \\
...LL     & [Ne\3]\,15.6\mic, [S\3]\,18.7\mic  & $\approx1.5\times10^{-19}$ \\
...LH     & [S\3]\,18.7\mic                   & $\approx1.2\times10^{-19}$ \\
\hline
\end{tabular}\\
\label{tab:scaling}
\end{table}

\begin{table}
\caption{Infrared line fluxes scaled to H$\beta=1000$. }
\begin{tabular}{llll}
\hline
\hline
                    &   \multicolumn{2}{c}{Normalized}           &    Adopted            \\
\hline
\textit{Herschel} \\
\hline          
$[$C\2$]$ 157\mic  & \multicolumn{2}{c}{$76\pm20$}    &  $76\pm20$  \\
$[$O\1$]$ 63\mic & \multicolumn{2}{c}{$60\pm20$}    &  $60\pm20$  \\
$[$O\3$]$ 88\mic & \multicolumn{2}{c}{$198\pm30$}    &  $200\pm30$ \\
\hline
\textit{Spitzer} &    SL/LL  &       SH/LH    &         \\
\hline
 H\1\ Hu$\alpha$ $12.37$\mic          & $ 8.5\pm3$ &   $   <7.5$      &   $ 8.5\pm3$ \\
 $[$O\4$]$ 25.9\mic   & $ 40\pm10 $  &   $37.5\pm8 $   &  $38\pm8 $  \\
                   &    (+[Fe\2] 26.0\mic) &         &       \\
 $[$Ne\2$]$ 12.8\mic        & $6\pm3   $ &   $\sim6$  & $6\pm3   $   \\
 $[$Ne\3$]$ 15.5\mic     & $ 48\pm10 $   &  $46\pm7.5 $  & $ 47\pm10 $   \\
 $[$Ne\3$]$ 36.0\mic      & $ <7       $   &    $   <21 $   &  $<7$    \\
 $[$Ne\5$]$ 14.3\mic         & $ <7     $   &    $     <2.3  $  &  $<2.3$ \\
 $[$Ne\5$]$ 24.3\mic          & $ <7     $   &  $     <4$    & $<4$   \\
 $[$Si\2$]$  34.8\mic          &  $ 63\pm13 $   &  $67\pm25 $   &  $65\pm12$  \\
 $[$S\3$]$ 18.7\mic         &   $ 27\pm7 $   &   $29\pm8 $   &  $28\pm7$ \\
 $[$S\3$]$ 33.5\mic         &  $ 37\pm7 $  &    $42\pm8 $  &  $40\pm7$   \\
 $[$S\4$]$ 10.5\mic           &  $43\pm5  $   &  $53.8\pm15.4 $   & $50\pm10$ \\
 $[$Ar\2$]$ 7.0\mic          &  $<5      $  &  ...    &  $<5$  \\
 $[$Ar\3$]$ 9.0\mic        &  $<10   $ &    ...    & $<10$  \\
 $[$Ar\3$]$ 21.8\mic       & $ <10     $  &    $     <8 $ &   $<8$ \\
 $[$Fe\2$]$ 17.9\mic         & $ <7     $   &  $    <3 $ &  $<3$  \\
 $[$Fe\2$]$ 26.0\mic       &  $40\pm10$    &  $ 12.5\pm8: $    &  $ 13\pm8: $ \\
                  &     (+[O\4] 25.9\mic) &             &   \\
 $[$Fe\3$]$ 23.0\mic        & $ <10     $  &    $     <6 $  &   $<6$ \\
 H$_2$ S(0)  28.2\mic          & $ <7 $  &   $      <8 $   &  $<7$ \\
 H$_2$ S(1)  17.0\mic             &  $<7$    &   $     <7.5$   &  $<7$ \\
 H$_2$ S(2)  12.29\mic           &  ...  &    $    <7.5$   & $<7.5$ \\
\hline   
\end{tabular}
\label{tab:finalfluxes}
\tablefoot{[O\4] 25.9\mic\ and [Fe\2] 26.0\mic\ are blended in the \textit{Spitzer} low-resolution modules.  }
\end{table}

\subsection{H\1\ region cooling lines}\label{sec:const_lines}

The maximum density in the H\1\ region, the outer radius (or else the minimum temperature) at which the model calculation is stopped, and the X-ray source parameters can be constrained to some extent by the [C\2], [O\1], [Si\2], and [Fe\2] fluxes measured in this study (Sect.\,\ref{sec:strat}). We expect these lines to be optically thin in \izw\ based on the low [O\1] $145$\mic/$63$\mic\ ratio found in other DGS sources \citep{Cormier15}. 

The [C\2] $157$\mic\ and  [O\1] $63$\mic\ lines are the dominant coolants in the neutral ISM as long as the metallicity is above some critical value ($\gtrsim10^{-3.5}$\,Z$_\odot$; e.g., \citealt{Santoro06}). The [Si\2] $34.8$\mic\ line has attracted relatively less attention, partly because its wavelength falls at the edge of the \textit{Spitzer}/IRS coverage and also because most of the silicon is usually depleted onto dust grains. In low-metallicity environments, however, depletion is weaker and [Si\2] (and to a lesser extent [Fe\2]) is an important coolant. Given the lack of depletion in \izw\ (Sect.\,\ref{sec:abundances}), we expect [Si\2] to be an important constraint to the models; [Fe\2] is observationally and theoretically less reliable.

\subsection{X-ray intrinsic spectrum}\label{sec:const_xrays}

X-rays are a fundamental ingredient in the models, but the X-ray luminosity and spectrum shape as derived from observations may not correspond directly to the radiation absorbed in the H\1\ region. The fact that the \textit{Spitzer} observations (December 2005) and \textit{Herschel} observations (May 2011 and October 2012) do not coincide in time with the X-ray observations (Sect.\,\ref{sec:xray}) is of little consequence since the light travel timescale across the shell is of order $10^3$ years. More importantly, however, the models indicate that typical cooling timescales in the temperature range $100-300$\,K are $\sim10^4$ years and $\sim10^5$ years for densities $500$\cc\ and $100$\cc,  respectively. Then, the X-ray spectrum we are considering in the models is nothing but an {\sl average over at least several $10^4$ years}, which may readily differ from the present observations.  

Moreover, the observed X-ray spectrum is itself subject to caution. 
There is observational evidence that the X-ray flux may vary considerably over a few years in \izw\ (Sect.\,\ref{sec:xray}; \citealt{Kaaret13}), which is compatible with state transitions in HMXBs between a low-luminosity hard spectrum state and a higher luminosity state with varying hardness. Such state transitions
seem to have been observed in \izw\ \citep{Kaaret13} and another BCD (\object{VII Zw 403}; \citealt{Brorby15}).

The X-ray spectrum \textit{shape} seen by the gas also bears some uncertainty, in particular for the soft X-rays that are absorbed in the H\1\ region. Because of the degeneracy induced by this absorption, whose value is not accurately known along the X-ray source line of sight, the observed soft X-ray spectrum is poorly constrained (see Appendix\,\ref{secapp:xspec}). 
Nonetheless, high-ionization lines such as [Ne\5], with the help of detailed photoionization models, can considerably reduce the uncertainty on the intrinsic soft X-ray flux (Appendix\,\ref{secapp:xspec}).

Despite these uncertainties, and ignoring possible strongly anisotropic X-ray emission (\citealt{Pakull02,Kaaret04}; see however \citealt{Bachetti15,King01,Koerding02}), it is relatively safe to assume that, within a factor of a few, the presently observed X-ray luminosity, $1.4\times10^{40}$\,erg\,s$^{-1}$, should represent the (average) luminosity seen by the H\1\ region in \izw. In order to appraise the generality of the conclusions, combinations of blackbodies that are compatible with the X-ray observations are considered together with the \texttt{diskbb} spectrum (see Appendix\,\ref{secapp:xspec}).

\subsection{Dust mass}\label{sec:const_dustmass}

There is evidence that, globally, the D/G in \izw\ is much lower than the value assuming a simple scaling with metallicity \citep{Galliano08b,Herrera12,Remy13b,Fisher2014,Remy15}. The dust mass derived by \cite{Remy15} for the entire galaxy is robust and used here for reference. The authors derived two values assuming the carbon-rich dust component is described either by graphite or amorphous carbon. To remain consistent with the dust prescription in Cloudy, we use the dust mass derived with the ``standard'' dust composition of \cite{Galliano11}, where the carbon component is made of graphite, i.e., $724^{+535}_{-308}$\,M$_\odot$. 
To put this value in perspective, the giant H\2\ region \object{LMC-N\,11}, despite a smaller physical size ($\approx150-200$\,pc) as compared to the \izw\ NW region, harbors $\sim3\times10^4$\,M$_\odot$ of dust \citep{Gordon14}. The total gas mass in the main body is $10^8$\,M$_\odot$ \citep{Lelli12}, leading to D/G$\approx1/1000$\,D/G$_{\rm MW}$, where D/G$_{\rm MW}$ is the Milky Way value\footnote{D/G$_{\rm MW}$ is $1/148$, close to the values in \cite{Zubko04}, $1/158$, and \cite{Jones13}, $1/156$. }. 

The global D/G value must be regarded with caution. On the one hand, the dust emission may be associated with only a fraction of the main body gas mass. On the other hand, there is evidence that the extended H\1\ in \izw\ may contain as much mass as the main body \citep{Lelli12}, in which case the global D/G may be driven to even lower values if the extended H\1\ is dust-free. Based on the non-detections of CO (e.g., \citealt{Leroy07b}), the mass of H$_2$ is ignored, and we keep in mind that D/G could be even lower if a significant fraction of gas exists as a CO-dark molecular gas. In our models we test several values for D/G. Apart from the dust mass, the dust SED shape is a qualitative constraint to our models.

\subsection{H$^0$ mass}\label{sec:const_himass}

The H$^0$ mass in the main body of \izw\ is $\approx10^8$\,M$_\odot$, which includes the H\1\ located in NW and in the high H\1\ column density cloud between NW and SE \citep{Lelli12}. Our observations show that the [C\2] and [O\1] emission is compatible with a compact source within the main body (Sect\,\ref{sec:optimal}), so that the corresponding H$^0$ mass should be less than $10^8$\,M$_\odot$. 

We use the $2\arcsec$ resolution H\1\ map from \cite{Lelli12} to estimate the H$^0$ mass associated with either the NW region itself (defined as a $\approx430$\,pc diameter region centered on the NW stellar cluster) or with the slightly different region observed with IR tracers (Sects.\,\ref{sec:obs} and \ref{sec:nwse}). For the NW region, the H\1\ column density lies between $3-6\times10^{21}$\,cm$^{-2}$, which translates into a mass $5\times10^6$\,M$_\odot$. If we now consider instead the H$^0$ mass associated with the H\1\ column density peak between NW and SE \citep{Lelli12}, we find a larger mass, $\approx2\times10^7$\,M$_\odot$. In the following we consider that H$^0$ masses around $\approx0.5-2\times10^7$\,M$_\odot$ are acceptable.

\subsection{Applicability of the tracers to the NW region}\label{sec:nwse}

The models are built for the NW H\2\ region of \izw\ and its surroundings. Owing to limited spatial resolution of IR observations, some degree of contamination by SE (gas and dust emission) and more diffuse regions is unavoidable. Nevertheless, our approach can be justified on both
observational and theoretical grounds:
\begin{itemize}
\item The emission in all IR gas tracers is compact with a size similar to the NW region ($\lesssim6\arcsec$; Sect.\,\ref{sec:optimal}). The centroid for [O\1] and [O\3] seems to be located closer to NW, possibly coinciding with a dust-rich ionized gas shell located between NW and SE (labeled NW-D3 in \citealt{Cannon02}), near the H\1\ peak column density.
\item The convergence to a solution produces a kind of an average model because the (normalized) IR observations are rather global. Nonetheless, since the NW nebular emission is three times stronger than the corresponding SE emission at most wavelengths (e.g., H$\beta$, Hu$\alpha$, and [S\4]; Sect.\,\ref{sec:spitzerobs}) and given the location of the X-ray source within NW, it can be safely surmised that the model more closely reflects (average) properties of the NW region.
\item The ionization structure and metallicity do not vary much between NW and SE \citep{Legrand00,Skillman93,Kehrig2016}. Since our line measurements are normalized to H$\beta$, this means that line constraints should not be significantly affected by SE. 
\end{itemize}
\begin{itemize}
\item All H\1\ regions will be subject to the X-ray radiation, whether they are associated with the NW or SE giant H\2\ regions. Since the ionization state is stable in the H\1\ gas, the fine-structure line emissivity depends essentially on local temperature, which is controlled by the local flux of X-rays. Geometrical details have little consequence.
\item The H\1\ gas can only exist where the ionizing photons from the star cluster are exhausted, that is beyond an H\2\ region. Diffuse H$\beta$ emission exists out of the main H\2\ regions of \izw\ and may be associated with diffuse [C\2] and [O\1] emission. Trial calculations show that the ratio of strong lines from the H\2\ region, such as H$\beta$ and [O\3] $5007$\AA, to [C\2] and [O\1] lines is a slow function of assumed parameters (density and distance to the source) over a large range of conditions. This further suggests that the NW shell model is representative of the more global emission (see also end of Sect.\,\ref{sec:abs}).
\end{itemize}

We conclude that, in practice, IR gas tracers can be safely assumed to arise in NW. A possible contamination by regions outside of NW (SE or diffuse) would not significantly alter our results.  
The H$^0$ mass, on the other hand was calculated specifically for NW, or more precisely for the IR-emitting region that appears to coincide with NW (Sect.\,\ref{sec:const_himass}). Finally, the dust mass inferred from observations is the total dust mass of the galaxy (Sect.\,\ref{sec:const_dustmass}), so we keep in mind that the dust mass computed in the NW model can be smaller than the observed value. Several D/G values are explored in the models.

In summary, we consider that the observational constraints presented here can be applied to the NW model. The specific contamination of SE on the dust SED and dust mass is discussed further in Section\,\ref{sec:nonunidgr}.

\section{H\2\ + H\1\ region modeling}\label{sec:cloudy}

\subsection{Modeling strategy}\label{sec:strat}

Armed with a satisfactory Cloudy model equivalent to the Nebu models (Sect.\,\ref{sec:modelprep}) and with new or updated observational constraints (Sect.\,\ref{sec:obsconst}), our strategy consists in exploring several parameters to evaluate their importance in the H\1\ gas heating. Free parameters for the gas are the maximum density and the minimum temperature assumed in the radiation-bounded sector (Table\,\ref{tab:modelparams}). The other free parameters examined are the X-ray luminosity (Sect.\,\ref{sec:xspec}), the photoelectric effect through D/G (Sect.\,\ref{sec:cloudype}), and the CR ionization rate (Sect.\,\ref{sec:cosmicrays}). 
Other processes such as mechanical heating are not considered.

For each set of free parameters, we monitor the predictions for the main observational constraints, namely the [C\2], [O\1], and [Si\2] line fluxes but also the dust mass and  H$^0$ mass (Sect.\,\ref{sec:const_himass}). In most cases, the minimum temperature in the radiation-bounded sector is constrained by the observed [C\2] flux, while the maximum density is constrained by the [O\1]/[C\2] ratio. In short, given the X-ray source, a larger maximum density implies a smaller H$^0$ mass and smaller line fluxes, where [C\2] and [O\1] are more sensitive to change than [Si\2] and [Fe\2]. A smaller minimum temperature implies a larger H$^0$ mass and stronger [C\2] and moderately stronger [O\1] lines. Increasing the X-ray luminosity and thus the H\1\ zone temperature, all IR line intensities are increased; [Si\2] and [Fe\2] are selectively enhanced, while smaller H$^0$ masses (larger minimum temperature) are required.

Relevant properties of our models are shown in Table\,\ref{tab:models}. Since the combination of maximum density and depth (hence temperature) is constrained by the observed [C\2] and [O\1] fluxes, the model are expected, by design, to reproduce both lines. This is not the case, however, for models with no X-rays ($\mathcal{M}0a$ and $\mathcal{M}0b$, without and with dust, respectively), which severely underestimate the line fluxes and the H$^0$ mass. The [Si\2] flux is comparatively better reproduced in these models since the line is partly emitted in the H\2\ region. The reasons why the H\1\ region line fluxes are underestimated are twofold, (1) the UV luminosity provided by the central cluster is not able to heat a large enough H$^0$ mass, even with a large D/G, and (2) we have only considered the OB stellar cluster as the heating source.

All other models include the X-ray source. In order to isolate the effect of each process, we explore models with no dust and no CR ($\mathcal{M}1$), with dust but no CR ($\mathcal{M}2$), and with CR but no dust ($\mathcal{M}3$). The objective is not to find the best combination but to test the impact of each parameter. The full model ($\mathcal{M}4$) is the only one that combines all physical processes. For each model we test different values of D/G and the CR rate, and for each set of values we allow the X-ray luminosity to vary. We then examine the output quantities to compare to observations, namely [C\2], [O\1], [Si\2], the dust mass and SED shape, and the H$^0$ mass. We monitor the relative importance of the different heating processes for each satisfactory model.

\subsection{Exploratory model $\mathcal{M}1$: X-ray heating}\label{sec:xspec}

As originally proposed by \citetalias{Pequignot08}, the heating provided by XR photoionization is a natural explanation to the H\1\ region line emission. 
The intrinsic X-ray spectrum used in our models is described in Section\,\ref{sec:modeltopo} (see also Appendix\,\ref{secapp:xspec}). 

\begin{figure}
\centering
\includegraphics[angle=0,width=9cm,height=14cm,clip=true,trim=0 0cm 0 0cm]{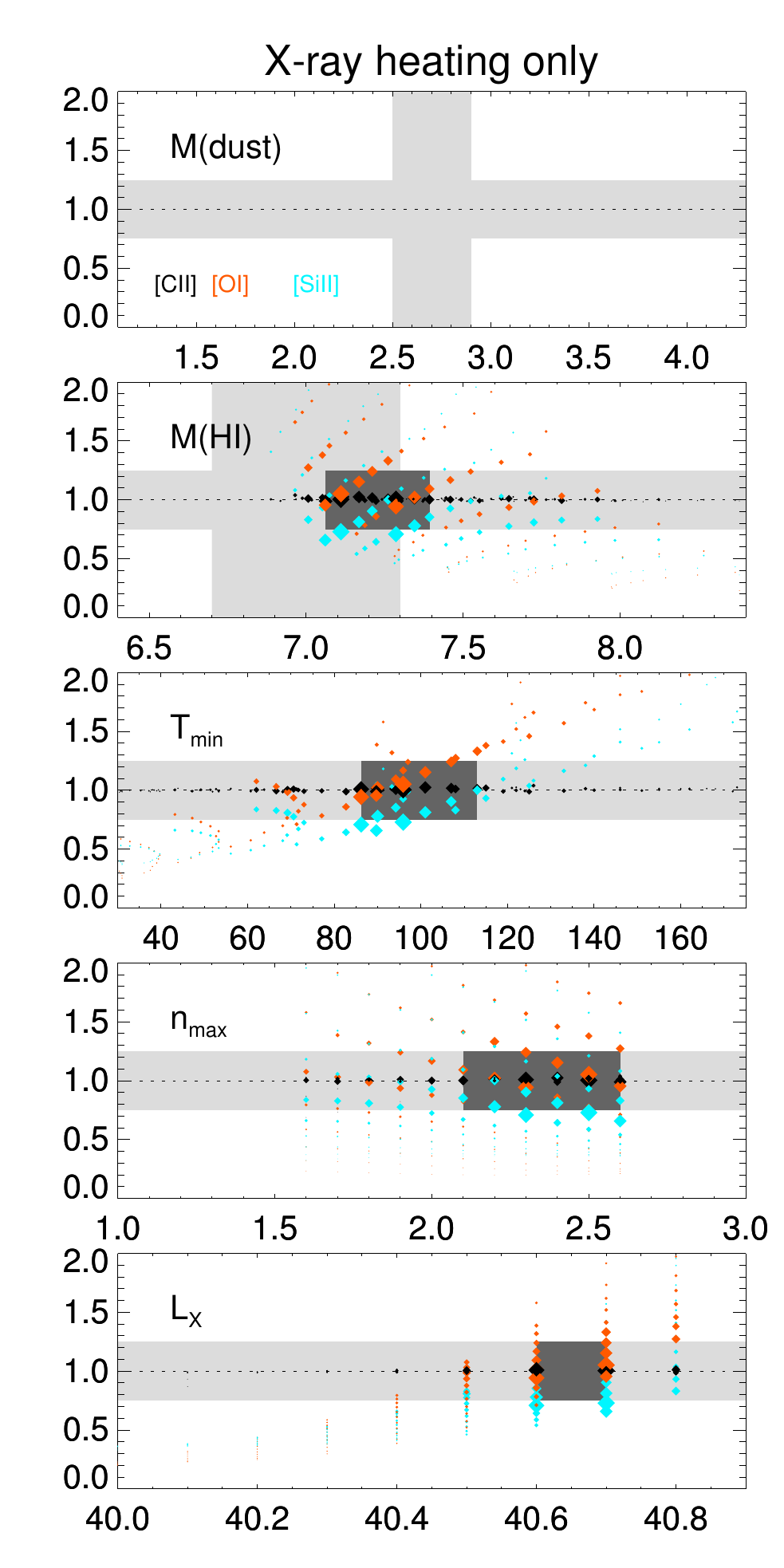}
\caption{Exploration of models with X-ray heating only (no dust and no CR). The predicted line fluxes (model/observation) are plotted against each model parameter with from top to bottom, the dust mass and H$^0$ mass ($\log$\,M$_\odot$), minimum temperature (K) and maximum density ($\log$\,cm$^{-3}$) in the radiation-bounded sector, and the X-ray luminosity ($\log$\,L$_\odot$). Each point represents a model, and the symbol size is inversely proportional to the residuals between model predictions and observations for [C\2], [O\1],  and [Si\2]. The light gray zone corresponds to the observed values and their associated error bar. The dark gray zone corresponds to the best models. }
\label{fig:conv_try}
\end{figure}

Figure\,\ref{fig:conv_try} shows the suite of models with only the X-ray heating considered. The predicted [O\1]/[C\2] line ratio increases with increasing X-ray luminosity and with increasing temperature. Both effects are due to the fact that [O\1] dominates the cooling at the inner edge of the H\1\ region while [C\2] dominates inside. The predicted [Si\2] emission tends to be somewhat weaker than the observed one. Improving [Si\2] using a larger $L_{\rm X}$ is at the expense of [O\1], which becomes too large. In addition to observational uncertainties, this moderate
discrepancy may suggest that either the adopted Si abundance was
underestimated or [O\1] is affected by self absorption.

The model that reproduces the observations (line emission and H$^0$ mass) best is labeled $\mathcal{M}1$. As expected, the [C\2], [O\1] $63$\mic, and [Si\2] transitions are the strongest coolants, followed by [O\1] $145$\mic\ and [Fe\2], in the H\1\ region. The required X-ray luminosity is $4\times10^{40}$\,erg\,s$^{-1}$, which is three times larger than the observed luminosity. According to the analysis of Section\,\ref{sec:const_xrays}, such a difference is acceptable.

In summary, model $\mathcal{M}1$ shows that X-ray heating alone is already sufficient to explain the FIR line emission, through a reasonable alteration of the observed X-ray spectrum. Inasmuch as X-ray heating dominates in the H\1\ gas, the H\1\ region cooling lines provide a reliable average of the average soft X-ray luminosity over the last few $10^4$ years in \izw. 
We now explore models involving alternative heating mechanisms.

\begin{landscape}
\begin{table}
\caption{Cloudy model predictions for NW. \label{tab:models}}
\centering
\begin{tabular}{ll|c|lll|ll|lllll|ll}
\hline\hline
   &      &   &   \multicolumn{3}{c|}{Input parameters} & \multicolumn{9}{c}{Output parameters} \\
Model  & Description & Conf.$^{\rm a}$  & $L_{\rm X}$  & CR$^{\rm b}$  & D/G  & $n_{\rm max}^{\rm c}$  & $T_{\rm stop}^{\rm d}$ & $M({\rm H}^0)$ & $M_{\rm dust}$  &  [C\2]   & [O\1]   & [Si\2]  & \multicolumn{2}{c}{$\Gamma_{\rm tot}({\rm HI})^{\rm f}$}  \\
    &       &          & ($\log$ erg\,s$^{-1}$)   & ($\times$ Gal.)  &  ($\times$ Gal.)  & (cm$^{-3}$)    &  (K)  &  ($\log$ M$_\odot$)      &   ($\log$ M$_\odot$)    &   (M/O$^{\rm e}$)  & (M/O$^{\rm e}$)  & (M/O$^{\rm e}$) & PE & CR  \\ 
\hline  
\multicolumn{2}{l}{Exploratory models}   & &&&&&&&&&& \\
\hline

$\mathcal{M}0a$   & ($\approx$P08's $M2$)  & (a)  & $0$  & $0$   & $0$    & $150$      &   $50^{\rm g}$  &   \underline{$3.7$}   &  $0$ &   \underline{$0.01$}  &   \underline{$0.01$}      &  \underline{$0.27$}  & $0$\% & $0$\%  \\

$......b$   & Dust (Z-scaled D/G) & (a)  & $0$  & $0$   & $1/50$    & $50$      &   $50^{\rm g}$ &   \underline{$4.7$}   &  $2.6$  &  \underline{$0.02$}         &   \underline{$0.06$}      &  \underline{$0.35$} & $12$\% & $0$\%  \\

\hline

$\mathcal{M}1$   & XR & (a)  & $40.6$  & 0   & 0   & $150$      &   $100$ & $7.2$   & $0$   &   $1.01$    &  $1.10$     &  $0.81$ & $0$\% & $0$\%  \\

\hline

$\mathcal{M}2a$ & XR + dust (observed D/G) & (a)   & $40.6$  & $0$   & $1/1000$  & $200$      &  $95$  & $7.2$ &  $2.2$ &    $1.00$    &   $1.07$    &  $0.77$  & $4$\% & $0$\%  \\

$......b$   & XR + dust  (Z-scaled D/G)   &   (a)   & $40.3$  & $0$  & $1/50$  &  $250$      &  $110$  & $7.2$ &  \underline{$3.4$}   &  $1.01$    &   $1.05$     &  \underline{$0.62$} & $57$\% & $0$\%  \\

$......c$   & XR + dust  (maximum D/G)   &   (a)   & $40.6$  & $0$  & $1/300$  &  $200$    &  $100$  & $7.2$ &  $2.6$   &  $1.00$    &   $1.08$     &  $0.77$ & $10$\% & $0$\%   \\

$......d$ & XR + dust  (radial D/G) & (a)   & $40.6$  & $0$   & $1/(150\searrow750)$ & $200$      &  $105$  & $7.2$  & $3.0$  & $1.02$  &  $1.13$   & $0.80$  & $12$\% & $0$\%   \\

$......e$   & XR + dust  (extra sector) & (b)   & $40.6$  & $0$   & $1/(1000,50)$  & $150$      &  $90$  & $7.4$  & $2.3$ & $1.02$  &  $1.02$    & $0.75$ & $4$\% & $0$\%    \\

\hline

$\mathcal{M}3a$   & XR + CR   &   (a)     & $40.6$ & $1$  & $0$    &  $200$     &  $110$ & $7.2$   & $0$ &  $1.01$  &  $1.13$ & $0.81$ & $0$\% & $13$\%  \\

$......b$   & XR + CR    &   (a)     & $40.1$ & $5$  & $0$    &  $250$     &  $125$ & $7.2$   & $0$ &  $1.06$  &  $1.00$ & \underline{$0.65$} & $0$\% & $65$\%  \\

$......c$   & CR only & (a)  & $0$  & $1$   & $0$    & $25$      &   $230$ &    \underline{$8.1$}    &  $0$  &  $1.04$         &    \underline{$1.23$}     &   $0.80$  & $0$\% & $96$\% \\

$......d$   & CR only & (a)  & $0$  & $5$   & $0$    & $250$      &   $150$  &   $7.3$   &  $0$  &  $1.02$         &   $1.01$      &  \underline{$0.66$} & $0$\% & $96$\%  \\
\hline
\multicolumn{2}{l}{Full models}   & &&&&&&&&&& \\
\hline

$\mathcal{M}4a$   & Best parameter set  &  (b)  & $40.6$ & $0.1$ & $1/(1000,50)$ & $250$    & $100$         &  $7.4$ & $2.3$  & $1.05$        &   $1.05$      &  $0.77$ & $3$\% & $0.7$\%  \\

$......b$   & $\mathcal{M}4a$ + molecules    &  (b)  & $40.6$ & $0.1$ & $1/(1000,50)$ &  $250$     &   $80$        &  $7.3$ & $2.2$  & $1.03$        &   $1.00$      &  $0.71$ & $3$\% & $1$\%  \\

$......c$   & $\mathcal{M}4b$ + clumps    &  (c)  & $40.5$ & $0.1$ & $1/1000$ & $150$    &    $90$      & $7.3$  & $2.6$  &  $1.00$       &  $1.12$       &  $0.84$  & $3$\%  & $2$\% \\ 
      &&&&&&&&&&&&& ($23$\%) & ($17$\%) \\
\hline
\end{tabular}\\
\tablefoot{The model $\mathcal{M}0a$ is equivalent to the model $M2$ in \citetalias{Pequignot08} accounting for a few modifications in the latter (Sect.\,\ref{sec:p08}). The model $\mathcal{M}1$ is similar to model $M2_{\rm X2}$ in \citetalias{Pequignot08} but with a new X-ray description (Sect.\,\ref{sec:const_xrays}). Some input parameters are fixed (CR rate, D/G) while others are scanned in order to fit the observations ($L_{\rm X}$, $n_{\rm max}$, $T_{\rm stop}$). The observed dust and H$^0$ mass is close to $10^{2.7}$\,M$_\odot$  (Sect.\,\ref{sec:const_dustmass}) and $10^{7.2}$\,M$_\odot$ (Sect.\,\ref{sec:const_himass}) respectively. The underlined values indicate that the model prediction deviates significantly from the observation. }
\tablefoottext{a}{The sector configuration refers to the topology illustrated in Figure\,\ref{fig:sectors}. }
\tablefoottext{b}{CR ionization rate. }
\tablefoottext{c}{Maximum density in the radiation-bounded sector (\#1).  }
\tablefoottext{d}{Stopping temperature for the radiation-bounded sector (\#1).  }
\tablefoottext{e}{Model/observation. }
\tablefoottext{f}{Fraction of the total heating in the H\1\ region due to the photoelectric effect (``PE'') and to cosmic rays (``CR''). The remaining fraction is mostly due to ionization of H and He. For model $\mathcal{M}4c$ we give the fractions for the main radiation-bounded sector and for the clump sector (between parentheses). }
\tablefoottext{g}{The final temperature corresponds to the fixed outer radius. }
\end{table}

\end{landscape}

\subsection{Exploratory models $\mathcal{M}2$: Photoelectric effect heating}\label{sec:cloudype}

In this section we examine the photoelectric effect on dust grains and PAH molecules in addition to the X-ray heating. The objectives are to reproduce qualitatively the observed dust SED shape, obtain a reasonable dust mass, and quantify the photoelectric heating rate. Since the observed dust properties correspond to the entire galaxy (Sect.\,\ref{sec:const_dustmass}), models that overestimate the dust mass or the dust SED are not deemed acceptable.

For lack of an accurate description of the dust properties in \izw\ (or in any other extremely metal-poor environment), we use the \object{Small Magellanic Cloud} opacity files in Cloudy that are based on the size distribution given in \cite{Weingartner01b}. Our results concerning the gas line emission are barely changed if we use grain properties reflecting the size distribution in the diffuse ISM of the Milky Way. Considering the lack of evidence for elemental depletion on dust grains and the low dust-to-metal ratio in \izw\ (Sect.\,\ref{sec:abundances}), the gas-phase abundances are the same for all models. We test several choices for D/G and we calculate a grid with varying X-ray luminosity for each choice.

\subsubsection{Uniform D/G ($\mathcal{M}2a$, $\mathcal{M}2b$, $\mathcal{M}2c$)}\label{sec:uniformdgr}

\begin{figure*}
\centering
\includegraphics[angle=0,width=9cm,height=14cm,clip=true,trim=0 0cm 0 0cm]{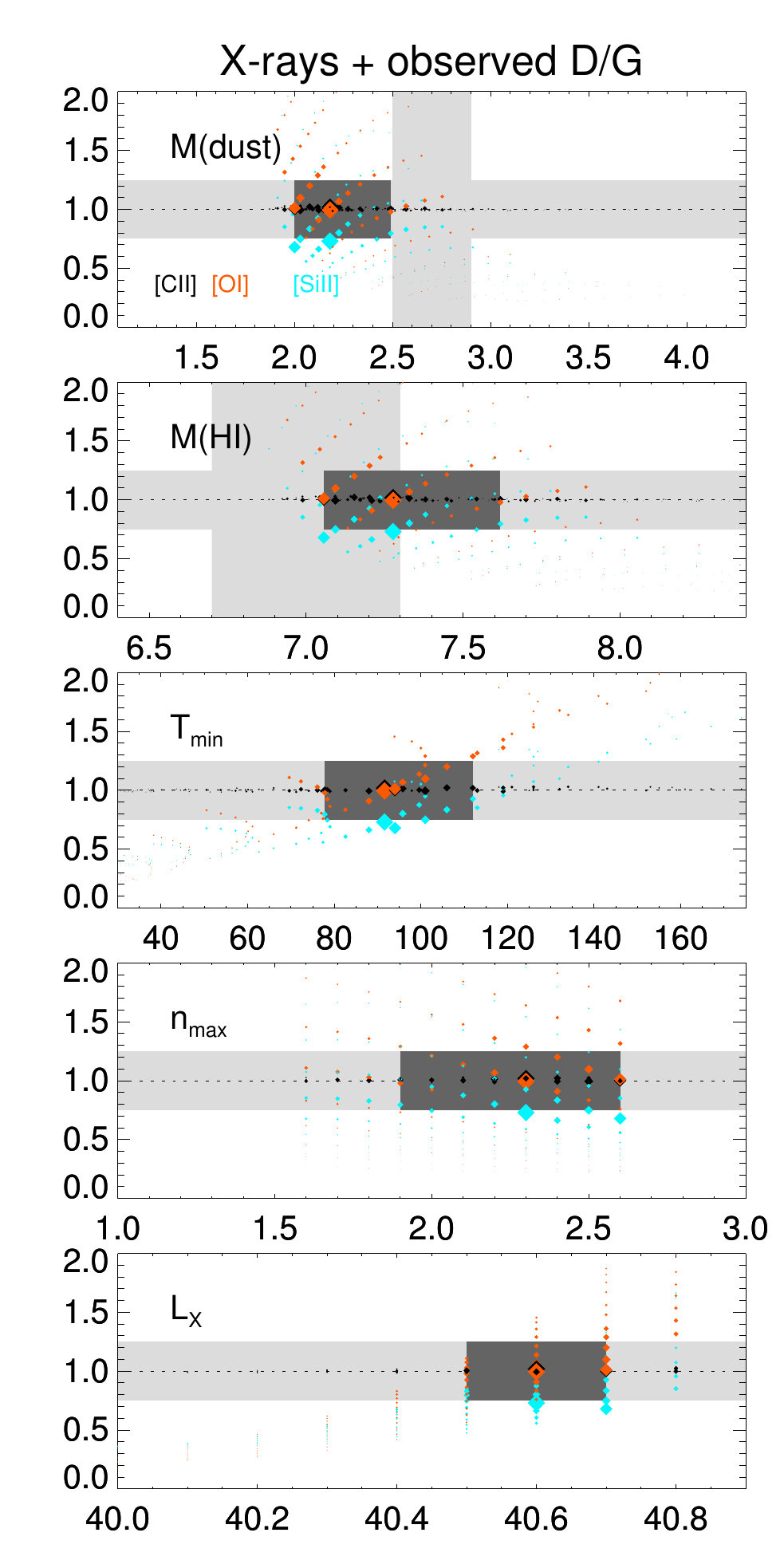}
\includegraphics[angle=0,width=9cm,height=14cm,clip=true,trim=0 0cm 0 0cm]{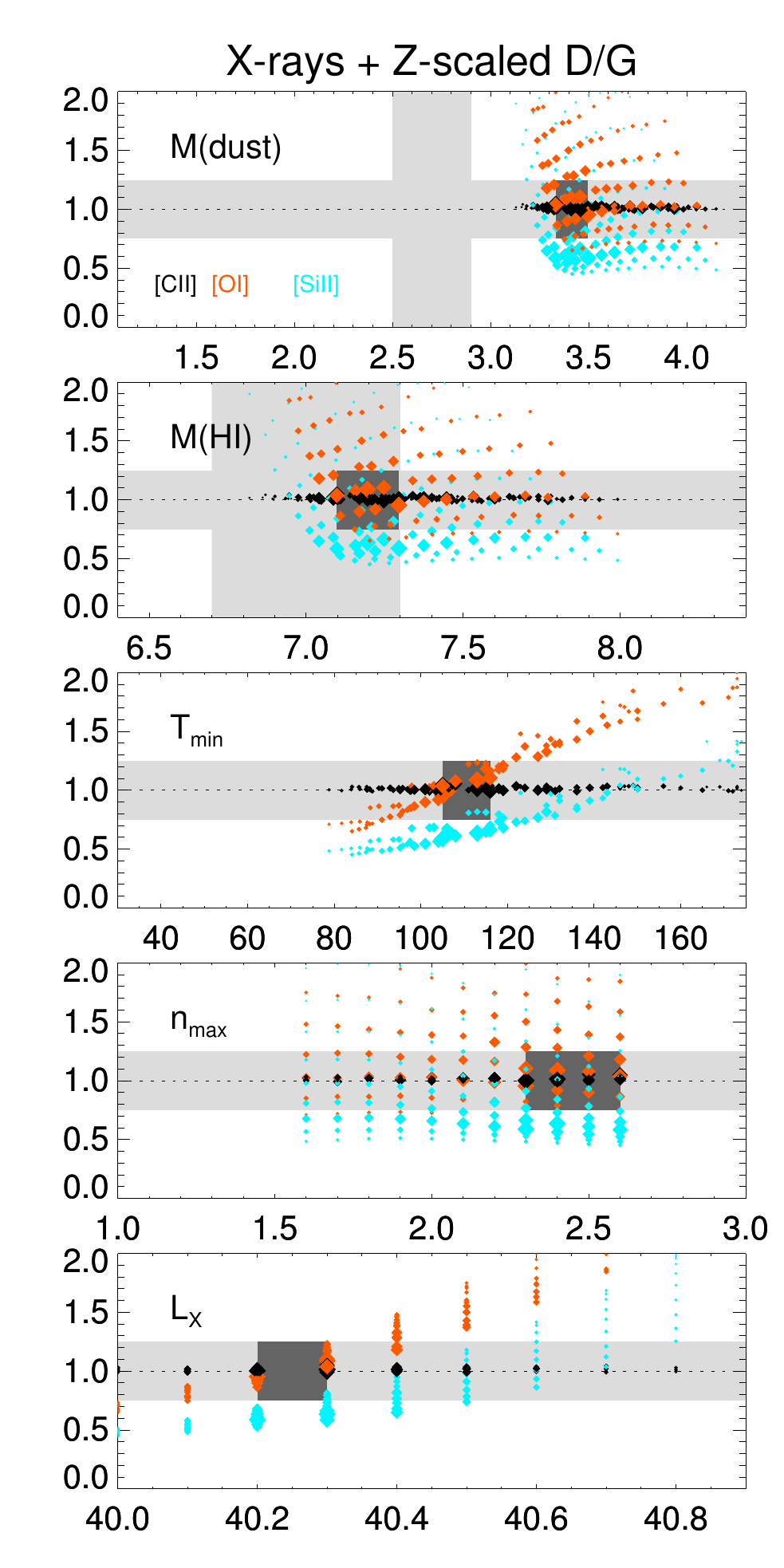}
\caption{Models with X-ray and photoelectric effect heating for D/G observed (\textit{left}) or scaled with metallicity (\textit{right}). See Fig.\,\ref{fig:conv_try} for the plot description. }
\label{fig:conv_dust}
\end{figure*}

First, a uniform D/G value is assumed across the galaxy. Model $\mathcal{M}2a$ uses a global D/G of $1/1000$\,D/G$_{\rm MW}$, corresponding to the observed value as constrained by \textit{Herschel} (Sect.\,\ref{sec:const_dustmass}). Model $\mathcal{M}2b$ uses $1/50$\,D/G$_{\rm MW}$, corresponding to the metallicity scaling. Figure\,\ref{fig:conv_dust} shows the results for both models. 

\begin{figure*}
\centering
\includegraphics[angle=0,width=9cm,height=12cm,clip=true,trim=0 0cm 0 0cm]{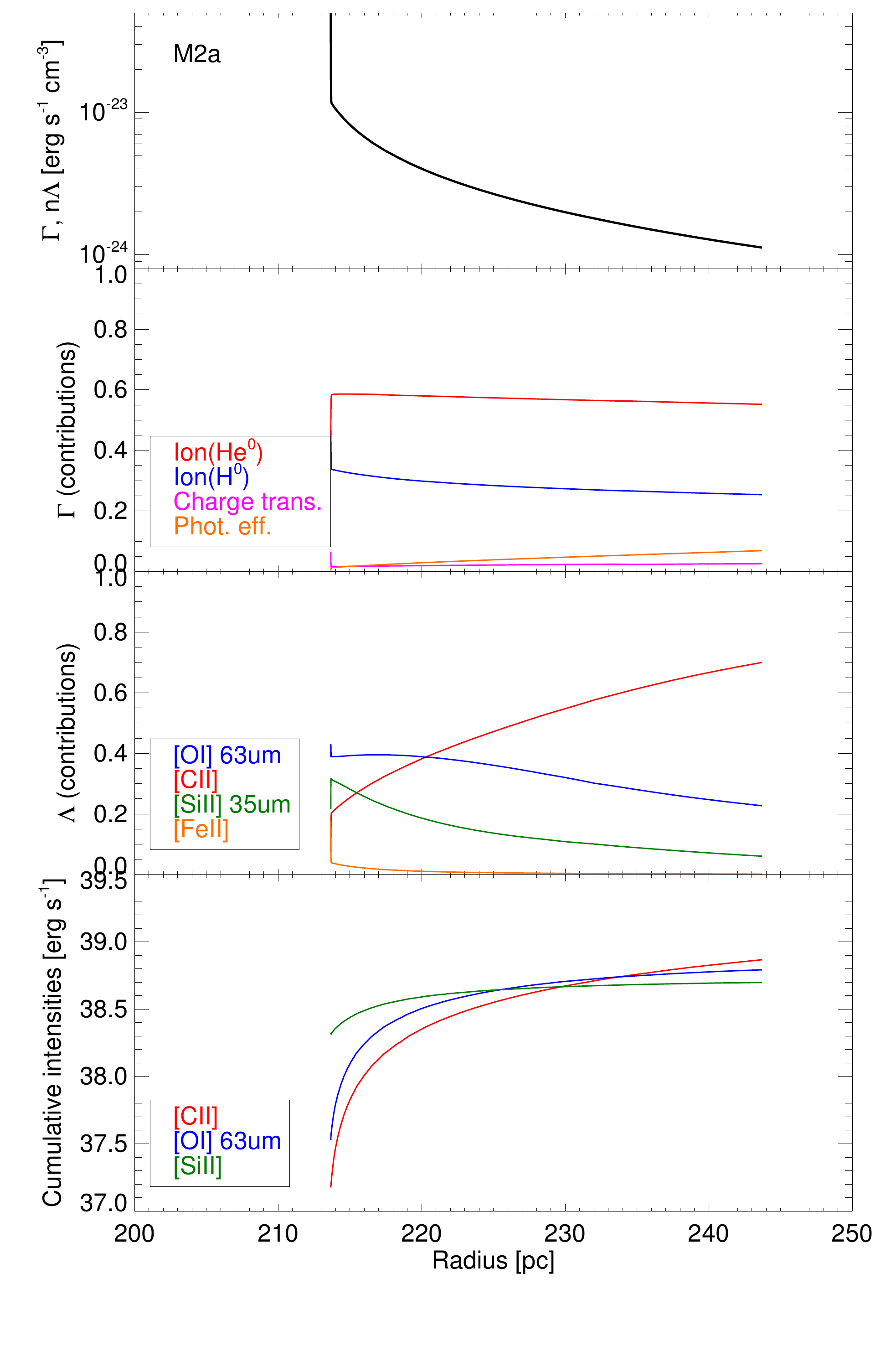}
\includegraphics[angle=0,width=9cm,height=12cm,clip=true,trim=0 0cm 0 0cm]{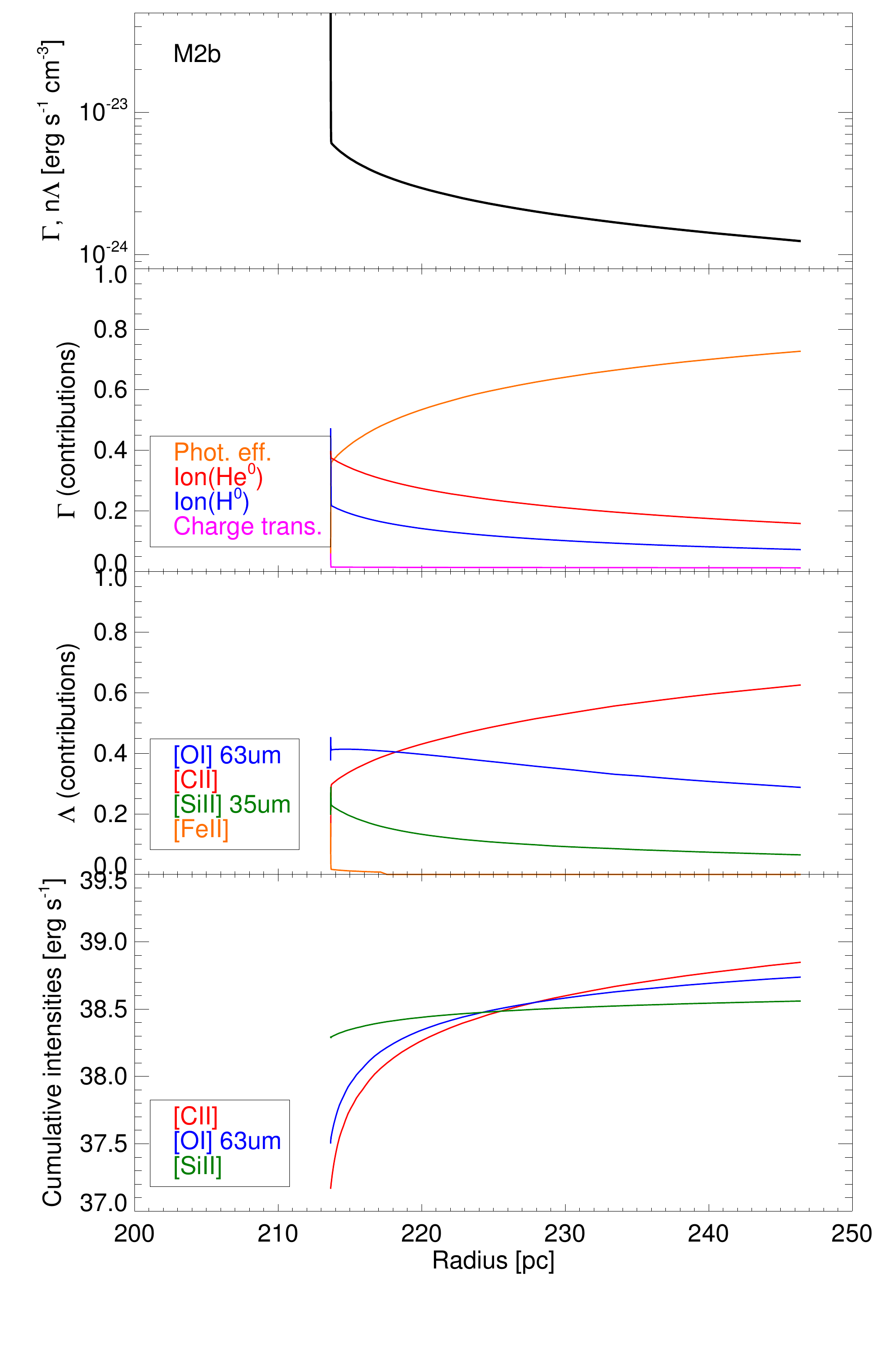}\\
\includegraphics[angle=0,width=9cm,height=150pt,clip=true,trim=0 0pt 0 22]{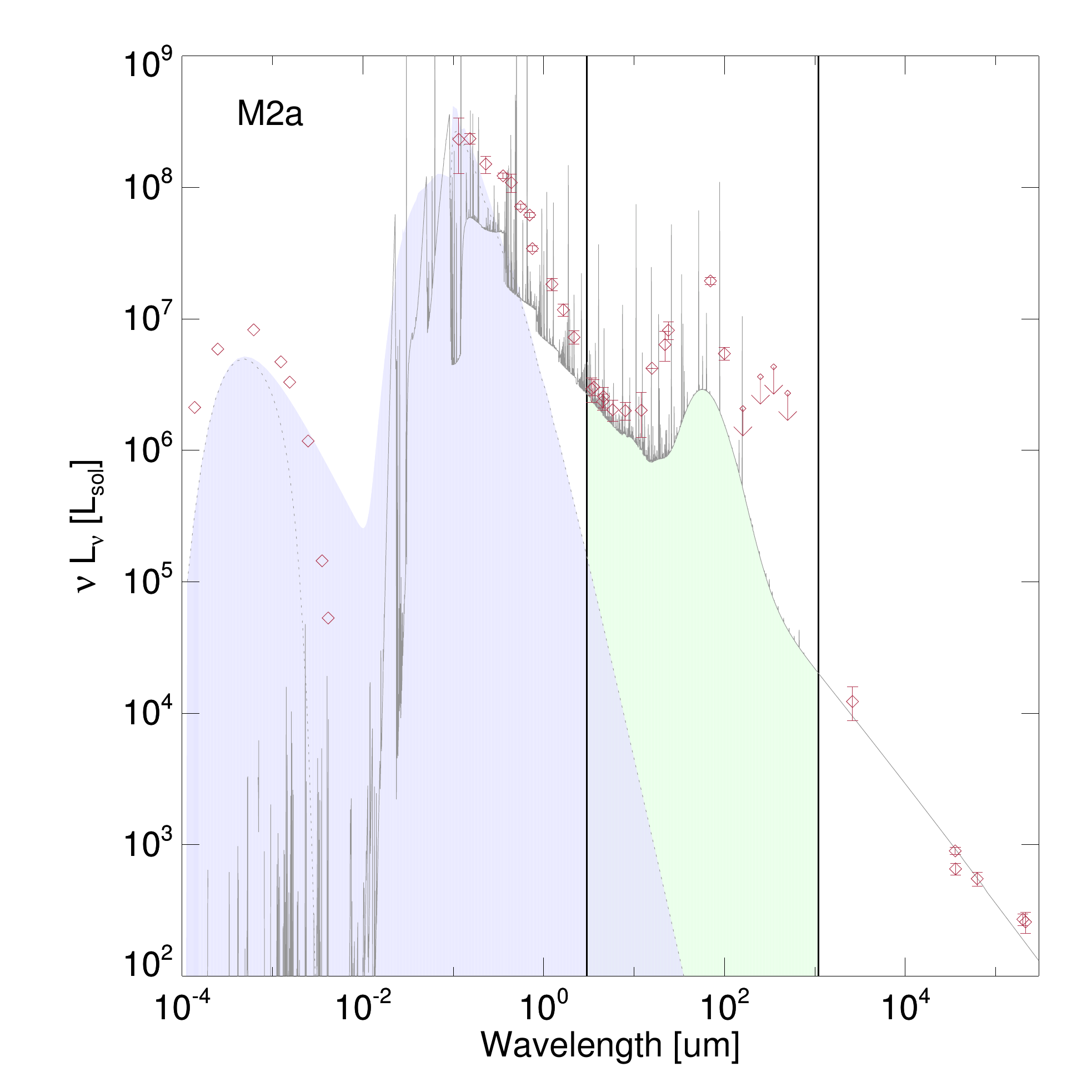}
\includegraphics[angle=0,width=9cm,height=150pt,clip=true,trim=0 0pt 0 22]{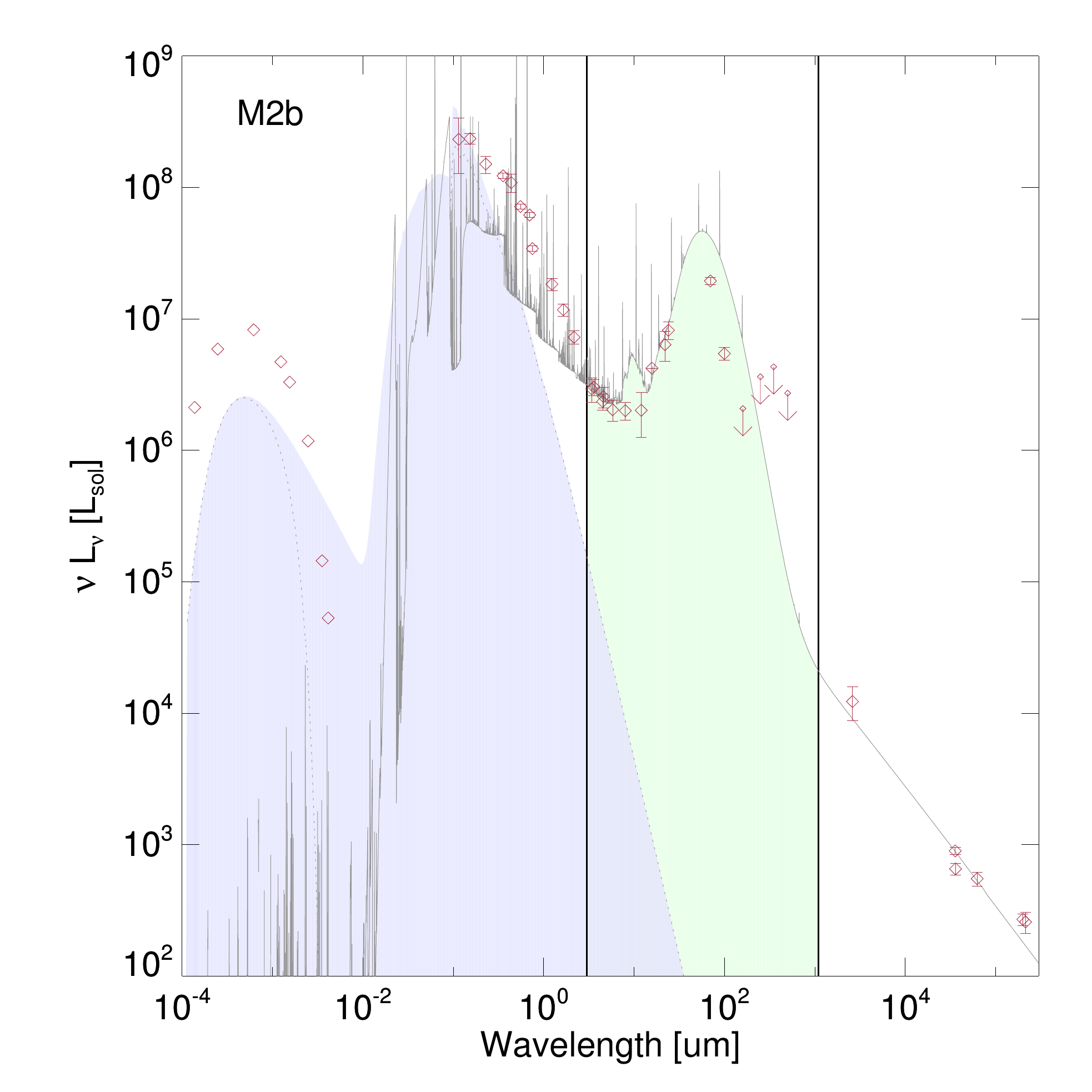}
\caption{Model results for $\mathcal{M}2a$ (D/G observed, \textit{left}) and $\mathcal{M}2b$ (D/G scaled with metallicity, \textit{right}). The plots on top show, from top to bottom, the total heating and cooling rate, the contributions to the heating rate, the contributions to the cooling rate, and the cumulative intensities of cooling lines for the radiation-bounded sector as a function of the depth within the cloud (the H\2-H\1\ transition lies at $\approx213$\,pc). For clarity, the heating and cooling mechanisms in the H\2\ region are not shown. The predicted dust SED is shown in the bottom panels for both models. The $20$\,cm measurement is thought to be dominated by synchrotron emission (see text). The vertical lines indicate the range used to calculate the total infrared luminosity ($3-1100$\mic) and the green shaded area shows the modeled emission in this range. The blue shaded area shows the input radiation field. The model $\mathcal{M}2b$ shows weak silicate emission bands at $10$\mic\ and $20$\mic\ that are not seen in the IRS spectrum. }
\label{fig:model_pe}
\end{figure*}

As shown in Table\,\ref{tab:models}, the solution for model $\mathcal{M}2a$ is almost the same as that of model with only X-ray heating ($\mathcal{M}1$), which is due to the fact that the photoelectric heating rate is much smaller than the soft X-ray ionization heating rate (Fig.\,\ref{fig:model_pe}). The photoelectric effect represent only $\approx4$\%\ of the total heating in the H\1\ region in model $\mathcal{M}2a$ (Table\,\ref{tab:models}), as compared to $58$\%\ and $30$\%\ for ionization by soft X-rays of He$^0$ and H$^0$, respectively. The dust SED is, however, very underestimated (Fig.\,\ref{fig:model_pe}) and the dust mass somewhat underestimated too. This can be at least partly ascribed to the fact that we model the NW region while the observed dust SED and dust mass correspond to the entire galaxy (Sect\,\ref{sec:const_dustmass}).

On the other hand, model $\mathcal{M}2b$ provides relatively more photoelectric effect heating ($57$\% of the total heating across the H\1\ region; Table\,\ref{tab:models}), which results in a lower required X-ray luminosity as compared to model $\mathcal{M}2a$. Nevertheless, the dust mass and dust SED are now significantly overestimated (Fig.\,\ref{fig:model_pe}) and [Si\2] is not as well reproduced as in model $\mathcal{M}2a$. 

The photometry data at $100$\mic\ and the upper limit at $160$\mic\ provide an upper limit on the D/G. We find that, with the current topology, a radiation-bounded cloud  necessarily produces too much dust emission if the D/G is larger than $1/300$\,D/G$_{\rm MW}$ (model $\mathcal{M}2c$; Table\,\ref{tab:models}). The corresponding dust mass is in relatively good agreement with what observations suggest, but the SED is not well reproduced shortward of $100$\mic\ (Fig.\,\ref{fig:sed2}), which could be due to a missing dust component (not contributing significantly to the dust mass). Model $\mathcal{M}2c$ has similar physical conditions in the H\1\ region (density, temperature) as model $\mathcal{M}2a$ and the heating fraction contributed for by the photoelectric effect is only $10$\%.

In summary, the H\1\ region cooling lines can be reproduced in models with X-rays together with a wide range of uniform D/G value, but the photoelectric effect is never a significant heating mechanism in the H\1\ region. 
Heating by soft X-rays dominate and the required X-ray luminosity is only twice as low when D/G increases from $1/1000$\,D/G$_{\rm MW}$ to $1/50$\,D/G$_{\rm MW}$.
Still, while a large D/G ratio does increase the photoelectric heating to significant amounts, the dust SED becomes overestimated if the D/G is larger than $1/300$\,D/G$_{\rm MW}$.
In any case, there is no solution that reproduces well the entire dust SED under a uniform D/G assumption. This leaves us with three possibilities: modify the grain size distribution, modify the topology of the model (e.g., inclusions of dust-rich ionized gas with a relatively large ionization parameter), or keep the topology unchanged but modify the D/G spatial distribution. We investigate in Section\,\ref{sec:nonunidgr} only the latter two possibilities because of the lack of constraints on the grain size distribution in such an extreme environment as \izw.

\subsubsection{Non-uniform D/G ($\mathcal{M}2d$, $\mathcal{M}2e$)}\label{sec:nonunidgr}

\begin{figure}
\includegraphics[angle=0,width=9cm,height=142pt,clip=true,trim=0 58pt 0 22]{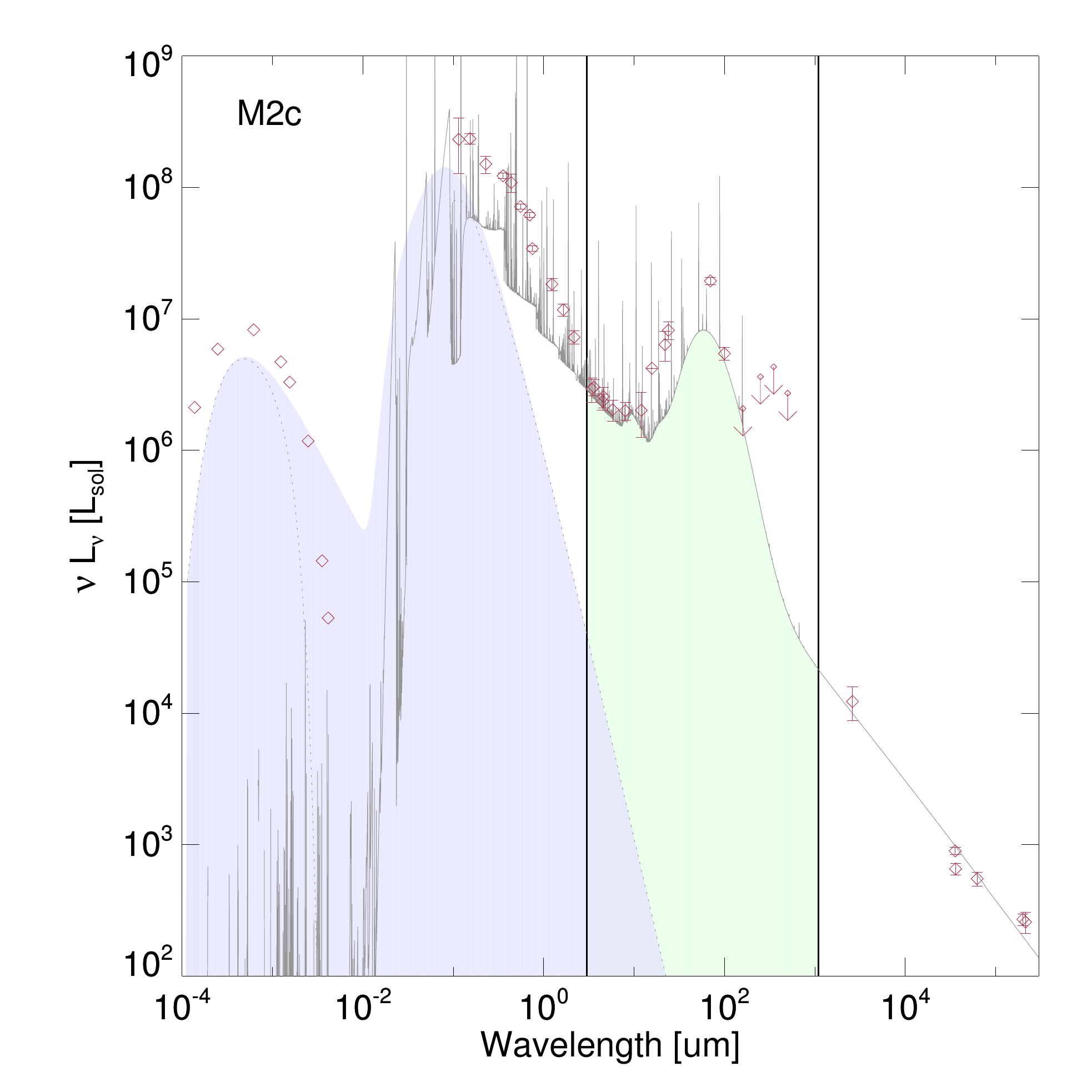}
\includegraphics[angle=0,width=9cm,height=142pt,clip=true,trim=0 58pt 0 22]{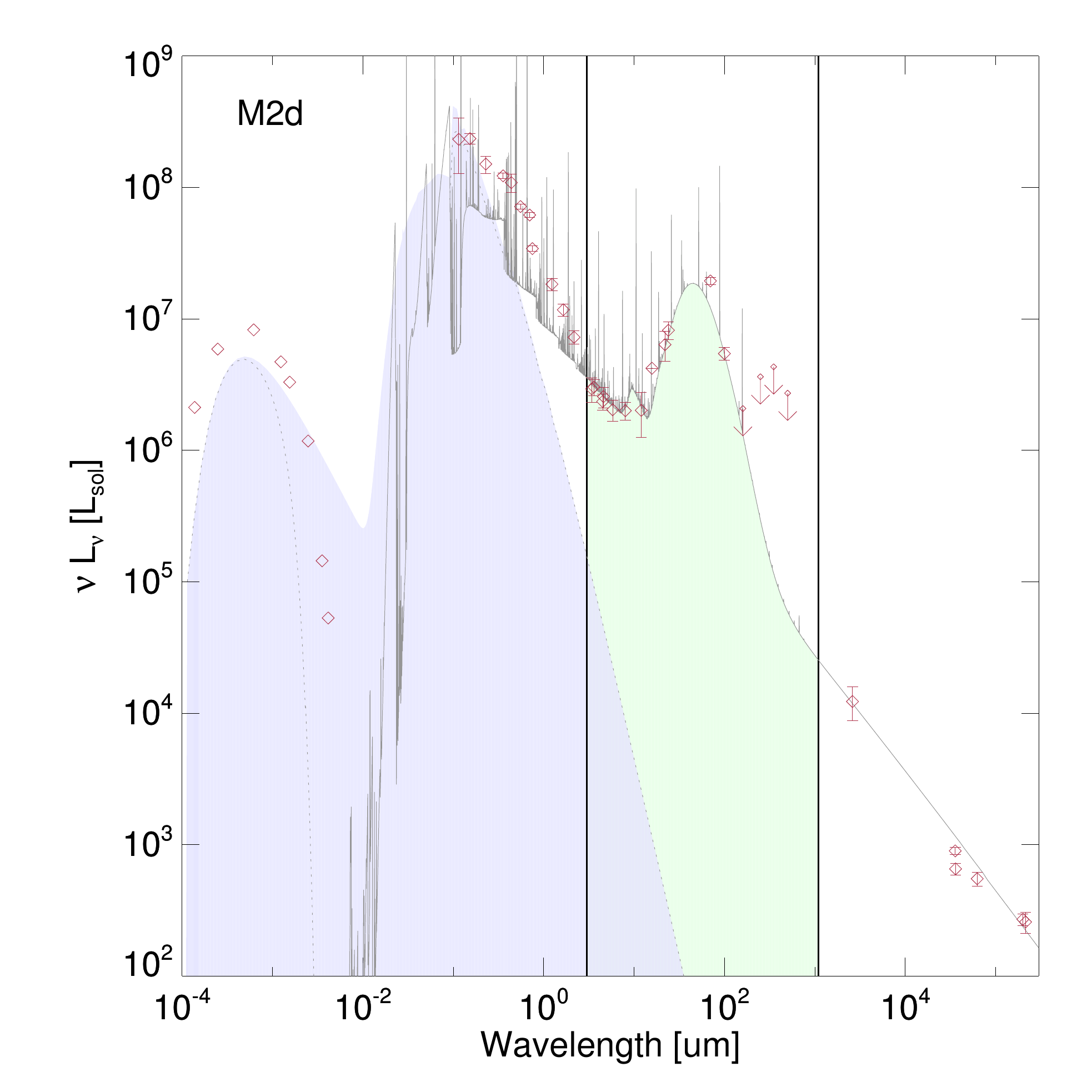}
\includegraphics[angle=0,width=9cm,height=150pt,clip=true,trim=0 18pt 0 22]{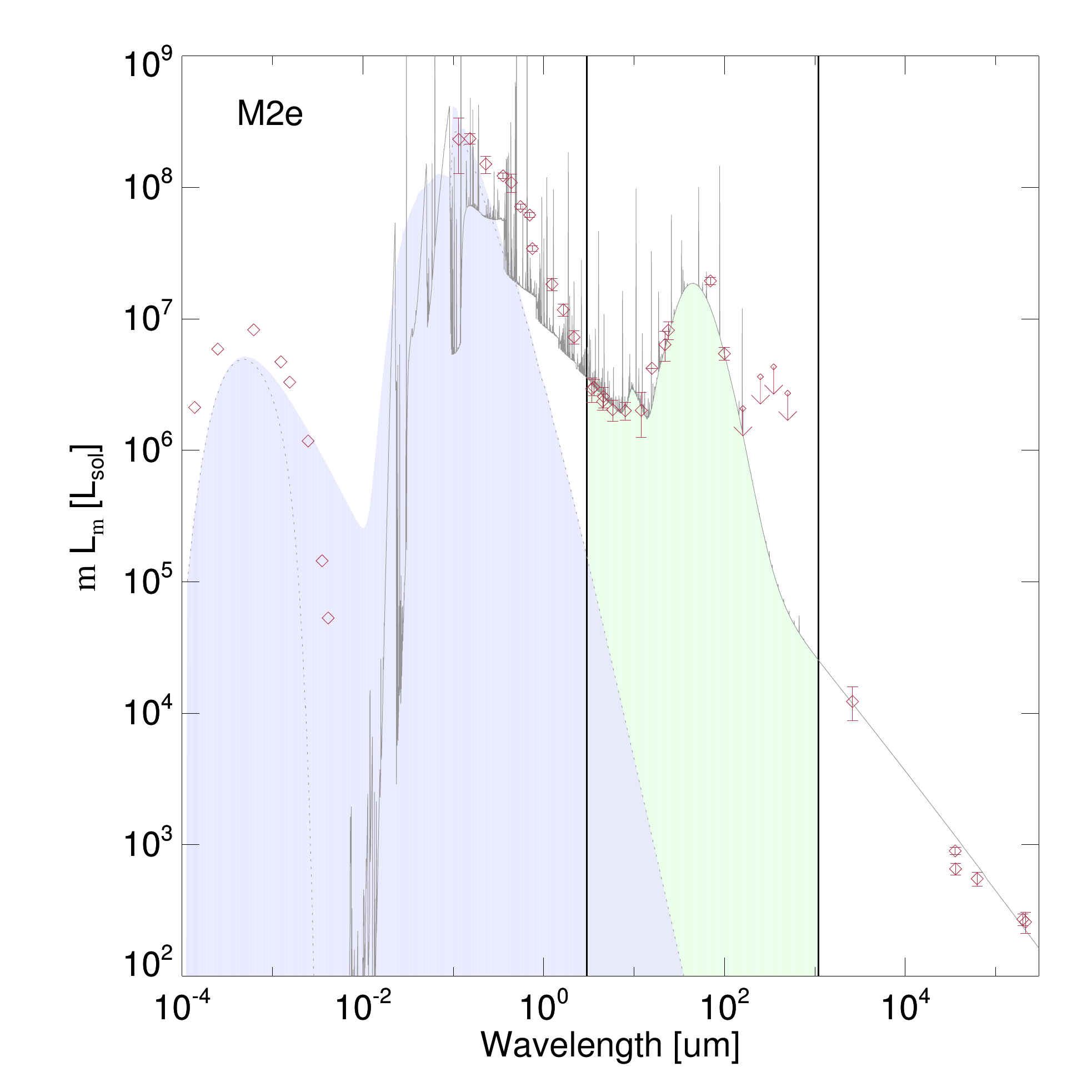}
\caption{Attempts to reproduce the dust SED, $\mathcal{M}2c$ (D/G maximum assuming standard topology), 
$\mathcal{M}2d$ (radial D/G), and $\mathcal{M}2e$ (extra sector with a relatively larger D/G). See Figure\,\ref{fig:model_pe} for the plot description.  }
\label{fig:sed2}
\end{figure}

In Section\,\ref{sec:uniformdgr} we assumed that the D/G was uniform across the galaxy. The measurement of \cite{Remy13b} only provides an average D/G. One way to reproduce the observed dust SED could be to keep the topology unchanged whilst modifying the D/G along some lines of sight. For instance, the dust SED assuming a radially decreasing D/G ($\mathcal{M}2d$) agrees better with the observations than models with uniform D/G (Fig.\,\ref{fig:sed2}). The photoelectric effect in model $\mathcal{M}2d$ represents only $\approx12$\%\ of the H\1\ heating (Table\,\ref{tab:models}). Even then, model $\mathcal{M}2d$ still somewhat overestimates the cold dust emission and therefore the total dust mass.

Another way to reproduce the observed SED is to alter the model topology. While in principle one could simply change the topology assuming a uniform D/G, in practice no satisfactory solution is found.
We modify model $\mathcal{M}2a$ ($1/1000$\,D/G$_{\rm MW}$) to include an extra sector with $1/50$\,D/G$_{\rm MW}$ and with a relatively small covering factor ($\mathcal{M}2e$). The extra sector can represent hypothetical dust-rich inclusions within the H\2\ region. \cite{Cannon02} found that several dust-rich regions were located between the SE and NW regions and also close to ionizing sources in the SE region (finding based on variations of the observed H$\alpha$/H$\beta$ ratio, even though part of these variations may be ascribed to collisional excitation of H$^0$ in the H\2\ region; see \citetalias{Pequignot08} for more details). A close look at the \textit{Spitzer} and \textit{Herschel} photometry maps reveals that dust emission peaks in between the SE and NW regions (Fig.\,\ref{fig:dustmaps}). In the nebular and stellar emission maps shown in \cite{Cannon02}, one can notice that compact nebular emission is associated with several stellar sources in SE\footnote{This is in contrast with the NW region where most of the nebular emission seems to be offset from the stars (except in the NW\,D1 H\2\ region), in filaments forming an incomplete shell.} and the corresponding dust emission could peak at relatively short wavelengths because of the compactness of the region. Whether the extra sector may correspond to the SE region or to dust-rich inclusions in the NW region makes little difference in our approach, and the only important constraints are the matter-bounded versus radiation-bounded nature of the new sector, the total H$\beta$ luminosity, and to a lesser extent, the radio free-free emission. If the X-ray source illuminates this new sector (i.e., sector within NW), then it must be matter bounded, otherwise too much cold dust emission would be produced. On the other hand, if the impinging X-ray flux is fainter (i.e., clouds in  SE), the extra sector does not necessarily need to be matter  bounded.

In model $\mathcal{M}2e$, the new sector has a uniform gas density of $20$\cc, extends from $50$\,pc to $\approx100$\,pc from the stellar cluster, and has a D/G of $1/50$\,D/G$_{\rm MW}$ and covering factor of $6$\%. The dust mass in this new sector is small, i.e., only $\approx17$\,M$_\odot$. The total dust mass is $\approx200$\,M$_\odot$, i.e., somewhat lower than the determination by \cite{Remy15}. The global D/G (all sectors combined) is $\approx1/1000$\,D/G$_{\rm MW}$. 
The dust-rich sector parameters were tuned to reproduce the observed dust SED and the solution is not unique. Still, while the dust SED is relatively well reproduced (Fig.\,\ref{fig:sed2}), the presence of this new sector or its origin in NW or SE has little importance for [C\2] and [O\1] since these lines mostly come from the radiation-bounded region dominated by X-ray heating.

Overall, considering the lack of observational constraints on the grain size distribution and on the small-scale distribution of  D/G in \izw, we cannot prefer any of the models that are able to reproduce the observed dust SED (in particular the mid-IR data points). 
At any rate, the evidence points toward a non-uniform D/G, with a D/G globally lower than $1/50$\,D/G$_{\rm MW}$. 
The various attempts at reproducing the observed dust SED shape actually have little impact on the predicted [C\2] and [O\1] line emission and on the required input parameters (in particular the X-ray luminosity). 
For all models that satisfactorily reproduce the SED shape and dust mass, the fraction of gas heating provided by the photoelectric effect is always $\lesssim12\%$.

\subsection{Exploratory models $\mathcal{M}3$: Cosmic ray heating}\label{sec:cosmicrays}

We now consider CR ionization as an additional parameter to the initial model $\mathcal{M}1$. For this test, dust is ignored. While the dependence of the CR heating rate on metallicity is uncertain, the CR ionization rate is expected to be proportional to the SN rate because CR are produced in SN remnants and increase their energy in SN shock fronts. The CR ionization rate is thus expected to scale roughly with the SFR (e.g., \citealt{Wolfe03a,Abdo10}). There is a generally good correlation between the IR luminosity and the radio emission in star-forming galaxies (e.g., \citealt{Yun01}), which seems to hold even in metal-poor BCDs \citep{Wu08a}. Although \izw\ is an outlier in the IR-radio relation, possibly hinting at an unusually large radio emission, this is also easily explained by the low D/G and/or by a low fraction of UV photons absorbed by dust. In fact, the SFR derived from $1.4$\,GHz and from H$\alpha$ agree within a factor of two (Sect.\,\ref{sec:sfrlit}) and since the integrated $1.4$\,GHz emission is dominated by synchrotron radiation \citep{Cannon05b}, this suggests that the CR ionization rate may indeed scale with the SFR in \izw. Nevertheless, the choice of SFR for \izw\ is not straightforward. Most SFR tracers (H$\alpha$, FUV+$24$\mic, or $1.4$\,GHz) agree with $0.05-0.1$\,M$_\odot$\,yr$^{-1}$, but a CMD analysis indicates a larger value of $\sim1$\,M$_\odot$\,yr$^{-1}$ over the last $10$\,Myr \citep{Annibali13} (Sect.\,\ref{sec:sfrlit}). 

Considering the lack of constraints on the actual CR ionization rate, several values are tested in the following. As for the test of the photoelectric effect (Sect.\,\ref{sec:cloudype}), we calculate a grid with varying X-ray luminosity and find the best model that reproduces the observations. The description of the CR heating properties in Cloudy is assumed to hold in \izw. We first consider a Galactic background ionization rate ($2\times10^{-16}$\,s$^{-1}$; e.g., \citealt{Indriolo15}) and let Cloudy calculate  the heating rate self-consistently (model $\mathcal{M}3a$ in Table\,\ref{tab:models}). The CR heating becomes significant only at large depths into the cloud. The required X-ray luminosity is marginally lower than the model with no CR, which is due to the additional heating from CR ionization. The heating rate contribution due to CR ionization in the H\1\ region is $\approx13$\%. 

In model $\mathcal{M}3b$ we test a CR rate that is five times the Galactic background rate. The heating rate in the H\1\ region due to CR ionization becomes $\approx65$\%\ and the required X-ray luminosity decreases to $\approx1.25\times10^{40}$\,erg\,s$^{-1}$. Although the H$^0$ mass is fairly well reproduced, we notice that as the CR ionization rate increases it becomes more and more difficult to find a satisfactory solution for [C\2], [O\1], and [Si\2], with either [O\1]/[C\2] or [Si\2]/[C\2] increasing with the CR ionization rate.

In order to isolate and test the heating efficiency of CR ionization alone, we briefly considered a model with no X-ray but with a CR ionization rate scaled to reproduce the observed [C\2] flux. For a Galactic background rate ($\mathcal{M}3c$), the H$^0$ mass required to reproduce [C\2] and the other H\1\ region lines is significantly larger than observations (Table\,\ref{tab:models}). For a rate five times larger ($\mathcal{M}3d$), observations are relatively well reproduced, although the prediction for [Si\2] is somewhat low. 
This shows that the heating provided by the X-ray source in \izw\ is in effect similar to what CR ionization could provide with a flat rate of $\approx10^{-15}$\,s$^{-1}$ across the H\1\ region. 
Since the characteristics of the X-ray source in \izw\ are well established, we argue that the large ionization fraction observed in the H\1\ region is the consequence of the X-ray source.

\subsection{Full model $\mathcal{M}4a$}\label{sec:finalmodel}

The full model $\mathcal{M}4a$ combines all heating mechanisms (photoelectric effect, soft X-ray and CR ionization).
We tentatively scale down the CR ionization rate by a factor of $10$ as compared to the Galactic background (i.e., assuming a scaling with SFR) and keep in mind that the global heating in the H\1\ region due to CR ionization remains under $13$\%\ even if a Galactic background rate is chosen (Sect.\,\ref{sec:cosmicrays}).
We include an additional dust-rich sector to reproduce the dust SED (Sect.\,\ref{sec:nonunidgr}). This model therefore reproduces well the observed [C\2], [O\1], and [Si\2] fluxes, the suite of \textit{Spitzer} and optical H\2\ region lines, and the H$^0$ mass. The dust mass is somewhat underestimated and the dust SED is qualitatively reproduced.

\begin{figure}[h!]
\centering
\includegraphics[angle=0,width=9cm,height=5cm,clip=true,trim=0 1.8cm 0 0.8cm]{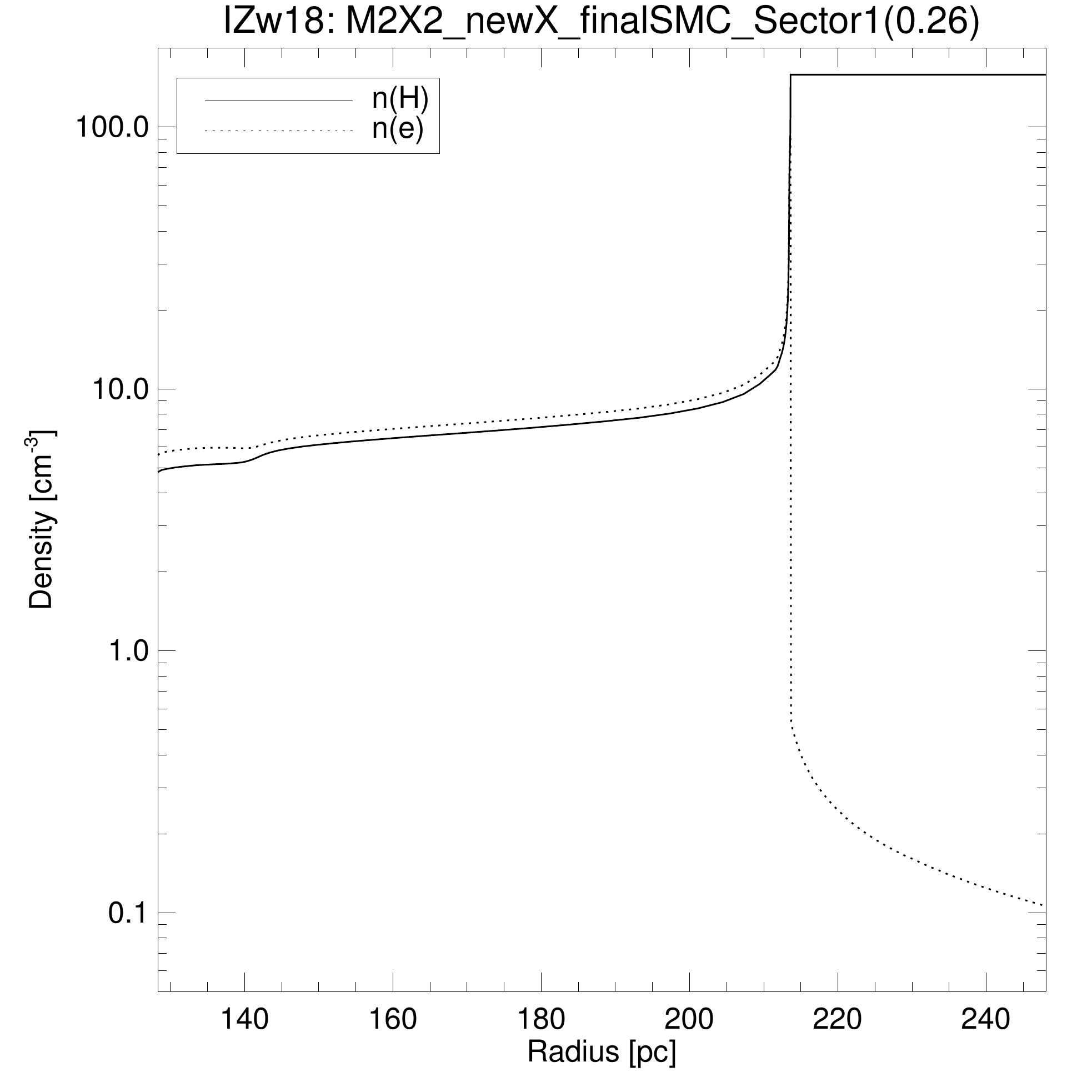}
\includegraphics[angle=0,width=9cm,height=5.5cm,clip=true,trim=0 0cm 0 0.8cm]{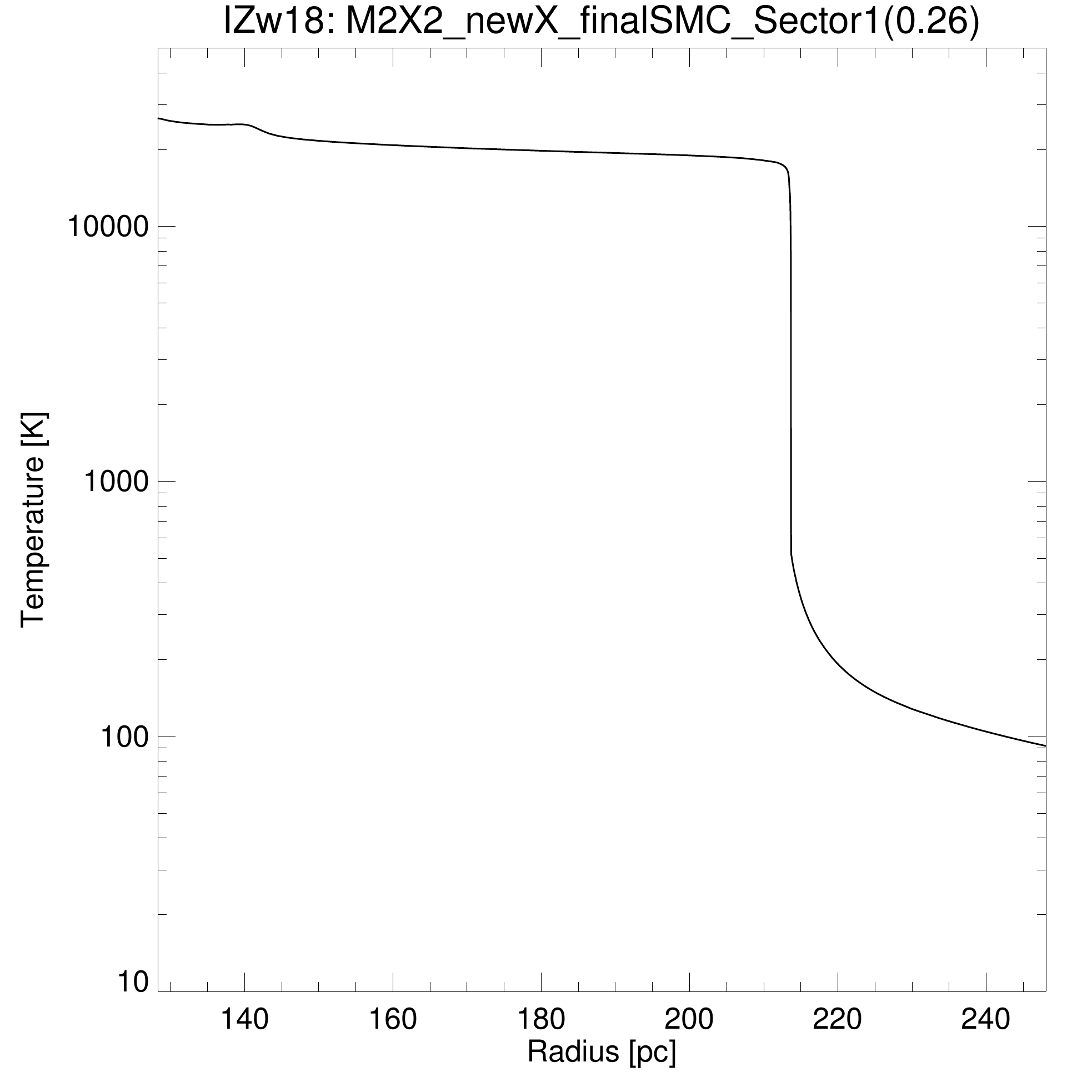}
\caption{Radial density and temperature in the radiation-bounded sector of the full model $\mathcal{M}4a$. The plots are shown for the radiation-bounded sector.  }
\label{fig:model_final1}
\end{figure}

\begin{figure}[h!]
\centering
\includegraphics[angle=0,width=9cm,height=12cm,clip=true,trim=0 0cm 0 0cm]{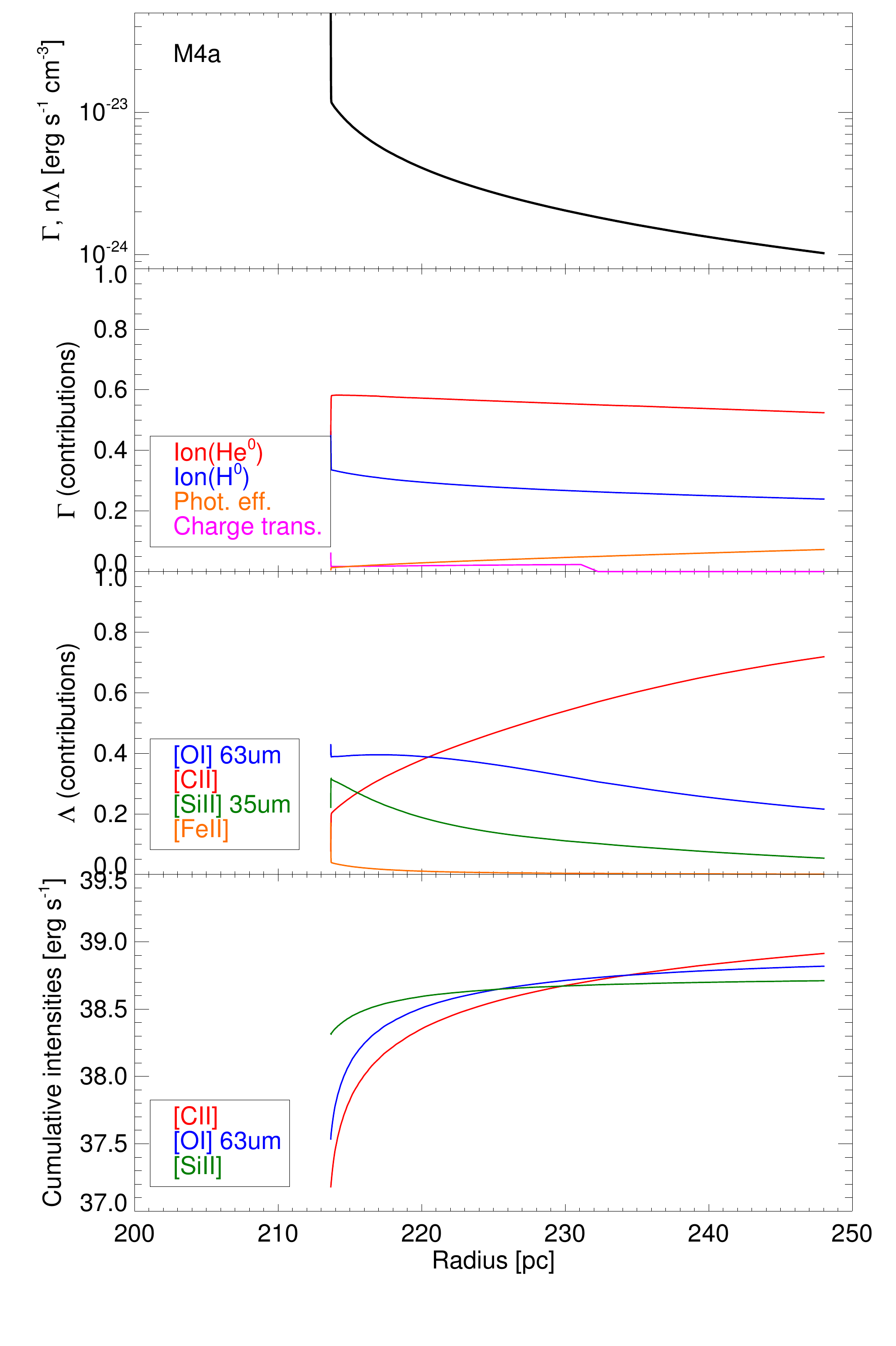}
\caption{Heating and cooling rates for the final model ($\mathcal{M}4a$). The plots are shown for the radiation-bounded sector. See Figure\,\ref{fig:model_pe} for the plot description. }
\label{fig:model_final2}
\end{figure}

The optical and infrared ionized gas line intensities predicted by model $\mathcal{M}4a$ are in good agreement with observations (Tables\,\ref{tab:cloudycompopt}, \ref{tab:cloudycompir}) and with Nebu (Sect.\,\ref{sec:p08}). The presence of dust has little impact on the predicted optical line intensities. The observed He\1\ intensities are affected by stellar line contamination (see \citetalias{Pequignot08} for more details). The [S\4] and [Ar\4] lines are also comparatively not well reproduced. The ionization balances of both S and Ar are thought to be affected by uncertain dielectronic recombination rates and we refrain at this stage from elaborating on these discrepancies.

Interestingly, the flux of [Ne\5] $14.3$\mic\ predicted by model $\mathcal{M}4a$ is only a factor of $\approx5$ below the current observed upper limit. 
Since the [Ne\5] emission traces the soft X-ray radiation, deeper observations with the \textit{James Webb} Space Telescope (JWST) may be useful to constrain the X-ray source emission properties in a hardly accessible spectral range (see also Appendix\,\ref{secapp:xspec}). The JWST will also be able to test model predictions for other faint neutral gas tracers, such as [Ar\2] and [Fe\2]. Nonetheless present collisional rates with H$^0$ are inaccurate for these lines.

Figure\,\ref{fig:model_final1} shows the density and temperature profiles for the radiation-bounded sector of model $\mathcal{M}4a$. The electron fraction ranges from $\sim0.5$\%\ down to $\sim0.05$\%\ and the electron temperature remains above $50$\,K throughout the H\1\ region. Figure\,\ref{fig:model_final2} shows the cooling and heating contributions. One can see that, as expected, [C\2], [O\1], and [Si\2] are the main coolants in the H\1\ region, with [O\1] and [Si\2] dominating at the surface of the cloud and [C\2] dominating deeper. The optical depth values of [C\2], [O\1], [Si\2], and [Fe\2] $26.0$\mic\ are $0.5$, $2.3$, $0.3$, and $0.3$, respectively; in a static model, i.e., no internal motions and assuming only thermal broadening for the lines. 

The observed low D/G (Sect.\,\ref{sec:const_dustmass}) and low PAH abundance in \izw\ (e.g., \citealt{Wu06,Remy13b}) already suggest that dust grains and PAHs do not contribute significantly to the gas heating. 
This is confirmed by our model in which the main heating source in the H\1\ region is the photoionization of He$^0$ and H$^0$ by the X-ray source. The photoelectric effect heating increases deeper into the cloud but never contributes to more than $\sim10$\%\ of the local heating rate. 

Our model is helpful to understand the origin of the total infrared (TIR) emission. About $85$\%\ of the TIR in the model comes from the ionized phase in the H\2\ region (i.e., the two matter-bounded sectors, the additional dust-rich sector, and the radiation-bounded sector before the ionization front). 
With regard to the physical processes accounting for TIR, about $60$\%\ of the TIR in the model arises from free-free emission in the ionized phase, with the remaining $40$\%\ arising from dust (mostly in the ionized phase). The important contribution from the ionized phase in TIR implies that the ratio $\epsilon'_{\rm TIR} =$([C\2]+[O\1])/TIR cannot be used to probe the photoelectric effect heating efficiency in the H\1\ region (see also discussion in Sect\,\ref{sec:ciitir}).

\begin{table*}
\caption{Comparison between the observed IR line fluxes and models.\label{tab:cloudycompir}}
\centering
\begin{tabular}{ll|l|llll}
\hline\hline
Line &  Observed &  Nebu$^\textrm{a}$ & $\mathcal{M}4a$ &  $\mathcal{M}4b$ & $\mathcal{M}4c$  \\
\hline
\textbf{Ionized gas} &   &   &  &  \\
 H\1\ Hu$\alpha$ $12.37$\mic & $8.5\pm3$   & 9.5 & 8.9 & 8.9 & 9.1 \\
$[$O\3$]$ 88\mic  & $200\pm30$    & 207.9 & 188.0 & 188.1 & 188.7  \\
$[$O\4$]$ 25.9\mic  &   $38\pm8$  &  49.6 & 45.3 & 45.3 & 43.6 \\
$[$Ne\2$]$ 12.8\mic   & $6\pm3$ &  3.0  & 2.0 & 2.0 & 2.1 \\
$[$Ne\3$]$ 15.6\mic  &  $47\pm10$ &  47.0  & 45.7 & 45.7 & 45.9 \\
$[$Ne\3$]$ 36.0\mic   & $<7$ & 4.2 &  4.1 & 4.1 & 4.2  \\
$[$Ne\5$]$ 14.3\mic   & $<2.3$ &  0.7 & 0.8 & 0.8 & 0.6 \\
$[$Ne\5$]$ 24.3\mic   & $<4$ & 0.8 & 1.0  & 1.0 & 0.7  \\
$[$S\3$]$ 18.7\mic   & $28\pm7$  & 28.3 &  23.3 & 23.3 & 24.3 \\
$[$S\3$]$ 33.5\mic  &  $40\pm7$ & 51.6 & 47.7 & 47.7 & 50.0 \\
$[$S\4$]$ 10.5\mic   & $49\pm10$ & 117.7   & 107.7 & 107.7 & 103.2 \\
$[$Ar\2$]$ 7.0\mic    &  $<5$ & 1.4  & 0.5  & 0.5 & 0.5 \\
$[$Ar\3$]$ 9.0\mic     & $<10$   & 8.1 & 6.3 & 6.3 & 6.6 \\
$[$Ar\3$]$ 21.8\mic    &   $<8$  & 0.6 &  0.5 & 0.5 & 0.5 \\
$[$Fe\3$]$ 22.9\mic  &   $<6$ & 3.2 &  2.8 & 2.8 & 3.1 \\
\hline
\textbf{Neutral gas} &   &   &  &  \\
$[$C\1$]$ 609\mic  & ... & ... &  $3\times10^{-4}$ &  $3\times10^{-4}$ & $0.2$  \\
$[$C\1$]$ 370\mic  &  ... & ... & $6\times10^{-4}$ & $6\times10^{-4}$ & $0.8$ \\
$[$C\2$]$ 157\mic  &  $76\pm20$ & $77.1$ & 79.0 & 77.0 & 77.9 \\ 
$[$O\1$]$ 63\mic  &  $60\pm20$    & $60.5$ & 63.3 & 63.2 & 68.3 \\
$[$Si\2$]$ 34.8\mic  &  $65\pm12$  &  $52.6$  & 49.9 & 49.9 & 54.7 \\
$[$Fe\2$]$ 17.9\mic  & $<3$   & $1.6$ &  1.1 & 1.1 & 1.1 \\
$[$Fe\2$]$ 26.0\mic   & $13\pm8$  & $9.6$ &  4.4 & 4.4 & 4.7 \\
\hline
\textbf{Molecular gas} &   &   &  &  \\
H$_2$ 0-0 S(0) 28.2\mic  &  $<7$      & ...  & ... &  $7\cdot10^{-6}$ & $2.7$\\
H$_2$ 0-0 S(1) 17.0\mic & $<7$         & ...  & ... &   $2\times10^{-4}$ & $0.1$ \\
H$_2$ 0-0 S(2) 12.3\mic & $<7.5$           & ... & ... & $2\times10^{-6}$ & $5\times10^{-5}$ \\
H$_2$ 1-0 S(1) 2.12\mic & $<2^\textrm{b}$ & ... & ..& $7\times10^{-4}$ &  $3\times10^{-3}$  \\
CO(1-0)   &   $<8\times10^{-3,\textrm{c}}$  & ... & ... & $2\times10^{-11}$ & $8\times10^{-3}$ \\ 
PAH (6-15\mic)     &  $<190$  & ... & ... & 200.0 & 212.4 \\ 
\hline
 \end{tabular}\\
\tablefoot{Fluxes in units H$\beta=1000$. }
\tablefoottext{a}{See Section\,\ref{sec:p08}. }
\tablefoottext{b}{From \cite{Izotov15}, assuming H$_2$ $2.122$\mic/Br$\gamma$ $\sim0.1$ and Br$\gamma$/H$\beta<1.7$\%.}
\tablefoottext{c}{\cite{Leroy07b}.}
\end{table*}

\section{Model discussion}\label{sec:discussion}

\subsection{Molecular gas content}\label{sec:dismol}

Our models, describing the H\1\ region physical properties of \izw, provide useful constraints to the molecular gas content. We consider two cases: molecular gas in the uniformly distributed gas of the H\1\ region model (Sect.\,\ref{sec:moluni}) and in putative small dense clumps (Sect.\ref{sec:molclump}).

\subsubsection{Molecular gas in the absence of dense clumps (variant $\mathcal{M}4b$)}\label{sec:moluni}

Molecular gas has not been detected yet in \izw, neither from diffuse H$_2$ absorption lines \citep{Vidal00}, H$_2$ near-IR emission lines \citep{Izotov15}, H$_2$ mid-IR emission lines (\citealt{Wu06}; this study), nor CO(1-0) \citep{Leroy07b}. Also, PAH emission has not been detected (\citealt{Wu06,Wu08b}; this study). 
This apparent lack of molecular gas is remarkable considering the recent formation of massive stars. While low observed CO luminosity leads to the hypotheses that either the molecular gas is more efficiently converted into stars or that CO does not trace the molecular gas, the overall scarcity of molecules could also imply that molecular gas may in fact not be a prerequisite for star formation to proceed (e.g., \citealt{Glover12a}). Following \cite{Krumholz12a}, star formation could proceed in atomic gas if the temperature is cold enough, with molecular gas forming  but only at latest stages, shortly before it is eventually destroyed by the FUV radiation from the newly formed massive stars. Alternatively, molecular gas could have been formed shortly before the star formation episode and then destroyed by UV radiation and shock from the newly born stars. 

In model $\mathcal{M}4b$ (Table\,\ref{tab:models}), molecular gas is uniformly distributed in the relatively diffuse gas ($\lesssim200$\cc) of the entire radiation-bounded sector (i.e., using the same sector configuration as model $\mathcal{M}4a$; Fig.\,\ref{fig:sectors}). 
The molecular gas that could be hidden in small clumps is discussed separately in Section\,\ref{sec:molclump}. 

The ``large'' H$_2$ molecule model is used in Cloudy (described in \citealt{Shaw05}), with fully consistent formation and destruction rates. Several iterations are performed in order for the line optical depth to be firmly established. The main processes for H$_2$ formation are the H$^-$ route (${\rm H}^-+{\rm H} \rightarrow {\rm H}_2+e^-$) and the dust route (e.g., \citealt{LeBourlot12}). The efficiency of the H$^-$ route is a strong function of the X-ray luminosity. If the X-ray source were absent, free electrons would be provided mostly by neutral carbon and iron ionization. The H$^-$ route prevails over the dust route in the following conditions:
\begin{itemize}
\item[$\bullet$] The electron fraction in the H\1\ region is significant, providing free electrons for producing H$^-$ through the reaction ${\rm H}+e^- \rightarrow {\rm H}^-+\gamma$ (e.g., \citealt{Jenkins97}). 
\item[$\bullet$] The medium density is $\gtrsim100$\,\cc, below which the H$^-$ photodetachment (${\rm H}^-+\gamma \rightarrow {\rm H}+e^-$) starts to dominate (e.g. \citealt{Kamaya01,Chuzhoy07}). Furthermore, regulation by electron recombination prevents a large molecular gas fraction through the H$^-$ route unless the density becomes much larger than $\sim10^4$\,\cc\ (for a metallicity adapted to \izw; \citealt{Omukai05}).  
\item[$\bullet$] The D/G falls below some critical value (see \citealt{Glover02}),
\end{itemize}

The formation of H$_2$ in dense clouds can therefore be X-ray-induced (e.g., \citealt{Haiman96}), with a significant number of free electrons provided by primary and secondary ionization. The number of secondary ionizations is $\approx25$ for a $1$\,keV photon\footnote{Number of secondary ionizations is $25\ (E\gamma/1\,{\rm keV})$ \citep{Shull85}}. The competing photodissociation of H$_2$ by Lyman-Werner photons may, however, still partly control the molecular gas fraction if the X-ray and UV sources are correlated through similar origin and timescale (e.g., \citealt{Machacek03}). The positive feedback of X-rays could manifest itself mostly if the UV field is weak but X-rays are still able to ionize interiors of dense clouds. The timescales are critical, and if the X-ray source survives long enough after the UV field weakens, H$_2$ formation in dense clouds may be enhanced, thereby providing an important cooling. 

Model $\mathcal{M}4b$ predicts an H$_2$ column density of $4\times10^{13}$\,cm$^{-2}$  that is lower than the observed upper limit $\lesssim10^{15}$\,cm$^{-2}$ measured by \cite{Vidal00} with FUV absorption lines. The authors also calculated a theoretical column density of $2\times10^{12}$\,cm$^{-2}$ but underestimated the importance of the H$_2$ formation process via the H$^-$ route by assuming that free electrons were only provided by ionization of carbon. 
In model $\mathcal{M}4b$, the H$^-$ formation route contributes to $\sim60$\%\ of the H$_2$ mass. H$_2$ is mostly formed through the H$^-$ route at the inner edge of the H\1\ shell, while the dust route eventually dominates deeper in the neutral cloud. 

We therefore propose that the diffuse (as opposed to the clumps discussed in Section\,\ref{sec:molclump}) H$_2$ column density may be close to a few $10^{13}$\,cm$^{-2}$ based on our models, bearing in mind that the latter may not account for an accurate description of the photodissociation in a clumpy medium.
Even gas with extremely low molecular gas fraction can produce a significant amount of self-shielding with a large enough total column density. Figure\,1 of \cite{Draine96} shows the self-shielding function without any dust. One can see that the self-shielding factor becomes important ($0.5$; i.e., half of the flux in the Lyman and Werner bands is absorbed) for $N({\rm H}_2)\approx2.5\times10^{14}$\,cm$^{-2}$. In \izw, the column density is thus barely sufficient, if at all, to enable H$_2$ self-shielding unless it is distributed into dense clumps (Sect.\,\ref{sec:molclump}).

The H$_2$ mass distributed within the radiation-bounded sector of model $\mathcal{M}4b$ is only $\sim0.1$\,M$_\odot$ (corresponding to a molecular gas mass fraction of $\sim5\times10^{-9}$). 
In the sector configuration of $\mathcal{M}4b$, the [C\2] and [O\1] emission lines thus trace an almost purely atomic medium.

The CO(1-0) intensity predicted by model $\mathcal{M}4b$ is $2\times10^{-30}$\,W\,m$^{-2}$, to be compared to the upper limit by \cite{Leroy07b} of $8\times10^{-22}$\,W\,m$^{-2}$ (with a beam of $\approx380$\,pc). Both the sensitivity and the beam filling factor could explain in part why the current observed upper limit is so large as compared to the model.

\subsubsection{Dense molecular clumps (variant $\mathcal{M}4c$)}\label{sec:molclump}

As mentioned in Section\,\ref{sec:moluni}, absorption-line studies did not succeed in detecting H$_2$ in the diffuse ISM along the lines of sight toward the massive stars. This is in agreement with model $\mathcal{M}4b$. Nonetheless, H$_2$ may still exist in significant amounts out of the available lines of sight toward the massive stars in NW, which are located within an H\1\ hole (Fig.\,\ref{fig:hipacs}), far from the dust-rich regions (Sect.\,\ref{sec:nonunidgr}). For example, H$_2$ may belong to undetected, relatively small, and dense clumps \citep{Vidal00}. \cite{Thuan05a} proposed the existence of such clumps in the BCD SBS\,0335$-$052 based on the lack of diffuse H$_2$ together with the detection of warm H$_2$ emission in the near IR. 
\cite{Rubio15} observed small CO clumps with ALMA in the low-metallicity galaxy \object{WLM} ($12+\log({\rm O/H})=7.8$) of $1.5$ to $6$\,pc in size, which begs the question of whether such clumps can exist in \izw.

Similar to \cite{Cormier12}, who modeled photodissociation regions (PDRs) in the dwarf galaxy \object{Haro 11}, we consider dense clumps in \izw\ with a small covering factor as compared to the H\1\ shell (Fig.\,\ref{fig:sectors}). In other words, only a small fraction of the radiation-bounded sector reaches the molecular phase. Since there are no constraints on the mere presence of clumps, we bear in mind that the following models are only exploratory. Apart from the  upper limits on CO, H$_2$, and PAH emission, another observational constraint is the dust SED and in particular the long wavelength emission corresponding to cold dust.

\begin{figure}
\includegraphics[angle=0,width=9cm,height=150pt,clip=true,trim=0 10pt 0 22]{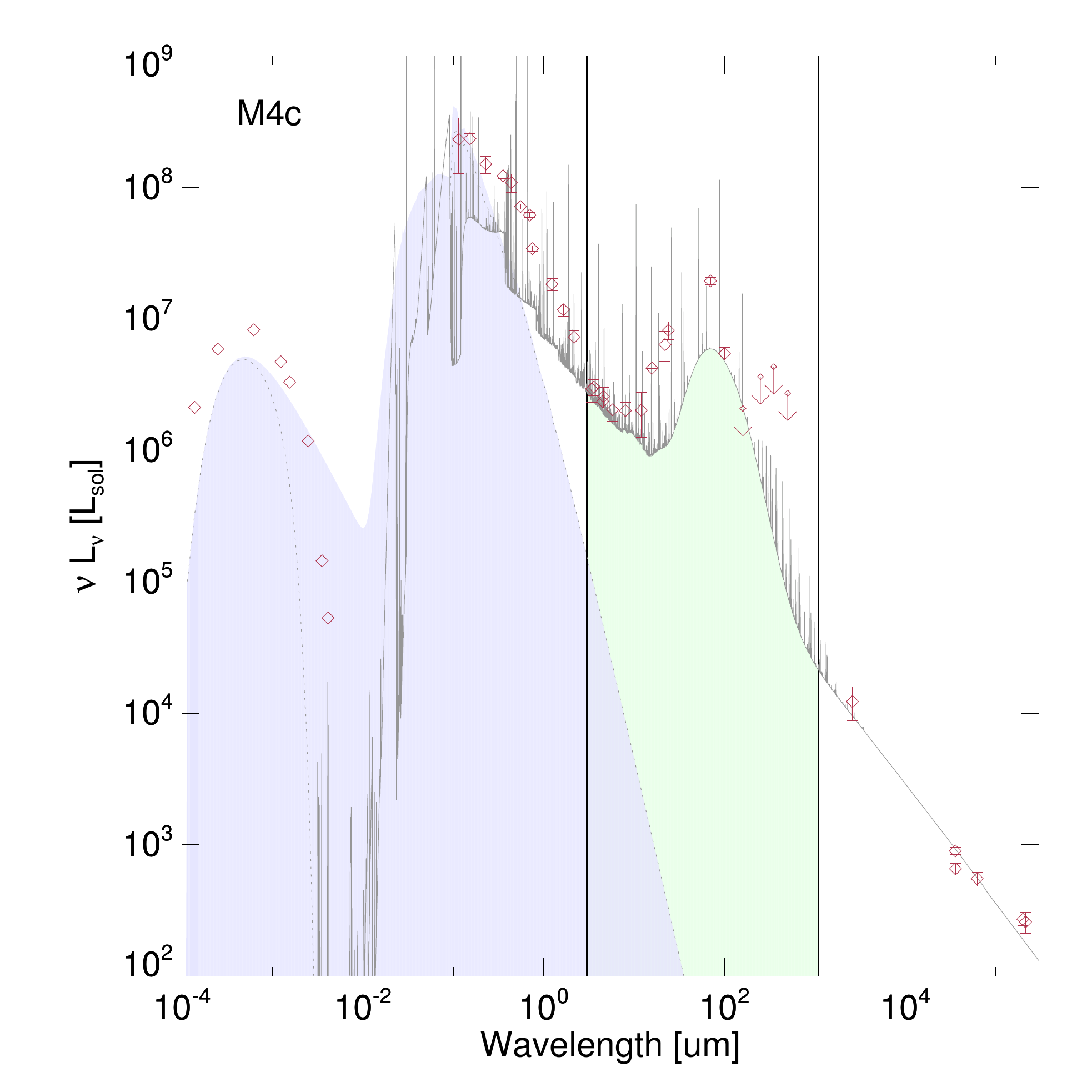}
\caption{Spectral energy distribution for model $\mathcal{M}4c$ (model with molecular clump). See Figure\,\ref{fig:model_pe} for the plot description. The dust-rich sector is ignored for this model. }
\label{fig:sedclump}
\end{figure}

We assume D/G$=1/1000$\,D/G$_{\rm MW}$ across the molecular sector. The dust-rich sector (added to reproduce the observed SED shortward of $100$\mic; Sect.\,\ref{sec:nonunidgr}) is ignored for this test since we wish to study the specific contribution of the clump sector. The density profile across the clump sector is calculated self-consistently by constraining a constant pressure ($P/k=8\times10^{6}$ CGS units). The density profile is thus similar to the radiation-bounded sector (\#1) used in our previous modeling up to the photoionization front, but the density then eventually reaches $\sim10^6$\cc, for which the medium becomes fully molecular (even with no dust grains). 

Various combinations of covering factor and stopping column density can lead to significant H$_2$ mass in the clump sector with more mass produced in variants with low covering factor and large stopping column density. As a test, we consider a sector reaching a total column density of $10^{25}$\,cm$^{-2}$ ($A_V\sim3$) with the covering factor scaled to accommodate the observed constraints (the CO upper limit, the observed dust SED, and the corresponding dust mass).  

In practice, the CO upper limit is the most stringent observational constraint. It is reached for a clump covering factor of $\approx0.1$\%\ (Table\,\ref{tab:cloudycompir}). For such a covering factor, the dust SED remains compatible with observations (Fig.\,\ref{fig:sedclump}) and the dust mass is also well reproduced (Table\,\ref{tab:models}). 
The resulting upper limit on the molecular volume is about $14^3$\,pc$^3$ ($\sim25\times25$\,pc$^2$ surface area and $\sim5$\,pc deep). Considering the depth scale and assuming spherical clumps, this should correspond to a couple dozen clumps at most.

\begin{figure}
\includegraphics[angle=0,clip=true,width=9cm]{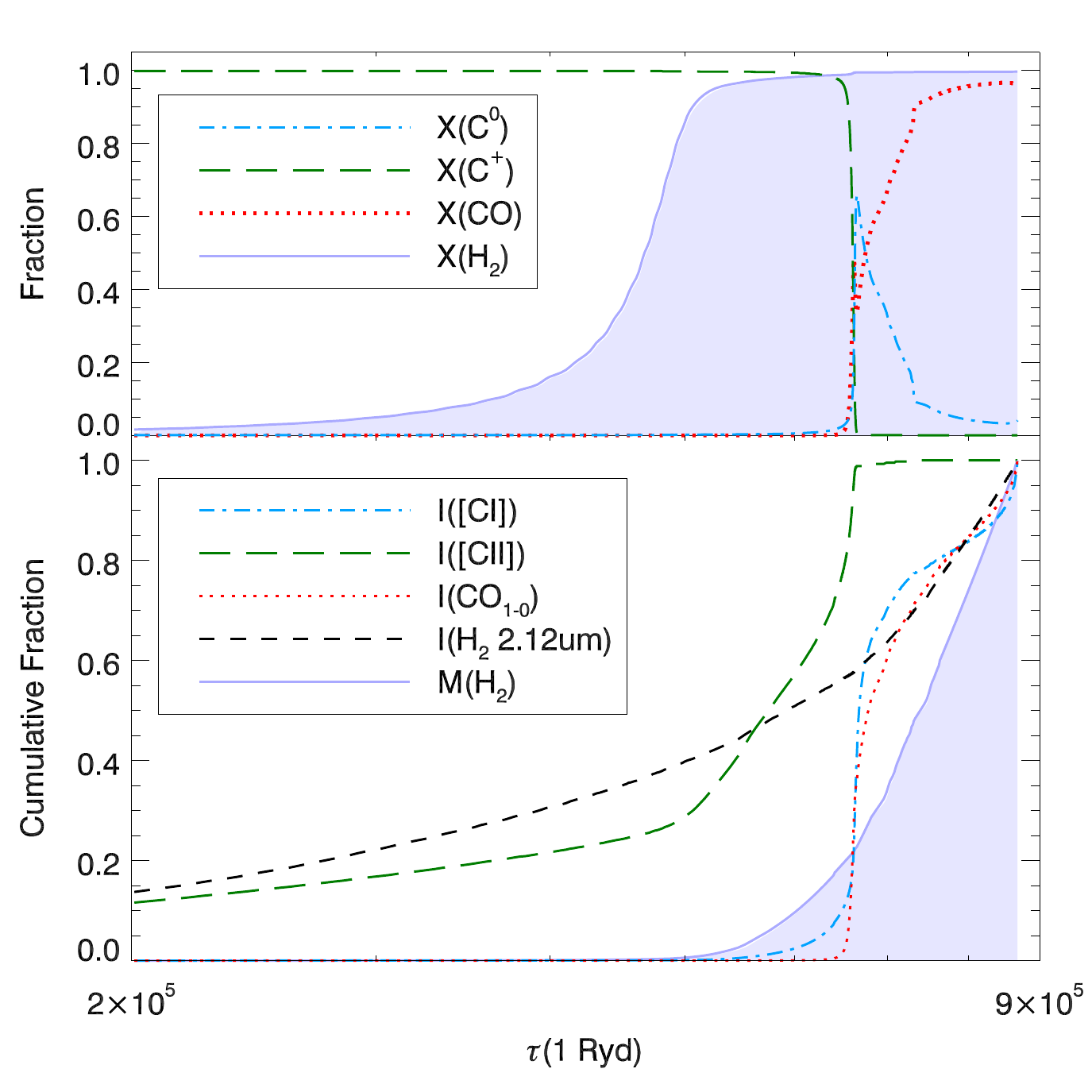}
\caption{\textit{Top}: Abundance fraction of C$^0$, C$^+$, CO, and H$_2$ plotted against the optical depth at $1$\,Ryd for the clump sector of model $\mathcal{M}4c$. \textit{Bottom}: Cumulative fraction of [C\1] $609$\mic, [C\2], CO(1-0), H$_2$ $2.12$\mic, and H$_2$ mass normalized to the maximum value. }
\label{fig:profclump}
\end{figure}

The main heating mechanisms in the clump sector are the photoionization of H and He at small depths, and the collisions with H$_2$, photoelectric effect, and CR ionization at larger depths.
The clump sector is almost fully ($98$\%) molecular in mass. As shown in Figure\,\ref{fig:profclump}, most of the [C\2] emission in the clump sector occurs at depths for which the gas is molecular ($\approx75$\%\ of [C\2] is associated with a molecular gas fraction $>50$\%). However, [C\2] does not trace well the H$_2$ \textit{mass}, since most ($80$\%) of the H$_2$ mass is produced at depths where C is into C$^0$ or CO. Comparatively, [C\1], CO(1-0), or H$_2$ $2.12$ emissions follow the H$_2$ mass profile more closely, although they emit mostly at depths where the H$_2$ mass is still small.

While [C\2] emission is associated with H$_2$ in the clump sector, the fraction of [C\2] arising from this sector is only about $15$\%\ of the total [C\2] in the model (the same fraction is found for [O\1]). Most of the emission arises from the relatively more diffuse H\1\ region in sector \#1 where [C\2] traces H$^0$.  

The H$_2$ mass in model $\mathcal{M}4c$ is $\sim4\times10^7$\,M$_\odot$, i.e., on the same order as the H$^0$ mass. From CO observations and using a point-source luminosity, \cite{Leroy07b} found $M$(H$_2$)$<7.4\times10^5$\,M$_\odot$\footnote{Value rescaled at $18.2$\,Mpc. with a Galactic CO/H$_2$ conversion factor.}. Assuming the star formation efficiency is the same as in the Milky Way, we can argue that $X_{\rm CO}$ is significantly larger. If $X_{\rm CO}$ is $100$ times larger than the Galactic value as proposed by \cite{Leroy07b} $-$ similar to the factor extrapolated from the trend by \cite{Schruba12} $-$ we obtain an upper limit on the H$_2$ mass on the same order as our estimate in $\mathcal{M}4c$. A large $X_{\rm CO}$ might be partly related to low cloud extinction due to low metallicity (i.e., enhanced photodissociation) rather than to a direct abundance effect \citep{Glover11}.

The molecular volume and clump size we estimate are upper limits. First, these values are calculated based on the CO(1-0) observed upper limit. Second, our model does not account for internal motions. The CO emission could be different if we considered turbulence or a velocity gradient, in which case both the molecular volume and clump size could in fact be much smaller.

We conclude that it is possible that significant amounts of molecular gas exist in \izw\ if it is distributed within a few small dense clumps of size $\lesssim5$\,pc ($\lesssim0.05$\arcsec\ at $18.2$\,Mpc). Parsec-size clouds are difficult to observe in absorption because it requires a line of sight intersecting the clump toward a FUV background source. Furthermore, absorption-line measurements of a group of FUV background sources privilege low-extinction lines of sight. While such clumps could be directly detected with high spatial resolution CO observations, it is also possible to observe relatively strong signatures from [C\1] and from the warm molecular layer in the mid-IR H$_2$ lines $-$ in particular S(0) (Table\,\ref{tab:cloudycompir}) $-$ as long as the molecular clouds remain excited by the X-ray source and by the stellar clusters (hence providing a significant gas temperature). 
The near-IR H$_2$ line 1-0 S(1) at $2.12$\mic\ is a few orders of magnitude below the current observed upper limit \citep{Izotov15}, suggesting that $8$\,m class telescopes could possibly reach such levels predicted by our models (e.g., \citealt{Vanzi11}). 
Clumps may also exist at significant distances from the clusters, where they can be near isothermal equilibrium with the CMB and CR, in which case only cold molecular gas can be detected. Another possible observational constraint to such dense clumps is provided by the dust emission longward of $500$\mic\ (Fig.\,\ref{fig:sedclump}).

The existence of molecular clumps in general in \izw\ would shed a new light on the importance of the molecular gas into the baryonic budget of low-metallicity dwarf galaxies; such galaxies have been generally thought to be dark-matter dominated although this is a disputed issue (e.g., \citealt{Swaters11,Ferrero12,Kuhlen13,Sawala14}). 
Such clumps are reminiscent of recurring and self-gravitating cold H$_2$ clumps that may exist in a hierarchical structure of various sizes \citep{Pfenniger94a,Pfenniger94b,Combes97,Combes06}.

\subsection{C$^+$ cooling rate in the dense versus diffuse medium}\label{sec:abs}

While our models focus on the relatively dense gas associated with the star-forming region of \izw, a significant fraction of the H$^0$ mass of the galaxy is distributed within the large H\1\ envelope \citep{Lelli12}. In this section we compare the cooling rates observed in both phases. 

\begin{figure}
\centering
\includegraphics[angle=0,width=9.5cm,height=14cm,clip=true]{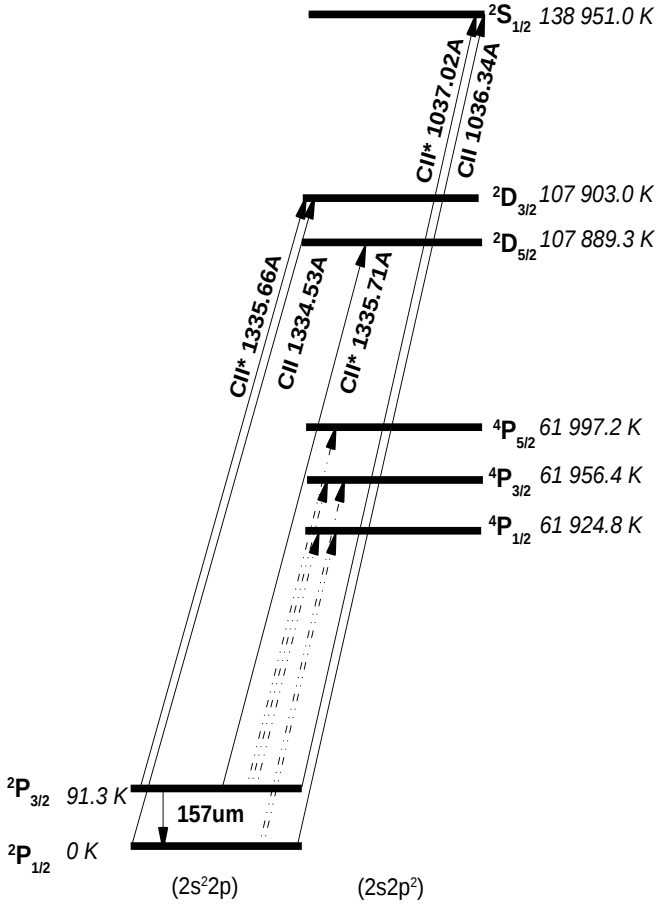}
\caption{Main levels of the C$^+$ ion. The dot-dashed lines correspond to weak transitions. }
\label{fig:levels}
\end{figure}

The C$^+$ cooling rate can be measured either using the [C\2] 157\mic\ emission line or   the C\2* $\lambda1335.7$ FUV absorption line doublet (Fig.\,\ref{fig:levels}). In the Milky Way, the $^2$P$_{3/2}$ fine-structure line population was first observed in absorption toward Galactic stars \citep{Pottasch79}. The 157\mic\ emission line was first observed shortly after in \object{NGC 2024} and the \object{Orion nebula} \citep{Russell80}. A comparison between the cooling rate inferred from the absorption and the emission was first performed for our Galaxy by \cite{Stacey85}, indicating a rough agreement. 
The same comparison in extragalactic targets is challenging as in most sources the material observed in emission cannot always be easily associated with that observed in absorption (e.g., quasar lines of sight). Furthermore, most IR-bright galaxies are difficult to observe in the FUV because of absorption by dust. Inversely, most FUV-bright galaxies are difficult to observe in the IR because of the lack of dust. The advent of \textit{Herschel} made it possible to detect relatively dust-poor galaxies at FIR wavelengths.

In that vein, the comparison between our results concerning [C\2] and those of \cite{Lebouteiller13a} concerning C\2* is informative and is, to our knowledge, the first comparison made in an extragalactic target. 
\cite{Lebouteiller13a} examined with the FUV absorption lines C\2\ $\lambda1334.5$ and C\2* $\lambda1335.7$, arising from the ground level and fine-structure level of C$^+$, respectively. The observations of C\2*, performed with \textit{Hubble}/COS, probe lines of sight toward bright young stars withing the $2.5\arcsec$ aperture. 
The authors inferred an electron fraction upper limit of $2$\%\ in the H\1\ region with a best guess of $\sim0.1$\%, which is within the range we have determined with \textit{Herschel}/PACS and \textit{Spitzer}/IRS line ratios (Sect.\,\ref{sec:finalmodel}) and with our models (Sect.\,\ref{sec:finalmodel}). Furthermore, the cooling rate as determined by the C\2* column density was found to be incompatible with the photoelectric effect on dust, which is again consistent with the results in the present study. 

We do not expect a priori an agreement between the C$^+$ cooling rate calculated in the present study and the results of \cite{Lebouteiller13a}. Firstly, only the most diffuse medium should be seen in absorption because the most dust-rich lines of sight suffer from UV absorption. However, we bear in mind that such a bias is likely negligible for \izw\ because of its low dust content. 
Secondly, and most importantly, the H\1\ region seen in emission corresponds to the entire main body while the H\1\ region seen in absorption by COS probes a region of $\approx2.5\arcsec$ across centered on NW and coinciding in fact with an H\1\ hole (Fig.\,\ref{fig:hipacs}). 

To quantify further the comparison between the cooling rate inferred in emission and absorption, we calculate the total cooling rate in erg\,s$^{-1}$. From the observed [C\2] 157\mic\ line intensity, $10^{-17}$\,W\,m$^{-2}$, we find a total rate of $3.9\times10^{31}$\,W. For the measure in absorption, we first use the cooling rate per H atom obtained by \cite{Lebouteiller13a}, $8.3\times10^{-28}$\,erg\,s$^{-1}$\,(H$^{-1}$), which we then multiply by the total mass of H$^0$ in the main body, $\approx10^8$\,M$_\odot$, assuming that the cooling rate per H atom is uniform across the H\1\ body. We find a rate of $10^{38}$\,erg\,s$^{-1} = 10^{31}$\,W, i.e., a factor of $\approx4$ difference with the rate measured in emission. 
Another way to show this discrepancy is to compare the COS measurement $8.3\times10^{-28}$\,erg\,s$^{-1}$\,(H$^{-1}$) to the cooling rate predicted by the model ($10^{-24}-10^{-23}$\,erg\,s$^{-1}$; Fig.\,\ref{fig:model_final2}). Considering the volume density in the C$^+$-emitting region in the model, the cooling rate becomes $\approx10^{-26}-10^{-25}$\,erg\,s$^{-1}$\,(H$^{-1}$), implying a factor of $\gtrsim10$ larger than the COS rate. 

\begin{figure}
\centering
\includegraphics[angle=0,width=9cm,clip=true]{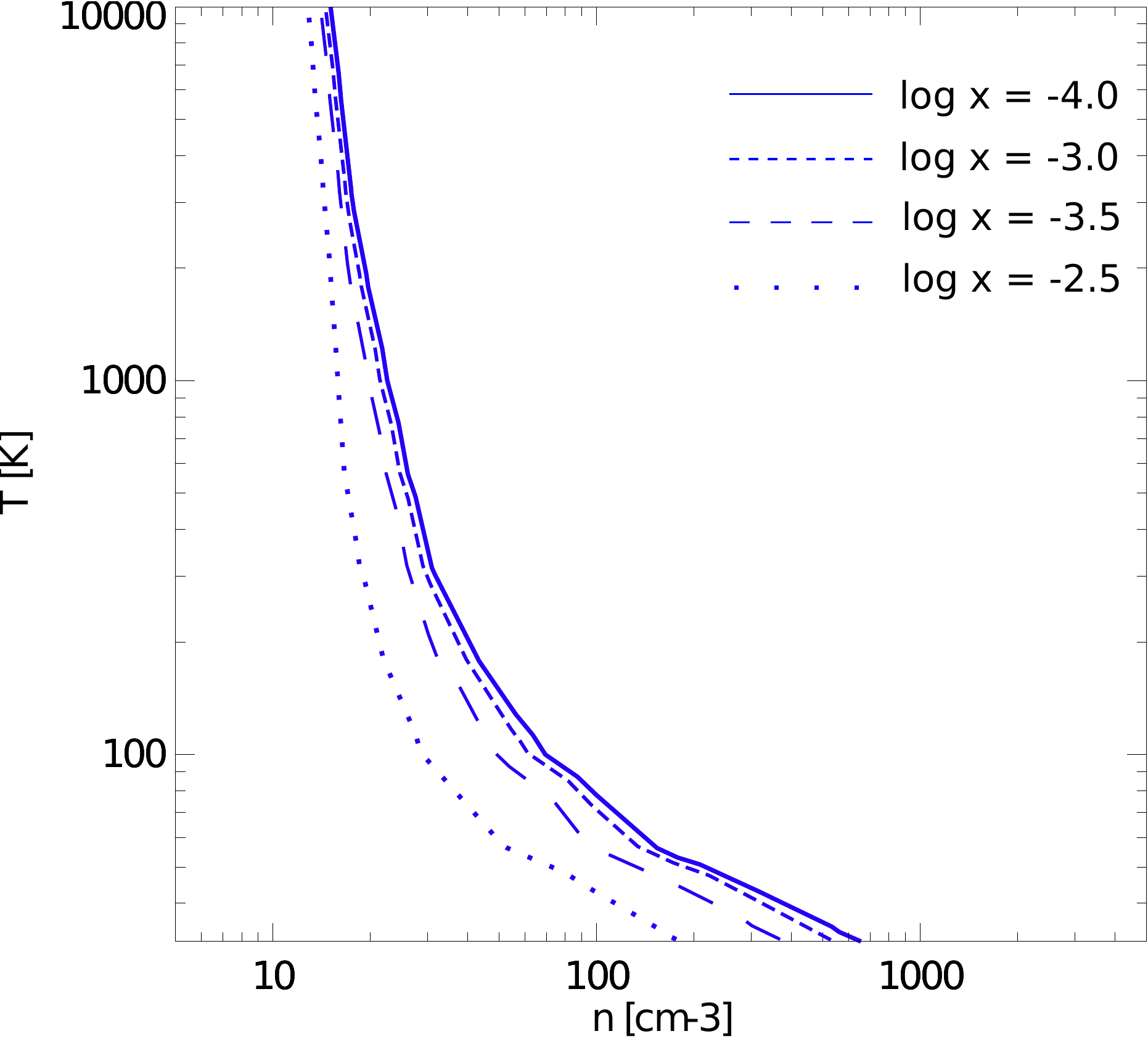}
\caption{Observed COS cooling line rate per H atom is shown for different temperature, density, and various electron fraction ($x$) values. }
\label{fig:lineratio}
\end{figure}

The discrepancy we observe cannot be solely attributed to the relatively uncertain H$^0$ mass of the main body used for the calculation and we argue that the cooling rate is genuinely lower in the H\1\ region probed toward the NW cluster. Taking at face value the cooling rates measured in absorption and in emission and the H$^0$ mass of the main body, it seems that most of the C$^+$ cooling in \izw\ occurs in the star-forming region with a relatively smaller contribution from the diffuse H\1\ envelope seen in absorption. 
The lower cooling rate in the diffuse H\1\ region could be at least partly due to the \textit{integrated} lower density of the ISM probed in absorption as compared to the H\1\ shells probed with \textit{Herschel} of about $300$\,\cc. \cite{Lebouteiller13a} derived an upper limit $\lesssim30$\,\cc\ on the density of the medium seen in absorption. We review this measurement by comparing the observed cooling rate to the theoretical value for various temperatures, densities, and electron fractions (Fig.\,\ref{fig:lineratio}). For a wide range of temperatures and electron fractions, we find densities on the order of $15-50$\cc. 

It is thus plausible that the C$^+$ cooling rate measured in emission and absorption do not correspond to the same physical regions. On the one hand, the [C\2] 157\mic\ emission line probes the neutral shell of the radiation-bounded H\2\ region with an H\1\ column density of $10^{22.2}$\,cm$^{-2}$. On the other hand, the C\2* absorption line appears to probe the more diffuse H\1\ region possibly extending to (or simply located) several kiloparsecs away from the stellar clusters, and providing only little emission in the [C\2] line. The temperature is unfortunately difficult to measure from absorption lines in extragalactic objects because fine-structure absorption lines other than C\2* are usually not detected and because absorption components corresponding to individual clouds cannot be disentangled. 
The total H\1\ column density seen in absorption is $10^{21.34}$\,cm$^{-2}$ \citep{Lebouteiller13a}. 
The lower chemical abundances measured in absorption by \cite{Lebouteiller13a} in the H\1\ envelope as compared to the H\2\ region abundances only reinforce the hypothesis of two distinct physical regions while at the same time providing another explanation for the relatively lower cooling rate measured in absorption. Even though the carbon abundance measured with COS is uncertain, \cite{Lebouteiller13a} concluded there is a general metal deficiency in the H\1\ envelope with abundances that are lower by a factor of $\approx2$ as compared to the H\2\ region. \cite{Lebouteiller13a} also hypothesized the presence of pristine gas along the lines of sight that would dilute abundances. If such metal-free gas does exist in the outskirts of the galaxy, it would also explain the measurement of the lower cooling rate per H atom. 

The diffuse H\1\ gas could, in principle, be considered an extra sector in our models, but we are limited by the unknown topology of this phase and by the lack of knowledge on the associated heating mechanisms. While clouds lying further out from neutral shells in the H\2\ region are supposedly shielded from the UV arising from the NW cluster, the clumpiness of the ISM and the scattering complicate this simple picture. Furthermore, the H\1\ structures seen several kiloparsecs away from the stellar clusters may be heated by an old stellar population and by CR, both of which are not well constrained. Other fine-structure absorption lines in the FUV, such as O\1* or Si\2* would be required to measure the physical conditions in greater detail and provide useful model constraints. 

Nevertheless, we have briefly explored a model in which a low-density ($10-50$\cc) H\1\ gas is located $\sim1$\,kpc away from the NW cluster. Such a model is in fact also motivated by H$\alpha$ emission detected at very large distances, which is likely due to photoionization by radiation escaping from the main star-forming regions (NW+SE). Then H\1\ zones beyond the ionization front of these far away (low-excitation) de facto ``H\2\ regions'' should exist, again probably heated by X-ray radiation. Zones deprived of significant H\2\ gas may as well be heated by X-rays. Interestingly, we find that the [C\2] and [O\1] line fluxes relative to H$\beta$ are not strongly affected. Thus, the inference from this tentative example (many other runs are needed to be more precise) is that the [C\2] and [O\1] emission arising from the XDR is more or less proportional to H$\alpha$ at large distances. [C\2] and [O\1] may be relatively enhanced due to shielding, which would primarily reduce the relative size of the H$^+$ region.  

In summary, the chemical and physical conditions of the medium inferred from absorption lines point at a different origin from the neutral filaments observed in emission around the NW H\2\ region. Detailed properties of this more diffuse phase would be of interest to investigate, as it must play an important role in the evolution of the galaxy, notably through gas infall, future widespread star-formation, or star-formation quenching from chemical heating.

\section{Implications of dominant X-ray heating}\label{sec:firdiag}

In this section we examine the consequences of our main result, which is chiefly that the main heating mechanism in the H\1\ region of \izw\ is probably photoionization by soft X-rays. 

\subsection{Relationship between [C\2] and SFR}\label{sec:ciisfr}

The [C\2] luminosity is often used as a SFR tracer (e.g., \citealt{deLooze14a} and references therein), which usually has the implicit assumption that the gas cooling down through [C\2] is heated by the UV field arising from young stars through the photoelectric effect. The use of [C\2] as a SFR tracer may become questionable when the heating mechanism of the H\1\ region is different. 

The SFR in \izw\ calculated from [C\2] is only a few factors below other measurements (Sect.\,\ref{sec:sfrlit}), while a larger discrepancy may have been expected considering that the heating is not dominated by the photoelectric effect. Nevertheless, we show below that there is a correlation between the X-ray luminosity (responsible for [C\2]) and the SFR for star-forming galaxies. 

The X-ray emission in \izw\ corresponds to an ultraluminous X-ray source (ULX, defined as $L_{\rm X}>10^{39}$\,erg\,s$^{-1}$). Ultraluminous X-ray sources are not associated with the nuclei of galaxies and are thought to be intermediate-mass black holes or to involve a HMXB system with either a neutron star and/or a stellar mass black hole (see, e.g., \citealt{Gilfanov04,Berghea13,King16}). As they are late stages of dying massive stars, the HMXBs are expected to be related to star formation, with progenitors $\gtrsim8$\,M$_\odot$, and with evolution timescales on the order of $10^{6-7}$\,yr. A correlation between SFR and $L_{\rm X}$ has been shown observationally by \cite{Grimm03} based on what seems to be a universal HMXB luminosity function in nearby star-forming galaxies (see also \citealt{Gilfanov04b}). The SFR-$L_{\rm X}$ relation is linear for high-SFR regime ($>4.5$\,M$_\odot$\,yr$^{-1}$) and non-linear below. Contaminations include active galactic nuclei and low-mass X-ray binaries (LMXBs), the latter being a more important contaminant for galaxies with low SFR. The non-linear behavior is thought to be a metallicity effect, as proposed by \cite{Douna15}, reflecting the enhancement of the HMXB population at low-metallicity. This enhancement has been expected from theory (e.g., \citealt{Linden10,Fragos13,Prestwich13,Mapelli10,Mapelli13}) and is supported observationally by the rising evidence of ULXs in nearby metal-poor BCDs (e.g., \citealt{Kaaret11,Kaaret13,Brorby14,Brorby15} and references therein). The measured values of SFR and $L_{\rm X}$ in \izw\ are consistent with such a metallicity dependence\footnote{\izw\ would fall closer to the linear relationship with SFR of $\sim1$\,M$_\odot$\,yr$^{-1}$ as suggested by the CMD analysis from \cite{Annibali13} (Sect.\,\ref{sec:sfrlit}). Photon escape probability and the use of indirect tracers to infer SFR may therefore play some role and explain in part the different behavior of SFR-$L_{\rm X}$ observed at low-metallicity (as the metallicity decreases, photon escape is presumably enhanced due to a lower D/G and to a more porous ISM). }. Besides, the dependence with metallicity is inferred indirectly from the deficit of ULX in metal-rich luminous IR galaxies \citep{Luangtip14}

In summary, the SFR-$L_{\rm X}$ relationship explains why [C\2] traces relatively well the SFR in \izw. It also explains why the value of the often-used diagnostic ratio ([C\2]+[O\1])/TIR is not unusual (Sect.\,\ref{sec:ciitir}).

\subsection{Relationship between [C\2] and TIR}\label{sec:ciitir}

The ([C\2]+[O\1])/TIR ($\epsilon_{\rm TIR}$; with TIR calculated between $3-1100$\mic) ratio is often used to estimate the photoelectric effect heating efficiency\footnote{$\epsilon_{\rm TIR}$ represents the gas cooling over the power absorbed by dust. Considering the low efficiency values (typically less than $1$\%), the dust IR emission is assumed to approximate the UV photon energy absorbed by dust.}, in which case it may also be used to calculate the physical scale of star formation \citep{Stacey10}. It is also used to probe the so-called ``[C\2] deficit'' in ultraluminous IR galaxies (e.g., \citealt{DiazSantos13} and references therein). Inversely, metal-poor galaxies show an elevated $\epsilon_{\rm TIR}$ \citep{Cormier15}. A larger $\epsilon_{\rm TIR}$ ratio is traditionally attributed to a larger photoelectric effect heating efficiency, which may be considered a consequence of a lower grain charging due to the relatively lower absorption of UV photons in a dust-poor environment \citep{Israel96,Israel11} or to a smaller dust grain size distribution \citep{Galliano05}.
Our result that the photoelectric effect heating is negligible in \izw\ allows us to test diagnostics held by $\epsilon_{\rm TIR}$. 

For \izw\ we find $\epsilon_{\rm TIR} = 0.6$\%. If we tentatively assume that half of the TIR emission arises in the H\2\ region (approximate fraction derived from the SED fitting; \citealt{Galliano08b}), we derive the following values for the PDR/H\1\ region component alone $\epsilon_{\rm TIR} = 1.2$\%. For comparison, \cite{Lebouteiller12b,Lebouteiller13a} found $\epsilon_{\rm TIR}=0.4-0.7$\%\ in the PDR-dominated regions of the \object{LMC-N11B} nebula. In Haro\,11, \cite{Cormier12} found $0.18$\%\ for the entire galaxy and $1.1$\%\ for the PDR component alone. In Galactic PDRs, \cite{Okada13} found $0.1-0.9$\%. We can also compare to other measurements of integrated galaxies (not corrected for the H\2\ region contribution from TIR). \cite{Cormier15} find $0.2-1.0$\%\ in dwarf galaxies of the DGS while \cite{Croxall12} find $0.2-0.7$\%\ in two super-solar metallicity galaxies. 

In summary, $\epsilon_{\rm TIR}$ in \izw\ is similar to the values in other galaxies, although somewhat larger. As shown in Section\,\ref{sec:cloudype}, this cannot be explained in terms of photoelectric efficiency for this galaxy as photoionization of the H\1\ region by X-rays can solely explain the emission of [C\2] and [O\1]. In order to understand why $\epsilon_{\rm TIR}$ is not extreme in \izw, we must take a closer look   at the tracers involved in the quantity $\epsilon_{\rm TIR}$.

First, the free-free contribution in the TIR range, TIR$_{\rm ff}$, could become important when D/G decreases to extremely low values. In \izw, using the number of ionizing photons in \citetalias{Pequignot08} (scaled at $18.2$\,Mpc), we find TIR$_{\rm ff} \approx6.9\times10^6$\,L$_\odot$, i.e., only about four times lower than the observed TIR luminosity $2.9\times10^7$\,L$_\odot$ \citep{Remy15}. The observed luminosity of [C\2]+[O\1] calculated at the same distance is $1.8\times10^5$\,L$_\odot$, which results in $\epsilon_{\rm TIR, ff} \sim 2.6$\%. Hence even if \izw\ were virtually dust free, $\epsilon_{\rm TIR}$ would remain on the order of a few percent, i.e., on the same order as the observed ratio in more metal-rich objects. 

Apart from the free-free emission, the observed TIR also includes an important contribution from the dust in the ionized phase. Such dust population does not participate significantly to the heating of the H\2\ region, and evidently does not contribute to the H\1\ region heating where [C\2] and [O\1] originate. In our models, the fraction of the observed TIR that arises by dust in the H\1\ region alone is only $15$\%. This shows that to calculate the actual photoelectric effect heating efficiency in the H\1\ region, we would need to apply a significant correction to TIR. 

The fact that $\epsilon_{\rm TIR}$ does not reach extreme values in a dust-free environment can be surprising at first, as it suggests a physical connection between cooling lines and TIR, even though the heating mechanisms are different as compared to more metal-rich sources. While TIR scales with the UV field intensity (through dust and  free-free emission), [C\2] and [O\1] scale with the X-ray emission through photoionization heating. It turns out that both energy sources (UV and X-rays) are correlated (Sect.\,\ref{sec:ciisfr}), implying that the cooling line emission and TIR scale together on first approximation.

Nevertheless, in a completely metal-free environment the free-free emission would dominate TIR, and [C\2]+[O\1] would be null, leading in effect to $\epsilon_{\rm TIR}=0$. Therefore, the fact that $\epsilon_{\rm TIR}$ does not reach extreme values remains valid as long as the main H\1\ region coolants are metallic species.

\subsection{Star-formation quenching from X-ray vs.\ photoelectric effect heating}

\cite{Forbes16} proposed that the gas depletion timescale in dwarf galaxies may be dominated by the photoelectric effect heating and not by SN feedback. The photoelectric heating could then be an important quenching mechanism for star formation for isolated systems, implying a significant metallicity-dependence for the star formation history. \cite{Hu17} find conflicting results, with SNe dominating the feedback, but the photoelectric effect still has some impact on the SFR.

The H\1\ region heating in \izw\ is in fact dominated by X-ray ionization. Our result, together with the enhanced abundance of HMXBs in low-metallicity galaxies (Sect.\,\ref{sec:ciisfr}), suggests that star formation may have been mostly suppressed, or at least influenced, by X-ray heating. Our hypothesis stands for a closed-box model of star formation, but interestingly, X-ray and EUV ionization generated by star formation have also been proposed to regulate star formation in galaxies through the reduced cooling rate of the halo gas to be accreted \citep{Cantalupo09}. 

Overall, there seems to exist a compensation between the heating provided by HXMBs and the photoelectric heating from dust, where both heating mechanisms vary with metallicity (or D/G) in opposite ways. Although it is plausible that the photoelectric effect dominates the gas heating in most sources in the DGS sample the enhancement of the HMXB population at low metallicity may play a role in extreme environments such as \izw. The key parameter is not metallicity but, to first order, the competition between D/G and $L_{\rm X}$ (modulated by the topology and the photon absorption). For instance, our models for \izw\ predict that the photoelectric effect would be a significant heating mechanism (accounting for $\sim31$\%\ of the total heating in the H\1\ region) if D/G were to be scaled with the metallicity (Sect.\,\ref{sec:uniformdgr}).

\section{Conclusions}\label{sec:conclusion}

In this study the extremely metal-poor ($1/30$\,Z$_\odot$) galaxy \izw\ is examined to identify and quantify the main gas heating mechanisms in the H\1\ region.

To this end, a detailed modeling of the H\2+H\1\ region with the photoionization code Cloudy is undertaken, building on a previous model by \citetalias{Pequignot08}. Our new \textit{Herschel}/PACS observations ([C\2] $157$\mic, [O\1] $63$\mic, and [O\3] $88$\mic) and updated \textit{Spitzer}/IRS measurements (particularly [Si\2] $34.8$\mic) are used as constraints. Other new constraints include observations of the dust SED, dust mass,  H$^0$ mass, and X-ray spectrum.

The photoelectric effect, CR, and X-ray heating are all considered in different consistent models. 
The selected models reproduce well the optical to FIR emission lines, dust SED, and other data. The main
conclusions are as follows:
\begin{itemize}
\item[$\bullet$] The photoelectric effect is not a significant heating mechanism in the H\1\ region, whether we use the D/G constrained by \textit{Herschel} or a larger D/G scaled with metallicity.
\item[$\bullet$] To order of magnitude, the weak photoionization of H$^0$ and He$^0$ by X-rays can provide enough heating to explain the observed low-ionization fine-structure line intensities. The dominant coolants are [O\1] and [Si\2] at the inner edge of the H\1\ region and [C\2] deeper inside. 
\item[$\bullet$] More precisely, the H\1\ zone fluxes imply a luminosity $L_{\rm X}\approx4\times10^{40}$\,erg\, s$^{-1}$ that is about three times larger than the value inferred from a \texttt{diskbb} model fitting of the XMM-\textit{Newton} observations. This difference is not critical, given that
(1) the source may not emit isotropically, (2) the X-ray source is variable, and (3) the model X-ray flux corresponds to a time average over the H\1\ gas cooling timescale of several $10^4$ years. This large average X-ray luminosity could imply the existence of a massive accreting black hole in \izw-NW.
\item[$\bullet$] The lack of constraints in an environment such as \izw\ prevents from any definite conclusion concerning CR heating. Nonetheless, CR heating is negligible if the ``standard'' description in Cloudy is adopted. 
\item[$\bullet$] The models predict only about $0.1$\,M$_\odot$ of H$_2$ in the region traced by [C\2] and [O\1] if the radiation-bounded sector has uniform properties (no clumps). Under such conditions, [C\2] and [O\1] are therefore not (CO-dark) molecular gas tracers. 
\item[$\bullet$] We also explore the possible existence of undetected molecular clumps. These H$_2$ clouds should reach a density of $10^6$\cc\ and have a small covering factor ($\lesssim0.1$\%). The main constraints come from the CO emission and the dust SED. This dense sector can shelter $\lesssim4\times10^7$\,M$_\odot$ of H$_2$, in which case H$_2$ must belong to a few clouds of a few parsecs in size ($\lesssim0.1''$). Most of the [C\2] and [O\1] emission in the clump sector is associated with a large molecular gas fraction, but theses lines do not trace well the H$_2$ mass. Even in the presence of clumps, most ($\gtrsim85$\%) of the [C\2] and [O\1] emission in the model is associated with atomic gas in the more diffuse sector.
\item[$\bullet$] The cooling rate derived from C\2* in absorption against UV-bright stars is somewhat less than the rate measured from [C\2] in the H\1\ shell adjoined to the H\2\ region. In absorption, the lines of sight may intersect a rather diffuse neutral ($\sim10$\cc) up to several kiloparsecs away from the stellar cluster. 
\item[$\bullet$] The dust SED is not well reproduced with uniform D/G values. A warm dusty component is required to fit the mid-IR part of the SED. Adding this component has no impact on the FIR line emission and on any other result.  
\item[$\bullet$] More than half of the observed TIR flux is free-free emission from the ionized phase. The rest is dust emission (mostly in the ionized phase). Only $15$\%\ of the TIR comes from dust in the H\1\ region. Diagnostics involving TIR, such as the photoelectric heating efficiency proxy [C\2]+[O\1]/TIR, should be used with caution. 
\end{itemize}

That the H\1\ region heating is dominated by X-ray ionization has important implications for extremely metal-poor galaxies in general. There seems to be a competition between the photoelectric effect (scaling with metallicity) and the X-ray ionization, the latter becoming more and more important in low-metallicity star-forming galaxies through the larger occurrence and luminosity of HMXBs (expected theoretically and checked observationally). Some uncertainty pertains to the difficulty in recovering the X-ray energy distribution actually seen by the gas. We show that some of this uncertainty can be lifted through the use of high-ionization lines such as [Ne\5] (Appendix\,\ref{secapp:xspec}).

In environments with both an extremely low D/G and the bright X-ray binaries, the physical conditions can no longer be easily derived from FIR diagnostics. In \izw, we notice that the [C\2]+[O\1]/TIR ratio is surprisingly normal, with a value similar to what is observed in more metal-rich sources. We argue that this ratio is kept stable because of a correlation between X-ray luminosity, which is responsible for the heating of the gas in which [C\2] and [O\1] originate, and UV radiation field through the SFR, which is responsible for most of the TIR emission. This implies that the value of [C\2]+[O\1]/TIR cannot be interpreted as a photoelectric effect heating efficiency proxy (or as an indirect metallicity tracer) in \izw. 
For the same reason (i.e., the relationship between SFR and X-ray luminosity in star-forming galaxies), H\1\ region cooling lines, and in particular [C\2], scale with the SFR even though the gas heating mechanism may be different. 

Finally, we propose that X-ray heating may be an important quenching mechanism for star formation. This quenching could be efficient at large scales, as shown by the comparison of the C$^+$ cooling rate measured in emission to that measured in absorption. Our models suggest that X-rays can heat the gas at large distances.

The study of \izw\ is interesting to put in perspective with the expected occurrence of binary systems made of compact objects in the first galaxies. The latter are star-forming H\1\ rich objects that share similarities with nearby BCDs, and the heating of the H\1\ region may be dominated by the same mechanisms. The number and luminosity of HMXBs is expected to be enhanced in low-metallicity galaxies, and it is therefore likely that such HMXBs contributed significantly to heating the early Universe (e.g., \citealt{Mirabel11}). \izw\ provides an interesting testbed, but the fraction of escaping X-ray photons needs to be estimated. The ISM is expected to be more porous in dust- and metal-poor environments (e.g., \citealt{Cormier15}), which would favor both the escape of FUV and X-ray photons.

\begin{acknowledgements}
We thank the anonymous referee for constructive feedback. 
V.\ L. was supported by a CEA/Marie Curie Eurotalents fellowship when this study was initiated. We acknowledge financial support from "Programme National de Cosmologie and
Galaxies" (PNCG) funded by CNRS/INSU-IN2P3-INP, CEA and CNES, France. M.\ C.\ gratefully acknowledges funding from the Deutsche Forschungsgemeinschaft (DFG) through an Emmy Noether Research Group, grant number KR4801/1-1. This research was partly supported by the Agence Nationale pour la Recherche (ANR) through the program SYMPATICO (Programme Blanc Projet ANR-11-BS56-0023). 
We wish to thank J\'er\^ome Rodriguez and Giulia Migliori for their help with the XMM-Newton spectra. 
PACS has been developed by a consortium of institutes led by MPE (Germany) and including UVIE (Austria); KU Leuven, CSL, IMEC (Belgium); CEA, LAM (France); MPIA (Germany); INAF-IFSI/OAA/OAP/OAT, LENS, SISSA (Italy); IAC (Spain). This development has been supported by the funding agencies BMVIT (Austria), ESA-PRODEX (Belgium), CEA/CNES (France), DLR (Germany), ASI/INAF (Italy), and CICYT/MCYT (Spain). HIPE is a joint development by the Herschel Science Ground Segment Consortium, consisting of ESA, the NASA Herschel Science Center, and the HIFI, PACS and SPIRE consortia.  This research has made use of the NASA/IPAC Extragalactic Database (NED) which is operated by the Jet Propulsion Laboratory, California Institute of Technology, under contract with the National Aeronautics and Space Administration. 
\end{acknowledgements}

\bibliography{/media/WORK//Science/Documents/ScienceLibrary/jab.bib}

\appendix

\section{\textit{Herschel}/PACS spectral extraction}\label{secapp:herschel}

\subsection{PACS optimal extraction of pointed observations}\label{secapp:optimal}

\begin{figure*}
\centering
\includegraphics[angle=0,width=11cm,clip=true]{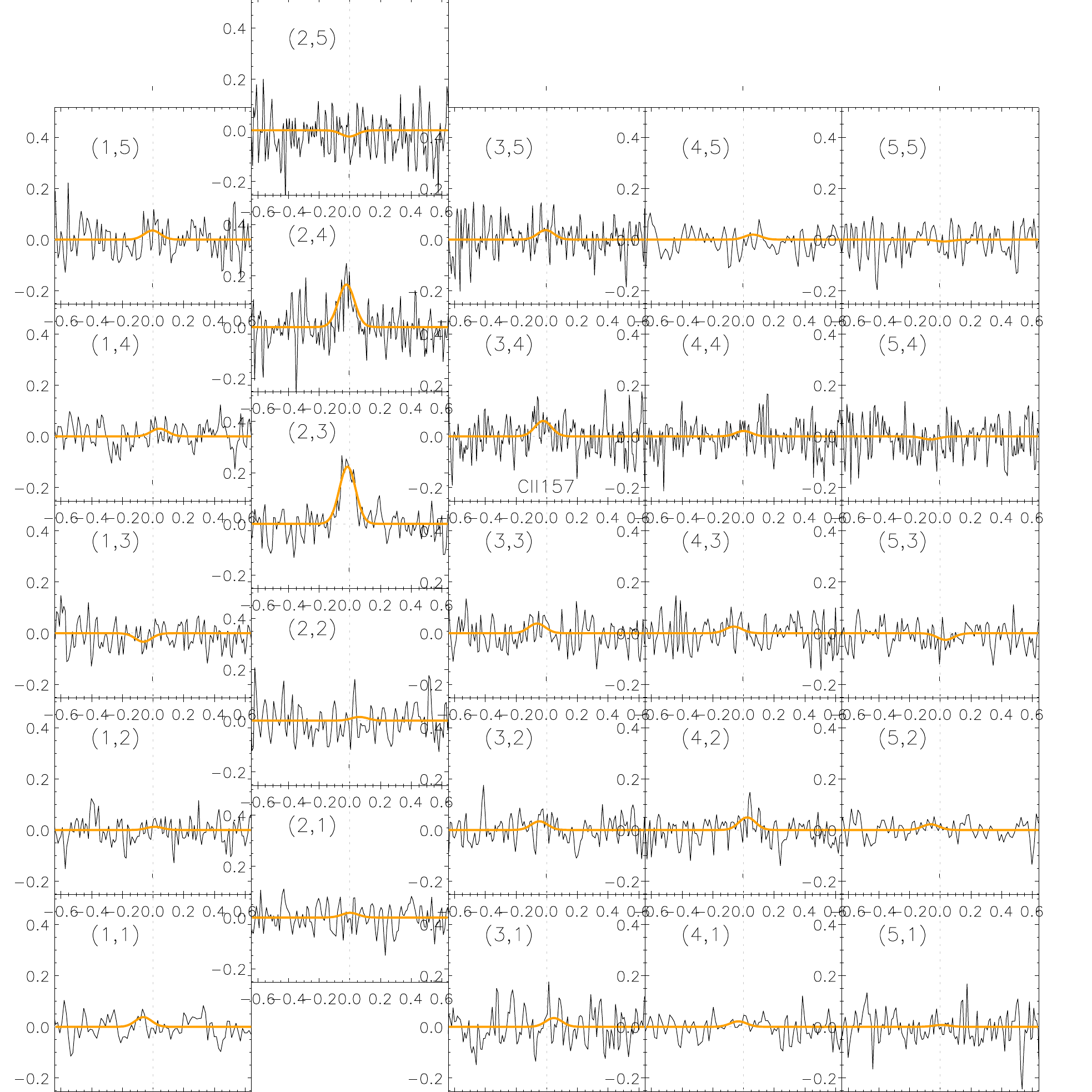}\\
\vspace*{0.5cm}
\includegraphics[angle=0,width=11cm,clip=true]{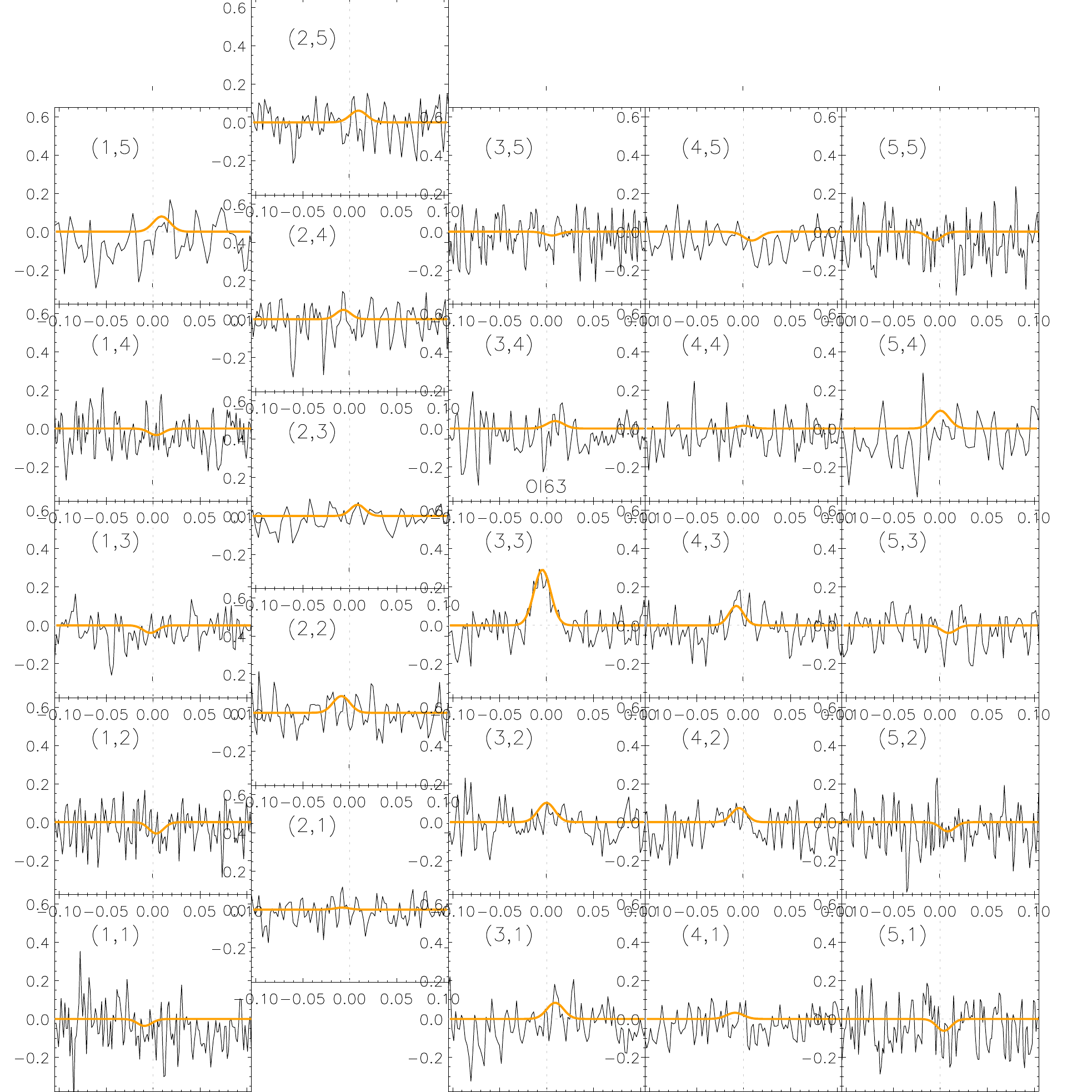}
\caption{Individual \textit{Herschel}/PACS spaxel background-subtracted spectra for [C\2] 157\mic\ (top) and [O\1] 63\mic\ (bottom). The observed spectrum is shown in black while the line fit is shown in orange. The number between parentheses indicates the spaxel coordinate within the footprint. The second column is shifted to reflect the PACS footprint projection on the sky. }
\label{fig:spec1}
\end{figure*}

\begin{figure*}
\centering
\includegraphics[angle=0,width=11cm,clip=true]{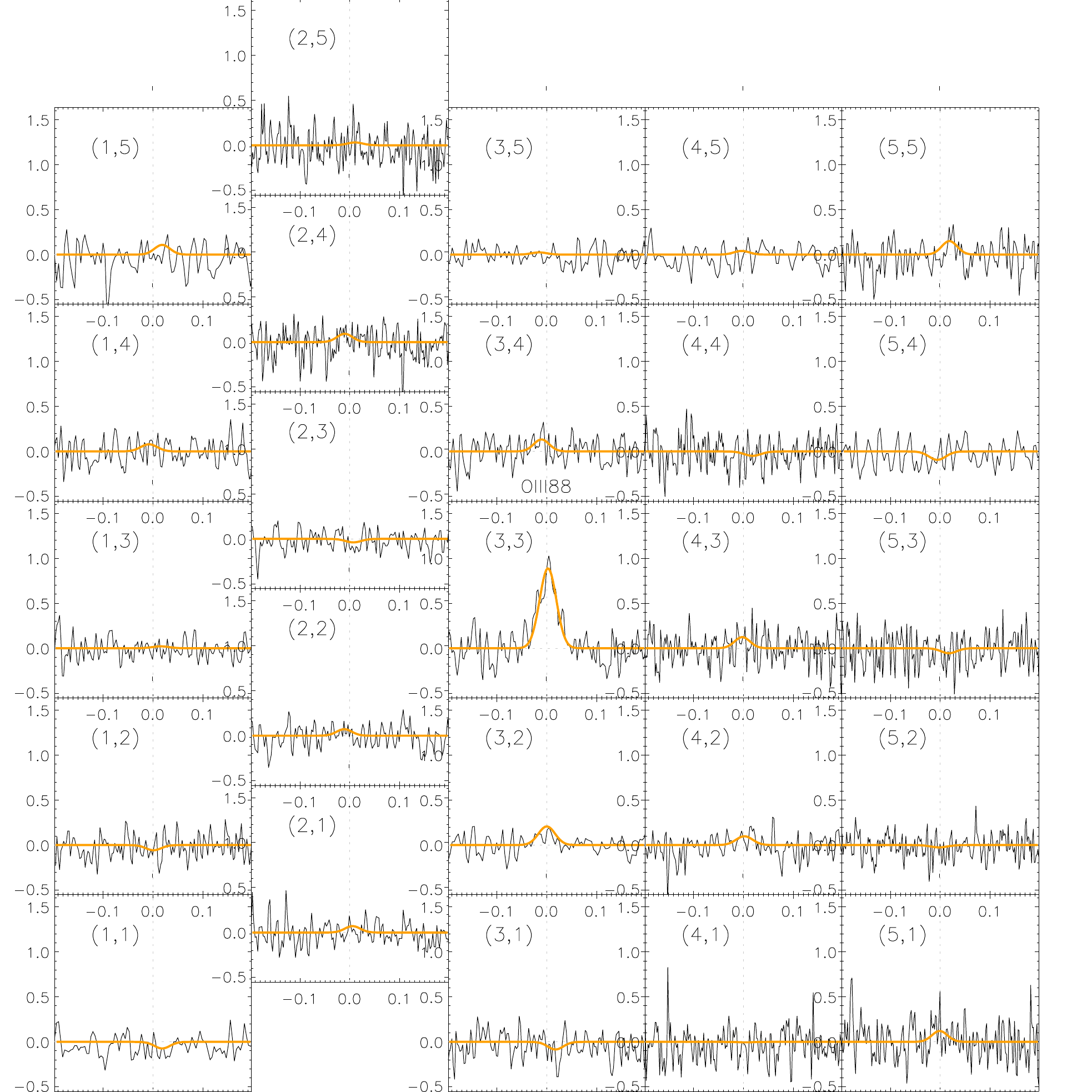}
\caption{Individual \textit{Herschel}/PACS spaxel background-subtracted spectra for [O\3] 88\mic. See Figure\,\ref{fig:spec1} for the description.  }
\label{fig:spec2}
\end{figure*}

The \textit{Herschel}/PACS spectra are shown in Figures\,\ref{fig:spec1} and \ref{fig:spec2}. We derive the following radial velocities, $795\pm100$\kms\ for [C\2], $760\pm40$\kms\ for [O\1], and $795\pm75$\kms\ for [O\3]. While the spectral resolution of \textit{Spitzer} is too low to provide a useful comparison ($>500$\kms; Sect.\,\ref{sec:spitzerobs}), \cite{Lelli12} modeled the H\1\ 21\,cm velocity distribution of \izw\ main body as a disk with a centroid close to the SE component and a central velocity of $\approx765$\kms. The velocity ranges from $\approx735$\kms\ to $\approx790$\kms\ from north to south. Interferometric H$\alpha$ measurements are in agreement with these values \citep{Petrosian97}. 

The optimal extraction uses the ``as-built'' telescope monochromatic PSF\footnote{\url{http://pacs.ster.kuleuven.ac.be/pubtool/PSF/}} projected on the PACS footprint. The flux is calibrated by ensuring that the encircled energy fraction is similar to that in the PACS photometer documentation. The algorithm does not take into account optical effects within the spectrometer. Nevertheless, the spectrometer PSF is to first order similar to the telescope PSF and it is considered to be a satisfactory model when used with a proper flux calibration. 
The flux is determined by scaling the normalized PSF to the data. The scaling is performed by using a multi-linear regression algorithm, following the method used for the \textit{Spitzer}/IRS spectra in \cite{Lebouteiller10}. The spaxels located far from the source centroid do not significantly affect the optimal extraction since the latter follows the PSF profile. We have validated the algorithm on point sources from the HERUS sample \citep{Farrah13}. 

\begin{figure*}
\centering
\includegraphics[angle=0,width=10cm, height=7.5cm,clip=true]{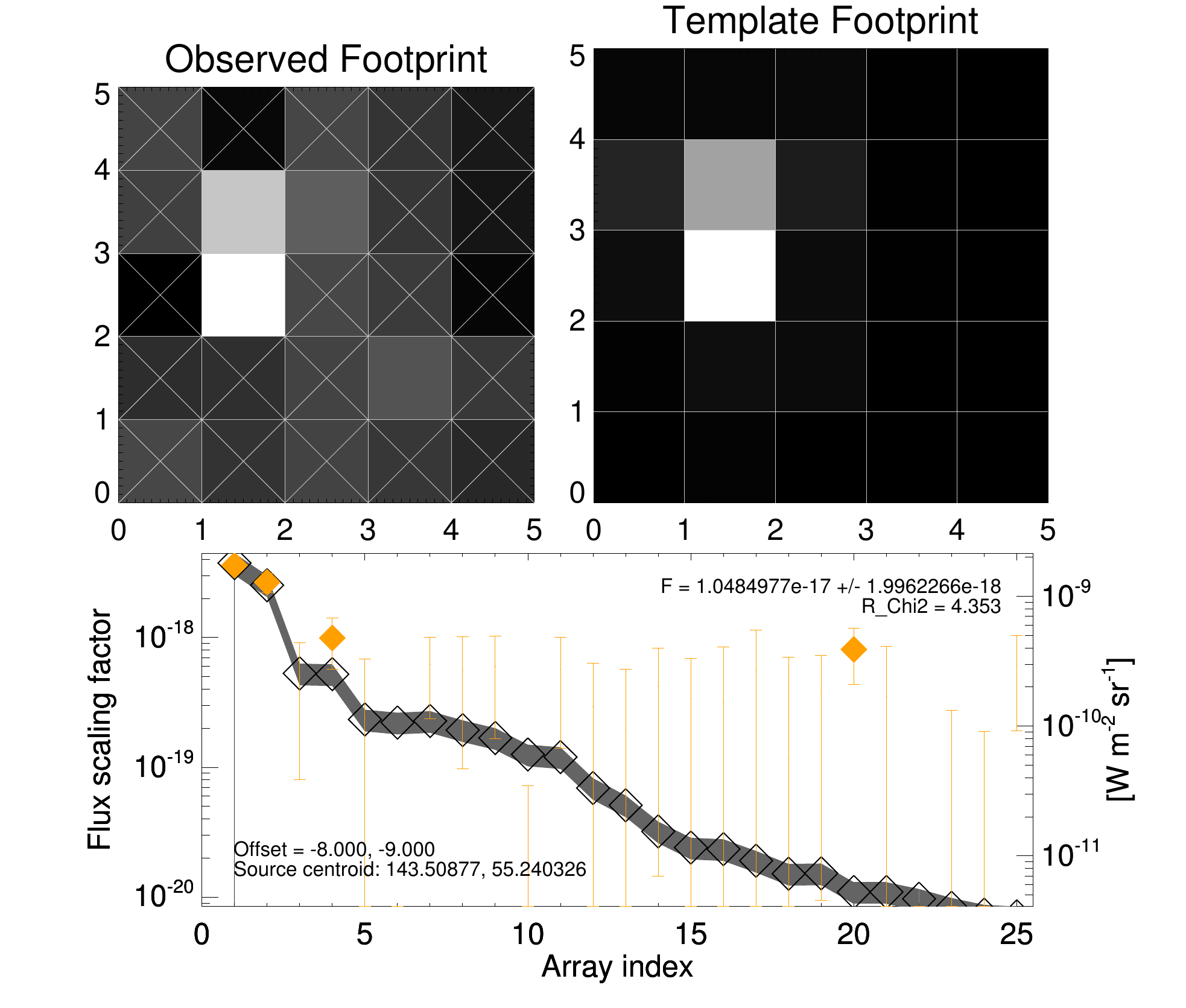}
\noindent\rule[0.5ex]{0.5\textwidth}{1pt}
\includegraphics[angle=0,width=10cm, height=7.5cm,clip=true]{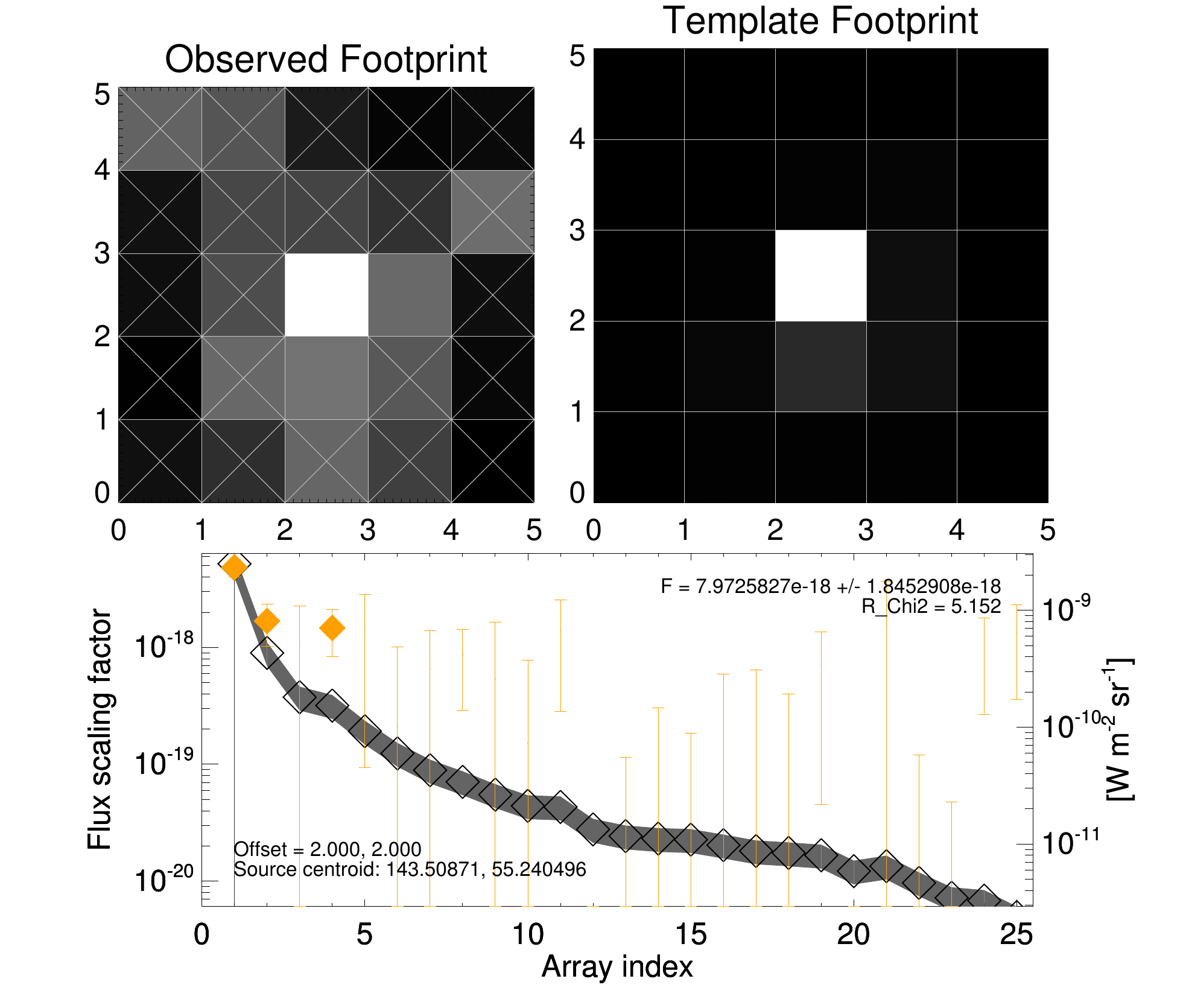}
\noindent\rule[0.5ex]{0.5\textwidth}{1pt}
\includegraphics[angle=0,width=10cm, height=7.5cm,clip=true]{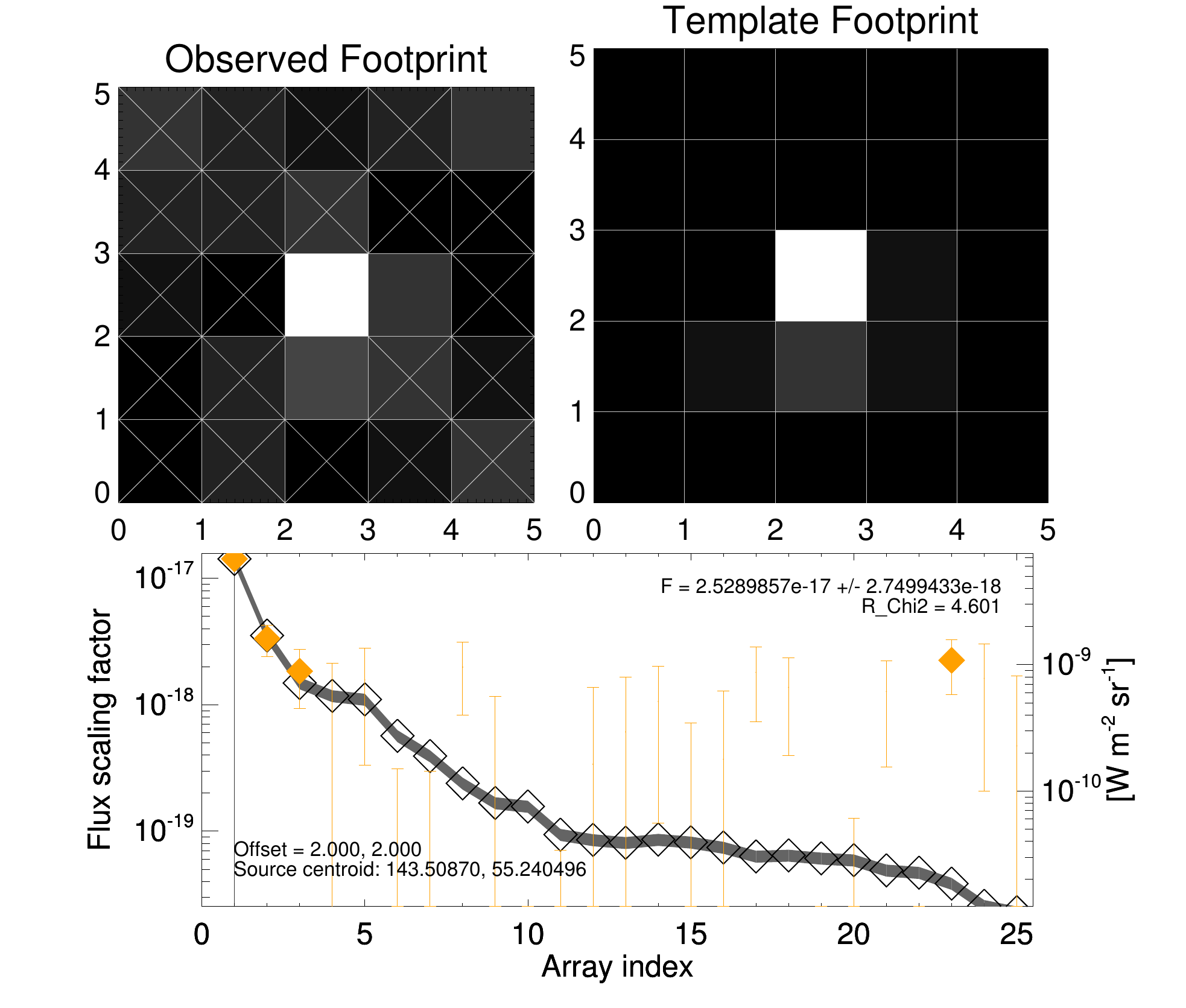}
\caption{Optimal extraction of [C\2] (\textit{top}), [O\1] (\textit{middle}), and [O\3] (\textit{bottom}). For each line the top panel shows the observed footprint (with crossed spaxels indicating non-detection) and the template footprint (chosen from a library of precomputed projected PSFs). The footprint is shown as a regular grid for display purposes only; in reality, the spaxels are not uniform in size and one row is shifted with respect to the others. The bottom panel shows the spaxel fluxes ordered by flux. Observed values are indicated in orange. The theoretical distribution from the template footprint is shown with the black curve. The $3$ brightest spaxels suffice to locate the peak. }
\label{fig:pacsopt}
\end{figure*}

The optimal extraction for [C\2], [O\1], and [O\3] in \izw\ is shown in Figure\,\ref{fig:pacsopt}. From the spatial profiles, we can estimate whether a source is point-like or not. Practically, only two or three detected spaxels are sufficient to show evidence of significantly broadened sources (for a single source in the footprint). The emission appears point-like for the three PACS lines. The intrinsic source extent is estimated to be $\lesssim6$\arcsec.  

We can also quantify the contribution from a large-scale low surface brightness component by adding a flat emission component in the optimal extraction algorithm. The large-scale component is fitted simultaneously with the point source. The flux of this large-scale component represents $\approx10$\%, $\approx18$\%, and $\approx7$\%\ of the total line flux for [C\2], [O\1], and [O\3], respectively. We conclude that our measurements of the point source are not significantly affected by a potential low surface brightness component. This finding is also consistent the point-like appearance of \izw\ in the PACS photometry $70$\mic\ and $100$\mic\ bands (with a similar PSF size as the PACS spectroscopy observations) \citep{Remy15}.

\subsection{Unchopped [O\3] 88\mic\ raster map}\label{sec:oiiimap}

\begin{figure}
\centering
\includegraphics[angle=0,width=9cm,clip,trim=0 0 0 0]{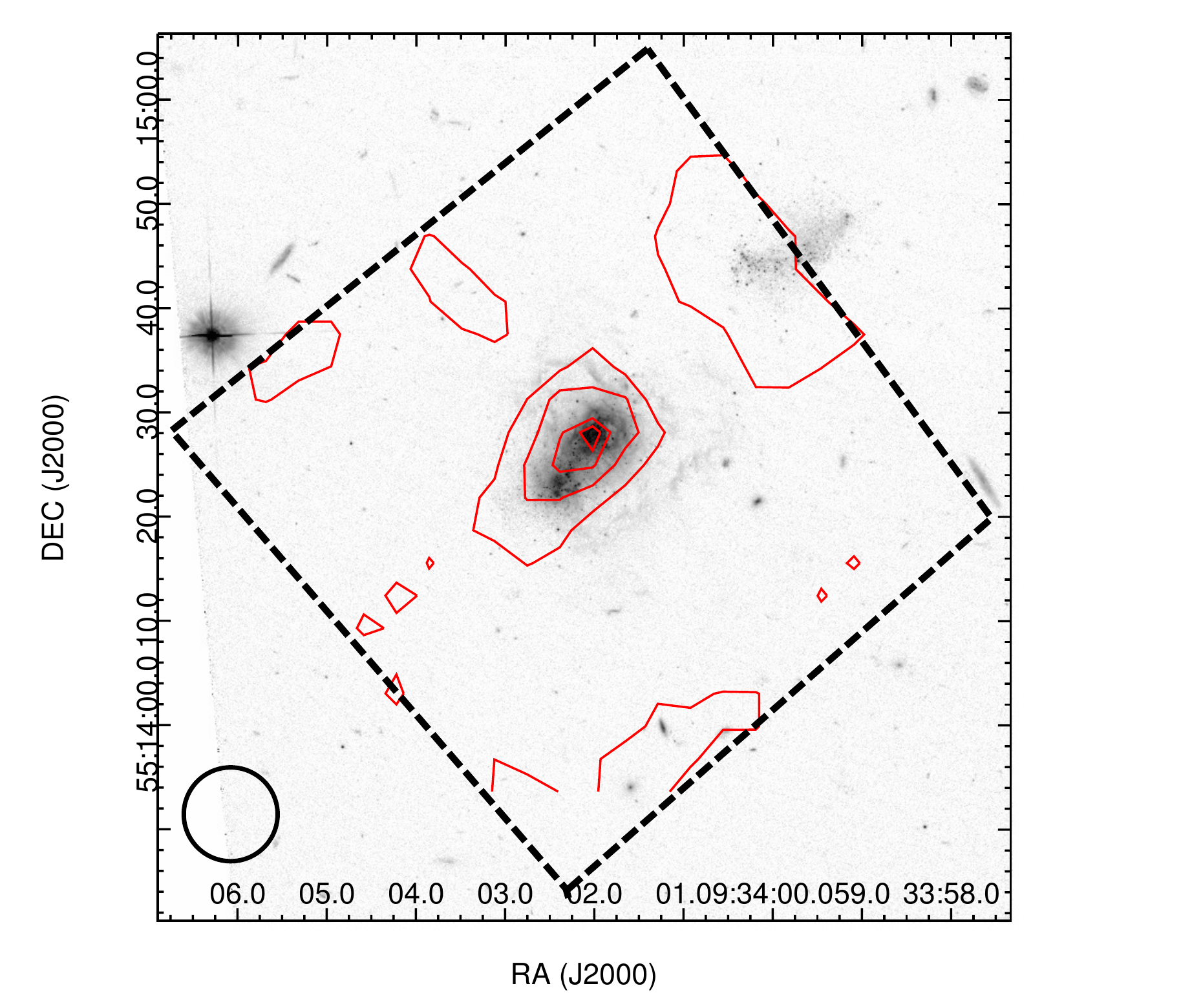}
\caption{The [O\3] map from the program OT2\_kcroxall\_1 (contours) obtained with \textit{Herschel}/PACS. The background is the \textit{Hubble}/ACS F555W image. The beam is shown in the bottom-left. }
\label{fig:oiiimap}
\end{figure}

\izw\ was observed independently with \textit{Herschel}/PACS on April 6th, 2013 as part of program OT2\_kcroxall\_1 (PI K.\ Croxall), for $2.7$\,ks. The observation (OBSID 1342269442) is a spectral map of the [O\3] $88$\mic\ line. Contrary to our DGS observations in which observations were performed using the chop/nod mode (with a fast removal of the telescope baseline, Sect.\,\ref{sec:pacsobs}), OBSID 1342269442 uses the unchopped grating scan mode in which an offset position was observed before and after the on-source exposure. The map consists of four raster positions shifted by half a spaxel, allowing a better spatial sampling of the line peak emission. 

We reduced the data and analyzed it with PACSman. The PACS spectral map projection algorithm is explained in \cite{Lebouteiller12b}. The final map is shown in Figure\,\ref{fig:oiiimap}. The emission corresponds to a compact source located at 09h34m02.12s/55$^\circ$14$\arcmin$27.4$\arcsec$, i.e., coinciding with the NW cluster (Fig.\,\ref{fig:oiiimap}). The source appears mostly point-like, with a slight elongation along the NW-SE axis. This result is consistent with the \textit{Spitzer}/IRS results showing that most of the IR ionized gas line emission arises in the NW region and that a relatively lower fraction arises in the SE region (Sect.\,\ref{sec:spitzerobs}). 

The integrated source flux in the reconstructed map is $2.05\pm0.25\times10^{-17}$\,W\,m$^{-2}$, close to the value derived from our observation (OBSID 1342253758; Table\,\ref{tab:fluxes}). For testing purposes we have also extracted the flux in each individual raster position of the map using the methods described in Section\,\ref{sec:optimal}. Only $F_{\rm opt}$ (optimal extraction) could not be used because only one spaxel is significantly detected in each raster position. Despite the lower S/N as compared to OBSID 1342253758, we find that the various determinations are overall in good agreement (Table\,\ref{tab:oiiimapfluxes}). The error bars on the [O\3] flux from the spectral map observation may be underestimated because of the presence of systematic uncertainties in the baseline used for the fit. Such uncertainties are better canceled in chop/nod observations than in unchopped scan observations because of the fast removal of the telescope background (see PACS Observer's manual\footnote{\url{http://herschel.esac.esa.int/Docs/PACS/html/pacs_om.html}}). 
In our analysis we only use the chop/nod [O\3] observations described in Section\,\ref{sec:pacsobs}.

\begin{table}
\caption{[O\3] $88$\mic\ line flux determinations from the spectral map.\label{tab:oiiimapfluxes}}
\centering
\begin{tabular}{llll}
\hline\hline
Obs. \& Method            & $F_1$                & $F_{3\times3}$               & $F'_{3\times3}$               \\
\hline
Raster 1      & $(>)\ 144\pm33$  & $133\pm104$       & $(>)\ 367\pm104$  \\
Raster 2     & $(>)\ 150\pm23$  & $335\pm192$ & $(>)\ 411\pm133$  \\
Raster 3      & $(>)\ 236\pm32$  & $239\pm173$ & $(>)\ 505\pm108$  \\ 
Raster 4   & $(>)\ 191\pm22$  & $270\pm203$ & $(>)\ 251\pm59$    \\
\hline
Map           & \multicolumn{3}{l}{Aperture extraction: $205\pm25$}      \\
\hline
\hline
 \end{tabular}\\
\tablefoot{Fluxes are given in $10^{-19}$\,\wm. The values obtained for methods $F_1$ and $F'_{3\times3}$ underestimate the actual flux by an uncertain factor, hence we only quote the values as lower limits. }
\end{table}

\section{\textit{Spitzer}/IRS line flux measurements}\label{secapp:spitzer}

\begin{figure}
\centering
\includegraphics[angle=0,width=9.9cm,clip=true]{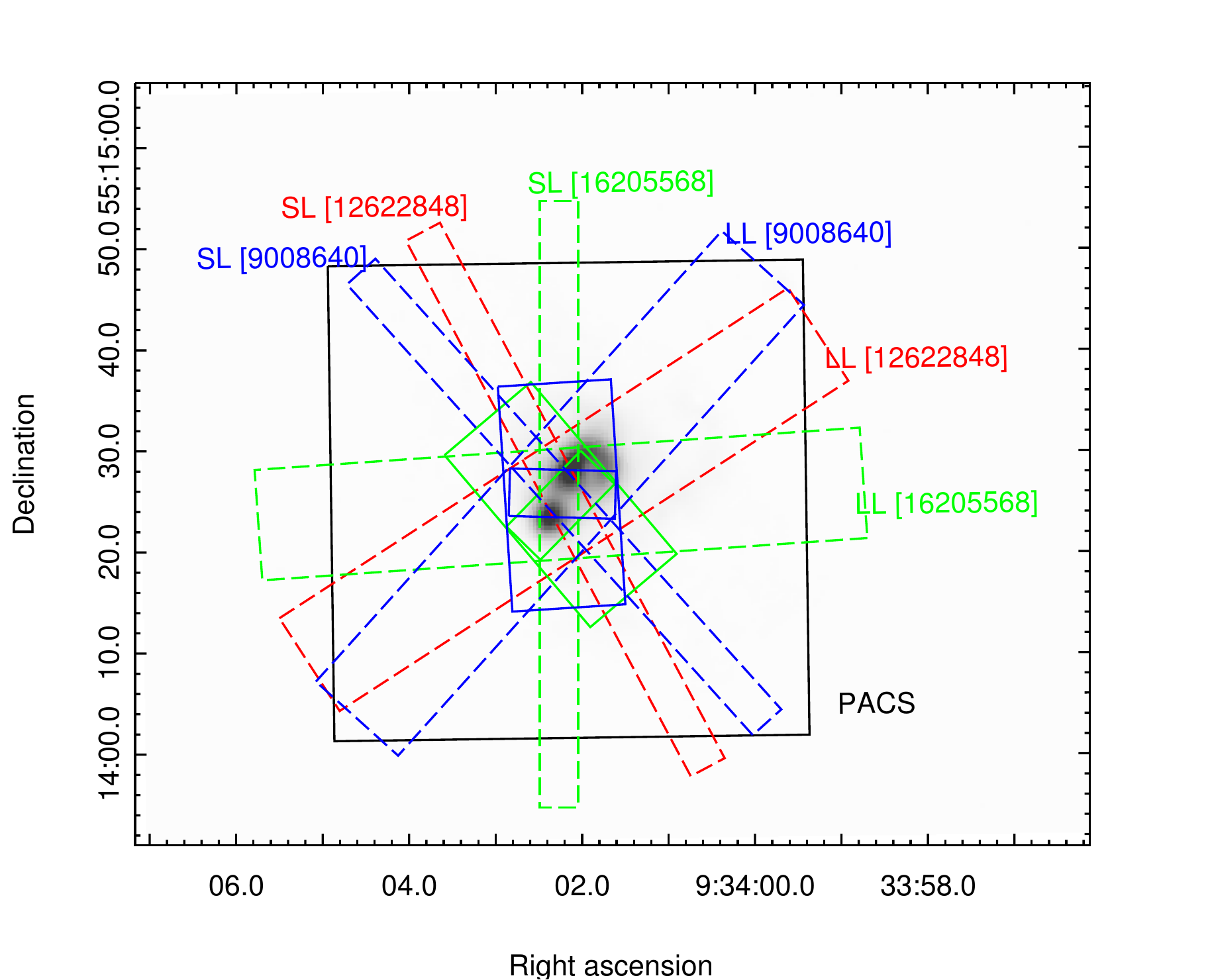}
\caption{Overlay of apertures on an H$\alpha$ image (from \citealt{depaz03}). Dashed colored rectangles correspond to the low-resolution apertures IRS SL ($3.6\arcsec\times57\arcsec$) and LL ($10.5\arcsec\times168\arcsec$). Solid colored rectangles correspond to the high-resolution apertures SH ($4.7\arcsec\times11.3\arcsec$) and LH ($11.1\arcsec\times22.3\arcsec$). The large black rectangle corresponds to the PACS footprint ($47\arcsec\times47\arcsec$).  }
\label{fig:irsslit}
\end{figure}

The apertures of the various \textit{Spitzer}/IRS observations\footnote{A small map was performed in SH (AORkey 21825792) with dedicated offset observations, but we do not consider it because of its low S/N. } are overlaid in Figure\,\ref{fig:irsslit}. It can be seen that the LL ($10.7\arcsec\times168\arcsec$) and LH ($11.1\arcsec\times22.3\arcsec$) observations include most of the emission from the two main H\2\ regions. Concerning the short wavelength observations, AORkey 16205568 is better oriented in both SL ($3.7\arcsec\times57\arcsec$) and SH ($4.7\arcsec\times11.3\arcsec$) as compared to the other observations. We therefore consider AORkey 16205568 as our reference observation for SL and SH, while all observations can be used in principle for LL and LH.

We use the extracted spectra from Cornell Atlas of Spitzer/IRS Sources (CASSIS), which is described in \cite{Lebouteiller11a} for low-resolution data and \cite{Lebouteiller15} for high-resolution data. For the low-resolution observations, the background sky emission was calculated via the image corresponding to the other nod position. For the high-resolution observations, no offset image was subtracted, but the background emission was removed during spectral extraction, as explained in \cite{Lebouteiller15}. 

In the following, we compare regular extraction (i.e., full aperture for high-resolution data and tapered column for low-resolution data) and optimal extraction methods. The regular method integrates the flux within the extraction window and uses either a point-source calibration (accounting for light losses outside of the aperture due to the PSF size) or extended source calibration (assuming uniformly extended emission within and outside the aperture). In practice, we find that the source is quasi-point-like in the long wavelength observations (LL, LH) and that it is somewhat extended in the short wavelength observations (SL, SH). Despite the noticeable extent, and for lack of a partially extended source calibration, the point-source calibration is still much more adequate than the extended source calibration. We consider hereafter that the regular extractions of LL and LH observations (with a point-source calibration) reflect the total emission from both NW and SE. On the other hand, we consider that the regular extraction of SL and SH might miss a fraction of the total flux. 

Optimal extraction is adapted to point sources only and, as such, it is a reliable method for the LL and LH observations for which only a minor correction due to the source extent may be necessary. The flux calibration already accounts for the aperture loss outside the slit. Therefore, the LL and LH optimal extraction are assumed to reflect the total emission from the NW and SE regions. For SL and SH, optimal extraction is only an approximation because the source is noticeably extended in these modules. 

We have also used a custom optimal extraction to extract the NW and SE regions simultaneously in the low-resolution observation 16205568 (6th column in Table\,\ref{tab:irs_lr}) using the SMART/AdOpt manual extraction tool described in \cite{Lebouteiller10}. With this special algorithm, the PSF is computed at the source position and broadened to reflect the intrinsic source broadening, so that the extracted spectra are calibrated in a reliable way and reflect the total source flux. The corresponding measurement therefore provides to first order the total NW+SE H\2\ region emission.

\begin{figure*}[!p]
\includegraphics[angle=0,width=8.5cm,height=5cm,clip=true]{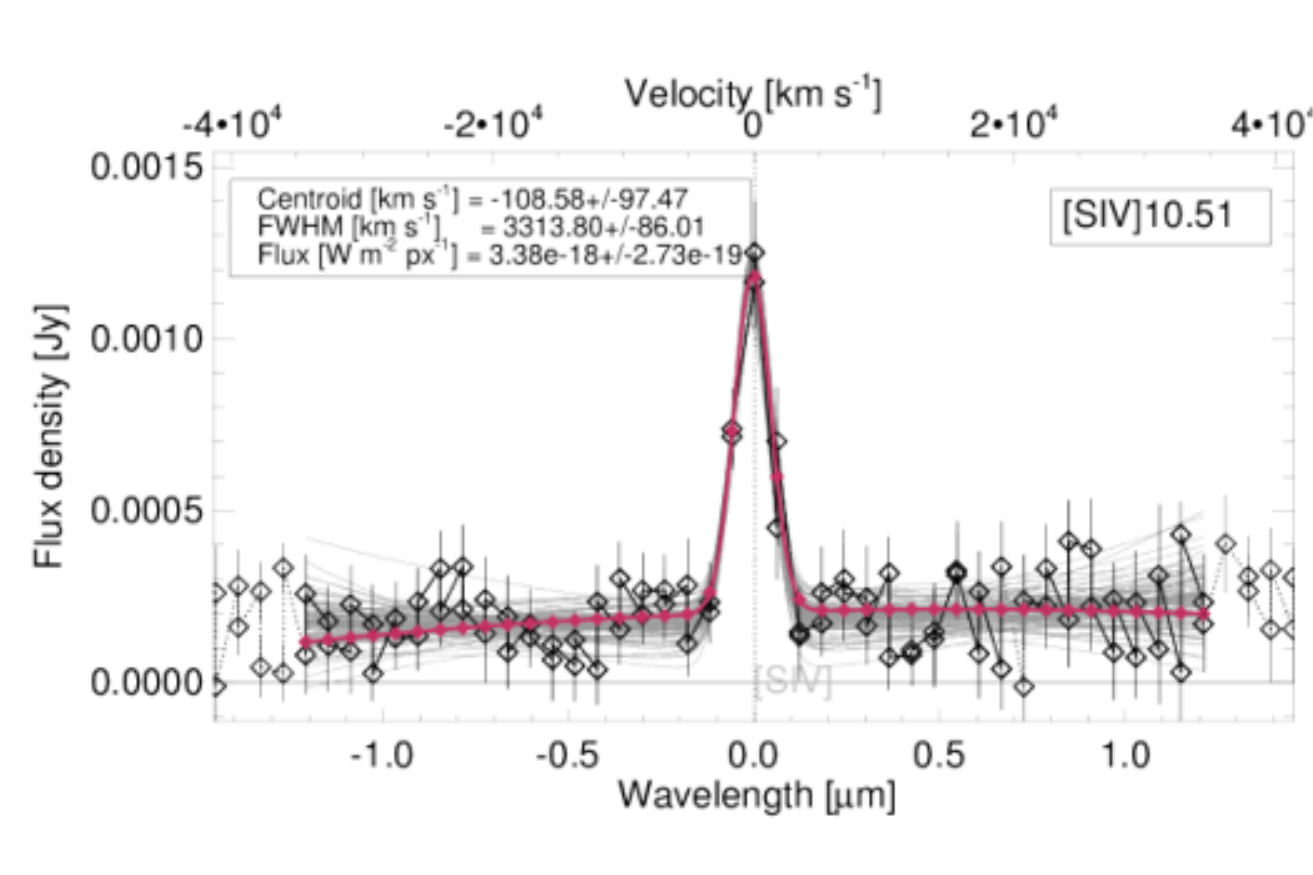}
\includegraphics[angle=0,width=8.5cm,height=5cm,clip=true]{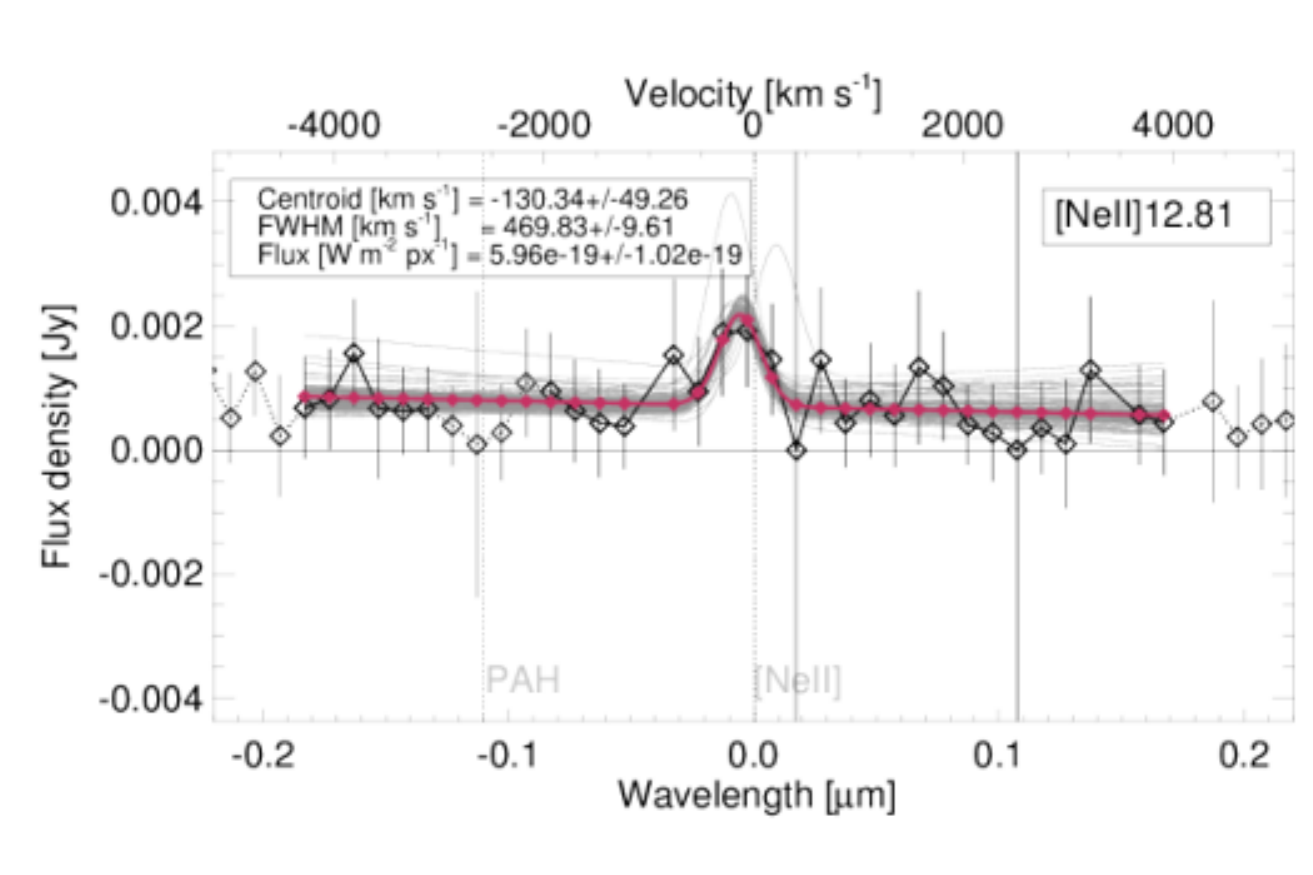}
\includegraphics[angle=0,width=8.5cm,height=5cm,clip=true]{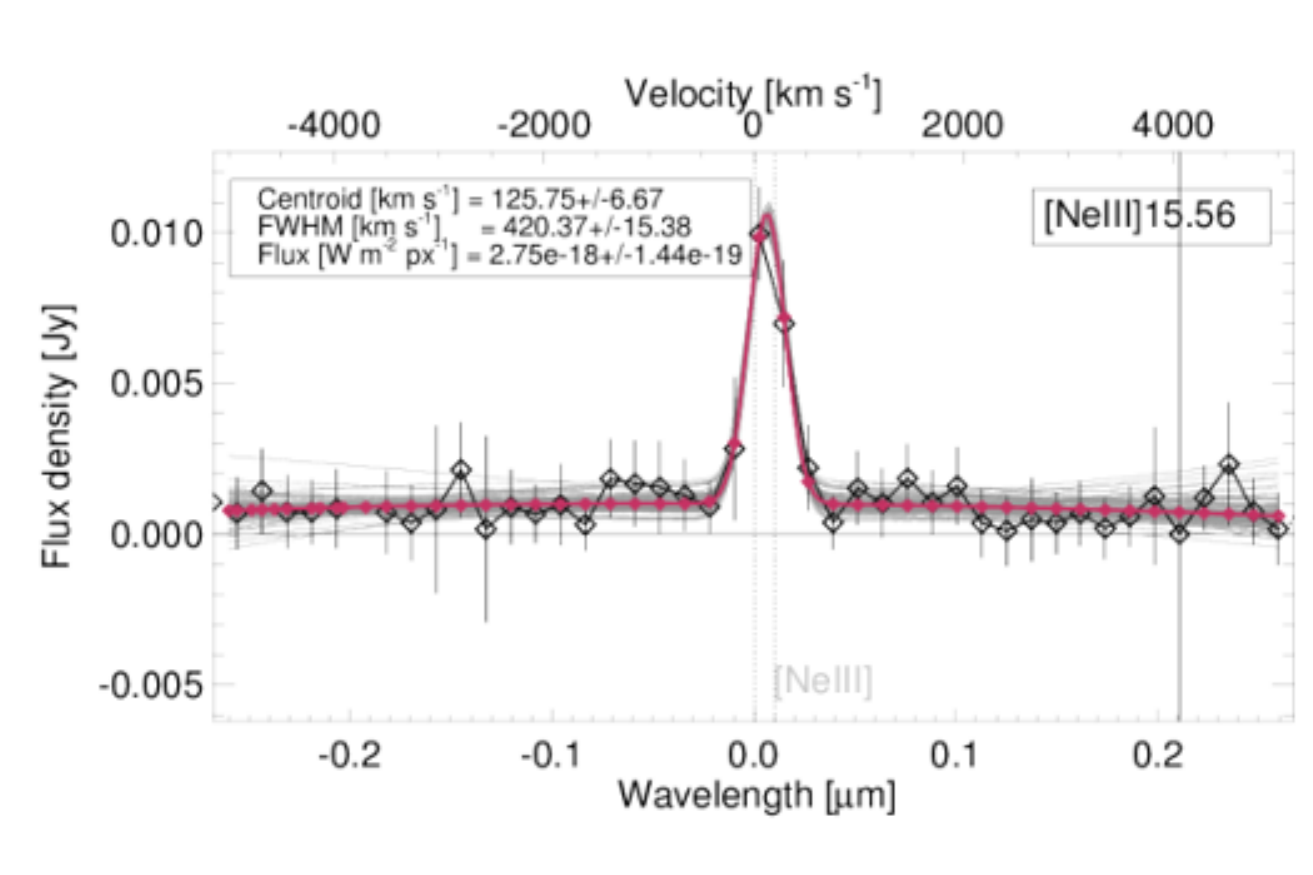}
\includegraphics[angle=0,width=8.5cm,height=5cm,clip=true]{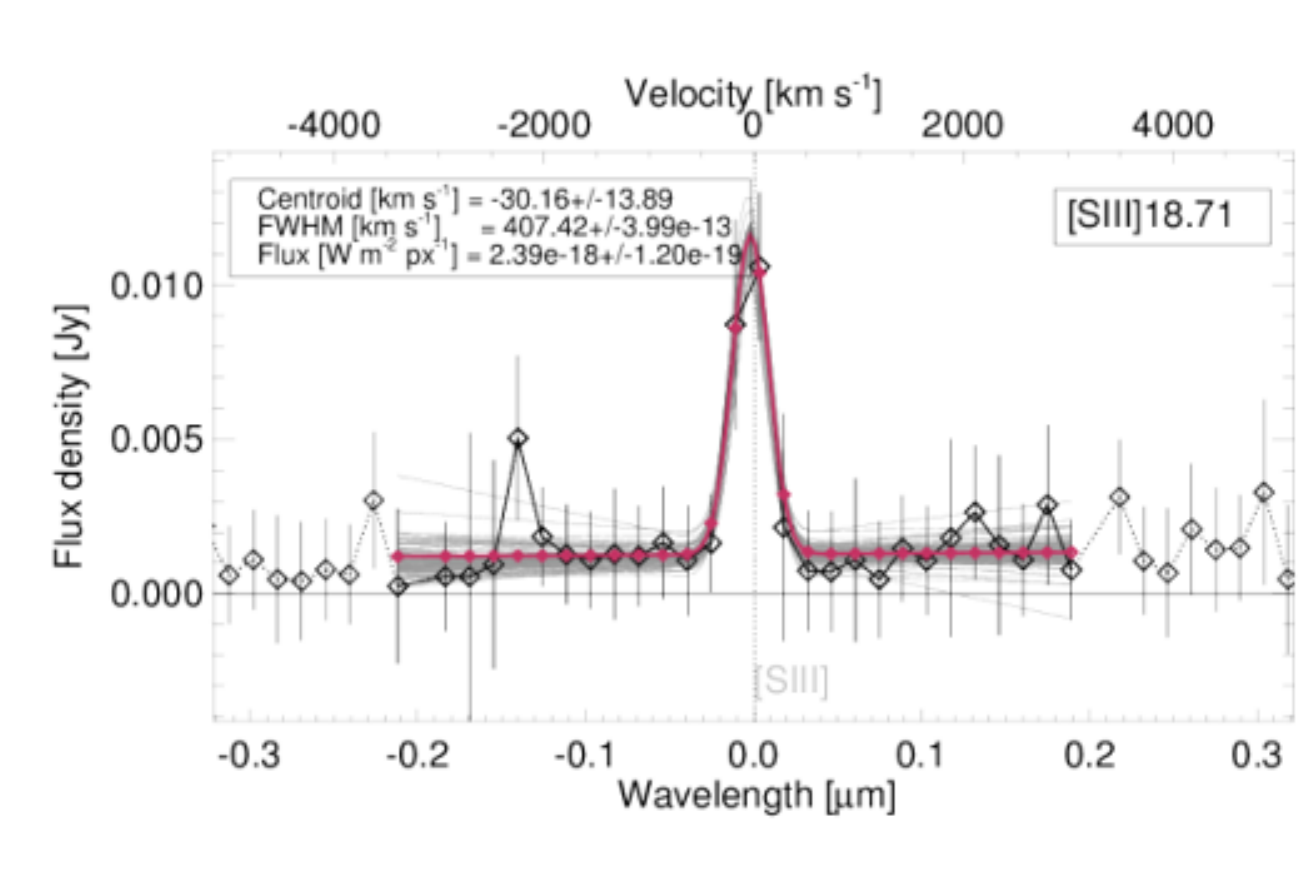}
\includegraphics[angle=0,width=8.5cm,height=5cm,clip=true]{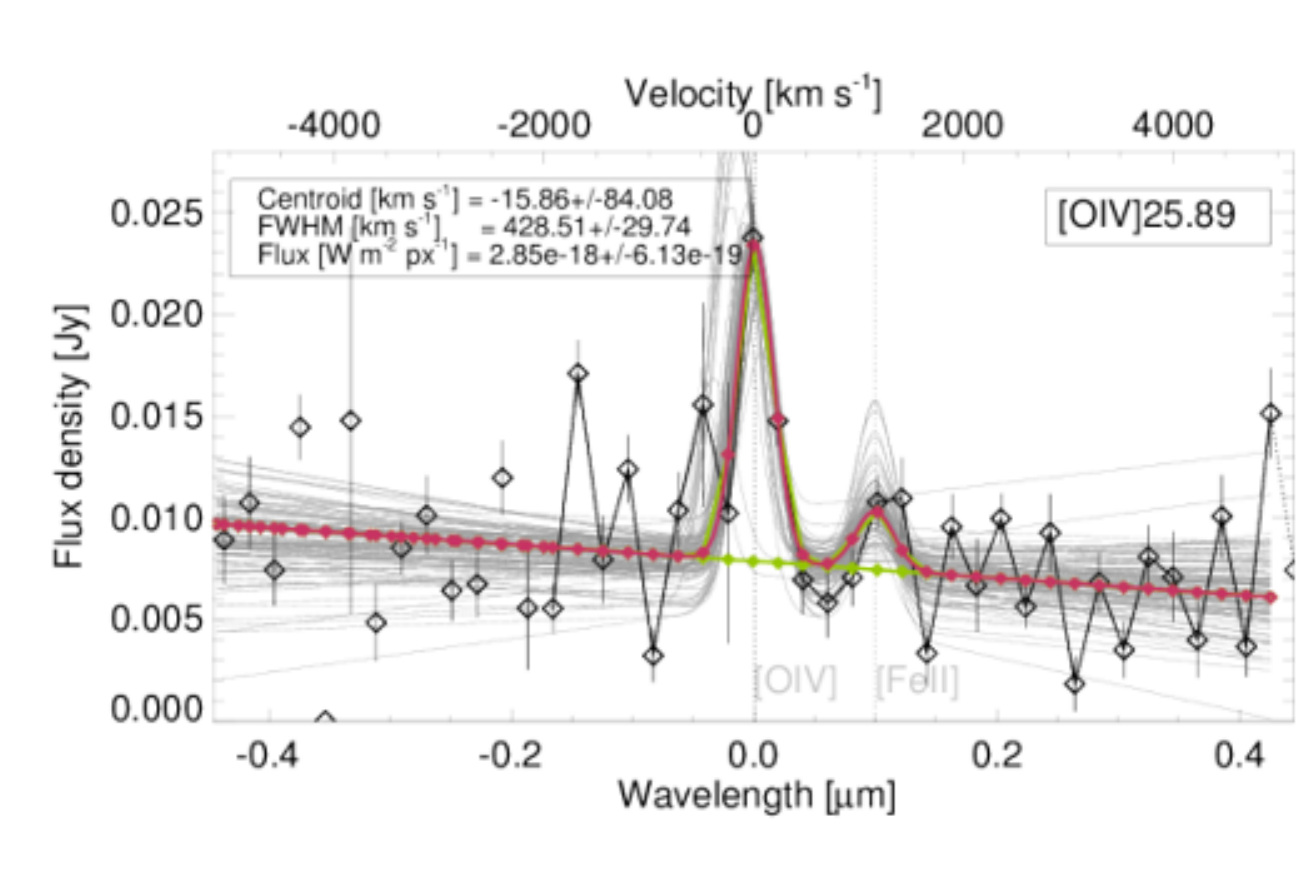}
\includegraphics[angle=0,width=8.5cm,height=5cm,clip=true]{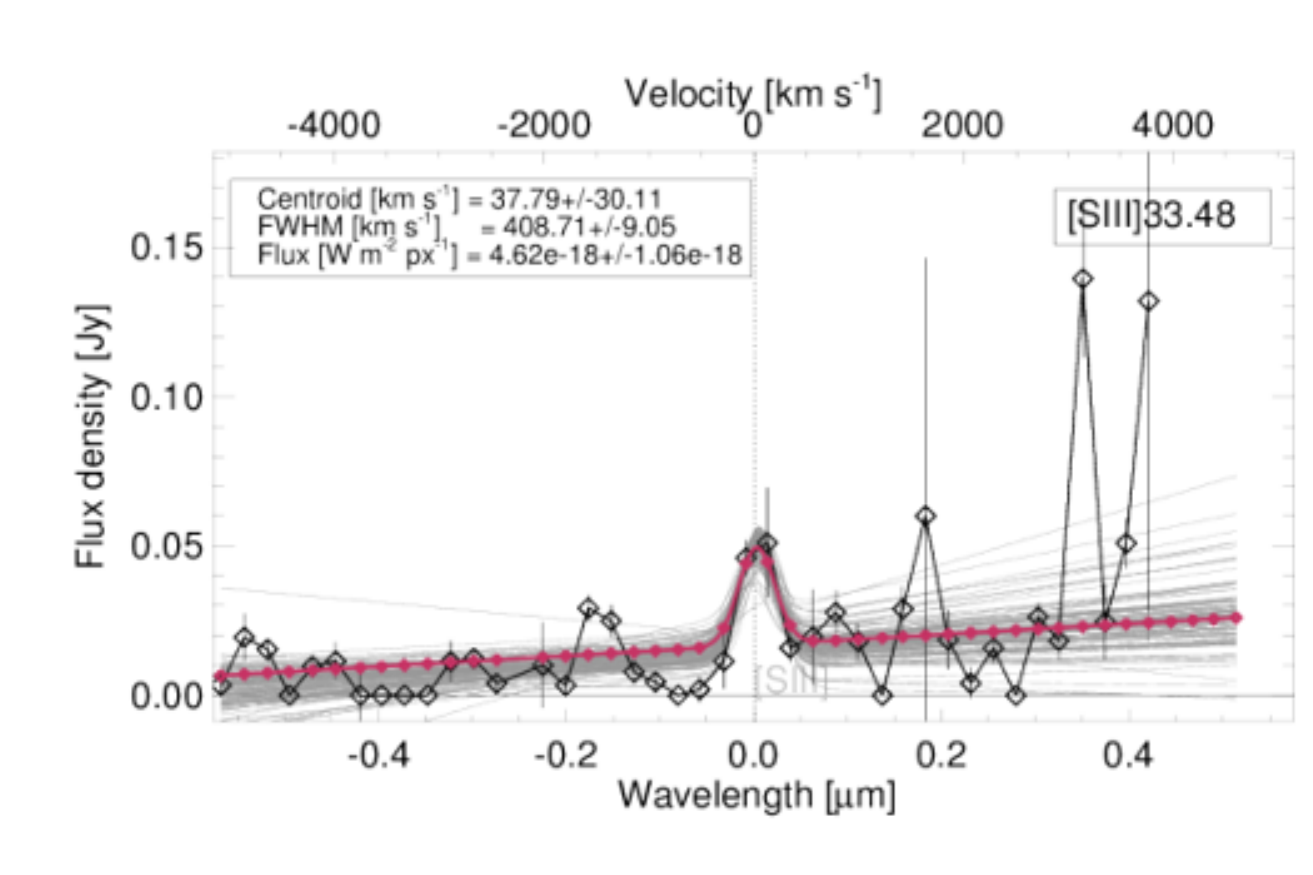}
\includegraphics[angle=0,width=8.5cm,height=5cm,clip=true]{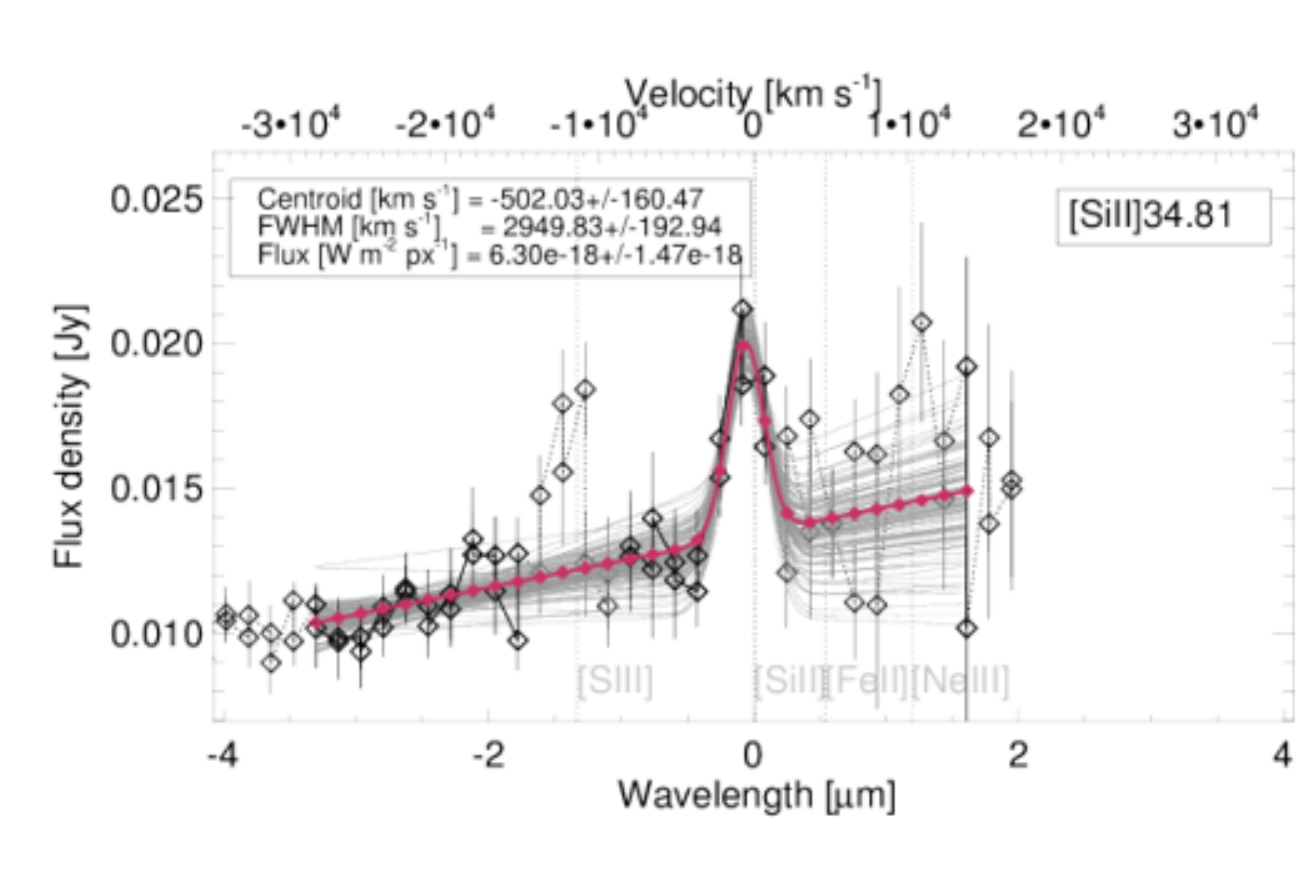}
\caption{Spectral fits for the most important lines used in this study. Several spectra are available for each \textit{Spitzer}/IRS line in both low- and high-resolution modes  with line widths of $\sim3000$\kms\ and $\sim400$\kms\ respectively. We selected one spectrum per line for illustrative purposes. The diamonds show the data and the connected diamonds represent the range used for fitting the line. The many light gray curves show Monte-Carlo iterations for determining the line flux uncertainty. The red curve shows the final fit. }
\label{fig:irsfits}
\end{figure*}

Figure\,\ref{fig:irsfits} shows some line fits for illustration. We list in Tables \ref{tab:irs_lr}.1 and \ref{tab:irs_hr}.2 the line measurements for each observation. The fact that we obtain similar results in both resolution modes, despite the different slit/aperture sizes, implies that we are successfully recovering the total flux of the NW+SE H\2\ regions. The final adopted fluxes are given in Table\,\ref{tab:irsfluxes}.3. along with a comparison to \citetalias{Wu07}.

\begin{table*}\label{tab:irs_lr}
\caption{\textit{Spitzer}/IRS low-resolution measurements. }
\begin{tabular}{lrrlllll}
\hline
\hline
   Method             &   Regular      &           Optimal   &   Regular       &  Optimal         &   Optimal custom$^\textrm{a}$   &   Regular & Optimal        \\
\hline
   SL                 &  9008640     &     9008640         &  \textbf{16205568}   &  16205568  &  \textbf{16205568}     &  12622848  &  12622848  \\
            & (Ap.)     &      (Ap., meth.)  &     (Pref.)             &   (Meth.)        &      (Pref.)       &  (Ap.)        & (Ap., meth.)             \\
\hline
 H\1\ Hu$\alpha$ $12.37$\mic                    &      $     <7  $   &   $              <5  $   &   $  8.2\pm2.5     $   &   $  4.1\pm1.5     $   &   $  2.6\pm2.0 , 5.5\pm1.8          $   &   $  7.0\pm3.5     $   &   $  6.5\pm3.3   $  \\
 $[$Ne\2$]$ 12.8\mic             &   $           <9  $   &   $              <6  $   &   $  <2            $   &   $  4.1\pm1.7     $   &   $  3.0\pm1.6 , 5.2\pm2.6          $   &   $  8.5\pm3.5     $   &   $  6.0\pm3.1  $   \\
 $[$S\4$]$ 10.5\mic              &   $          <10  $   &   $              <6  $   &   $  40\pm3        $   &   $  19\pm2        $   &   $  14\pm2 , 34\pm2                $   &   $  39\pm4        $   &   $  29\pm3     $   \\
 $[$Ar\2$]$ 7.0\mic            &   $            ...  $   &   $               ...  $   &   $  <5            $   &   $  <3            $   &   $  ...                              $   &   $  <5            $   &   $  <3         $   \\
 $[$Ar\3$]$ 9.0\mic           &   $            <5  $   &   $               <7  $   &   $  <8             $   &   $  <3             $   &   $  <10                              $   &   $  <15             $   &   $  <3          $   \\
\hline
LL     &  \textbf{9008640}     &     \textbf{9008640}         &  \textbf{16205568}   &  \textbf{16205568}  &  \textbf{16205568}     &  \textbf{12622848}  &  \textbf{12622848}  \\
 	              &    (Pref.)         &   (Pref., meth.)      &     (Pref.)              &     (Pref., meth.)      &     (Pref.)                &  (Pref.)       &      (Pref., meth.)          \\
\hline
 $[$O\4$]$+$[$Fe\2$]$     &   $      66\pm19  $   &   $         52\pm12  $   &   $  53\pm22       $   &   $  43\pm11       $   &   $  53\pm11                        $   &   $  79\pm20       $   &   $  60\pm15  $     \\
 $[$Ne\3$]$ 15.5\mic        &   $      71\pm13  $   &   $         66\pm17  $   &   $  83\pm8        $   &   $  65\pm8        $   &   $  76\pm10                        $   &   $  104\pm41*     $   &   $  59\pm13$   \\
 $[$Ne\3$]$ 36.0\mic        &   $            ...  $   &   $             <14  $   &   $  <11           $   &   $  <51           $   &   $  46\pm24*                        $   &   $  135\pm70*      $   &   $  60\pm24*     $  \\
 $[$Ne\5$]$ 14.3\mic            &   $          <14  $   &   $             <21  $   &   $  <7            $   &   $  <8            $   &   $  <8                             $   &   $  <21           $   &   $  <19         $  \\
 $[$Ne\5$]$ 24.3\mic            &   $          <23  $   &   $             <14  $   &   $  <10           $   &   $  <8            $   &   $  <9                             $   &   $  <18           $   &   $  <13         $  \\
 $[$S\3$]$ 18.7\mic           &   $      39\pm17  $   &   $        46\pm15  $   &   $  43\pm6        $   &   $  40\pm6        $   &   $  47\pm8                         $   &   $  42\pm17       $   &   $  56\pm16     $  \\
 $[$S\3$]$ 33.5\mic           &   $          <45  $   &   $            65\pm16  $   &   $  60\pm10       $   &   $  45\pm13       $   &   $  48\pm12                        $   &   $  91\pm30       $   &   $  57\pm11     $  \\
 $[$Ar\3$]$ 21.8\mic          &   $          <29  $   &   $             <21  $   &   $  <13           $   &   $  <10           $   &   $  <13                            $   &   $  <23           $   &   $  <17         $  \\
 $[$Fe\3$]$ 23.0\mic          &   $          <28  $   &   $             <19  $   &   $  <12           $   &   $  <10           $   &   $  <12                            $   &   $  <21           $   &   $  <15         $  \\
 $[$Si\2$]$  34.8\mic            &   $          <75  $   &   $         96\pm30  $   &   $  96\pm14       $   &   $  50\pm11       $   &   $  65\pm12                        $   &   $  130\pm63      $   &   $  115\pm24    $  \\
 $[$Fe\2$]$ 17.9\mic           &   $          <27  $   &   $             <24  $   &   $  <11           $   &   $  <9            $   &   $  <11                            $   &   $  <20           $   &   $  <15         $  \\
 H$_2$ S(0) 28.2\mic              &   $          <17  $   &   $             <14  $   &   $  <10           $   &   $  <7            $   &   $  <7                             $   &   $  <14           $   &   $  <11       $    \\
 H$_2$ S(1) 17.0\mic                &   $          <17  $   &   $             <24  $   &   $  <10           $   &   $  <9            $   &   $  <9                             $   &   $  <21           $   &   $  <17       $    \\
\hline
\end{tabular}
\tablefoot{Fluxes are in units of $10^{-19}$\wm. Upper limits are $1\sigma$. The line below the AORkey indicates whether the total emission from NW+SE is underestimated due to aperture mispointing (``Ap.'') or underestimated because the optimal extraction method is adapted to point sources (``Meth.''). The preferred determination is indicated with ``Pref.''. Line fluxes followed by asterisks are deemed unreliable due to the presence of spurious features. Question marks indicate that the line flux cannot be determined (because of numerous spurious features).  }
\tablefoottext{a}{NW and SE are extracted simultaneously in SL, with their spatial extent accounted for. In LL, a single slightly extended source is used. }
\end{table*}

\begin{table*}\label{tab:irs_hr}
\caption{\textit{Spitzer}/IRS high-resolution measurements. }
\begin{tabular}{llllll}
\hline
\hline
Method                     &   Regular (full)          &      Optimal   &   Regular (full)   &     Optimal   \\
\hline
SH                    &  9008640         &    9008640         &  \textbf{16205568}     &    16205568     \\
                      & (Ap.)     &  (Ap., meth.)  &     (Pref.)    &    (Meth.)  \\
\hline
 H\1\ Hu$\alpha$ $12.37$\mic                       &  $  12\pm6, 18\pm10          $    &  $  <2, <2                       $   &  $  <9, <10             $    &  $  <1, <2        $               \\
 $[$Ne\2$]$  12.8\mic            &  $  17\pm3*, 13\pm3*  $     &  $  6\pm1, <1                    $   &  $  ..., ...                $     &  $  <2, <1          $             \\
 $[$S\4$]$   10.5\mic            &  $  56\pm5, 37\pm4          $      &  $  14\pm3, 22\pm4               $   &  $  66\pm22, 83\pm25    $      &  $  23\pm2, 26\pm1  $             \\
 $[$Ne\5$]$ 14.3\mic            &  $  <3, <3                  $      &  $  <1, <1                       $   &  $  <3, <3              $      &  $  <1, <1          $             \\
 $[$Ne\3$]$ 15.5\mic        &  $  47\pm4, 44\pm4          $      &  $  28\pm1, 28\pm1               $   &  $  51\pm7, 78\pm7      $      &  $  21\pm3, 28\pm7  $             \\
 $[$S\3$]$ 18.7\mic           &  $  25\pm3, 32\pm4          $      &  $  13\pm2*, 13\pm2*       $   &  $  37\pm4, 31\pm4      $      &  $  21\pm3, 25\pm2  $             \\
 $[$Fe\2$]$ 17.9\mic           &  $  <4, <4                  $      &  $  <2, <2                       $   &  $  ..., <42              $      &  $  <23, 4\pm1      $             \\
 H$_2$ S(1) 17.0\mic                &  $  <3, <3               $     &  $  <1, <1                       $   &  $  ..., ...                $      &  $  ..., ...          $               \\
 H$_2$ S(2) 12.3\mic              &  $  15\pm7*, <7          $      &  $  <2, ...                        $   &  $  <6, <6              $     &  $  <4, <3        $               \\
\hline
LH                    &  \textbf{9008640}         &    \textbf{9008640}         &  \textbf{16205568}     &    \textbf{16205568}     \\
                      &  (Pref.)   &     (Pref., meth.)          &    (Pref.)     &    (Pref., meth.)  \\
\hline
 $[$O\4$]$ 25.89\mic      &  $  35\pm8, 49\pm11*     $   &    $  29\pm6, 25\pm4               $   &  $  <91, <86            $     &  $  <53*, <31* $            \\
 $[$Ne\3$]$ 36.0\mic        &  $  <32, <23                $      &  $  ..., ...                         $   &  $  <28, <32            $    &  $  <10, <50         $            \\
 $[$Ne\5$]$ 24.3\mic            &  $  <10, <10                $     &  $  <5, 8\pm4                    $   &  $  <6, <5              $      &  $  <3, <2           $            \\
 $[$S\3$]$ 33.5\mic           &  $  59\pm23, 46\pm22        $     &  $  37\pm21, <9                  $   &  $  38\pm11, 52\pm16    $      &  $  58\pm8, 43\pm8   $            \\
 $[$Ar\3$]$ 21.8\mic          &  $  <8, <9                  $     &  $  <5, <4                       $   &  $  <7, 16\pm7          $      &  $  <3, <2           $            \\
 $[$Fe\3$]$ 23.0\mic          &  $  <6, <11                 $    &  $  <5, <4                       $   &  $  <7, <4              $      &  $  <3, <2           $            \\
 $[$Si\2$]$  34.8\mic            &  $  64\pm14, 88\pm16        $    &  $65\pm6,52\pm10                         $   &  $  111\pm50, 117\pm32  $   &  $  95\pm2*, 61\pm10* $   \\
 $[$Fe\2$]$ 26.0\mic        &  $  34\pm10, 33\pm11          $      &  $  <5, <6                       $   &  $  16\pm7, 13\pm5           $      &  $  <2, <2                    $   \\
 H$_2$ S(0) 28.2\mic              &  $  ..., ...                    $     &  $  <4, <4                       $   &  $  <12, <10            $    &  $  <4, <3                  $     \\
\hline
\end{tabular}
\tablefoot{Fluxes are in units of $10^{-19}$\wm. Upper limits are $1\sigma$. See Table\,\ref{tab:irs_lr}.1 for the description. For the regular extraction, fluxes are given for the two nod positions and a point source calibration is used. For optimal extraction, we provide two determinations, one using the simultaneous nod extraction and one using the combined nod spectrum (see \citealt{Lebouteiller15} for more details on the high-resolution optimal extraction algorithm). The preferred sets of observations and methods are marked in bold. }
\end{table*}

\begin{table}
\label{tab:irsfluxes}
\caption{\textit{Spitzer}/IRS line fluxes. }
\begin{tabular}{ll|ll|ll}
\hline
\hline
                    &  \citetalias{Wu07}    &  \multicolumn{2}{l|}{Low-resolution}      &  \multicolumn{2}{l}{High-resolution}          \\
\hline
 H\1\ Hu$\alpha$ $12.37$\mic          & ... & $ 8.5\pm3^\textrm{a}$ & SL &   $   <10$  & SH     \\
 $[$O\4$]$ 25.89\mic   & $49\pm3$  & $ 60\pm15 $  & LL &  $45\pm10 $  & LH   \\
                      &             &  (+[Fe\2]) &              \\
 $[$Ne\2$]$ 12.8\mic          & $9\pm1$ & $6\pm3   $ & SL  &  $\sim8^\textrm{b}$ & SH  \\
 $[$Ne\3$]$ 15.5\mic     & $46\pm2$ & $ 70\pm15 $ & LL  &  $60\pm10 $ & SH  \\
 $[$Ne\3$]$ 36.0\mic     & ... & $ <10       $ & LL  &    $   <25 $ & LH      \\
 $[$Ne\5$]$ 14.3\mic         & ... & $ <10     $ & LL  &    $     <3  $ & SH  \\
 $[$Ne\5$]$ 24.3\mic         & ... & $ <10     $ & LL  &  $     <5 $    & LH  \\
 $[$Si\2$]$  34.8\mic         & $158\pm8$ &  $ 95\pm20 $ & LL  &  $80\pm30 $  & LH    \\
 $[$S\3$]$ 18.7\mic        &  $23\pm2$ &   $ 40\pm10 $ & LL  &   $35\pm10 $  & LH  \\
 $[$S\3$]$ 33.5\mic        & $120\pm12$ &  $ 55\pm10 $ & LL  &    $50\pm10 $ & LH   \\
 $[$S\4$]$  10.5\mic          & $48\pm3$ &  $43\pm5  $ & SL  &  $70\pm20 $   & SH \\
 $[$Ar\2$]$ 7.0\mic          & ... &  $<5      $  & SL  &  ...  & ...    \\
 $[$Ar\3$]$ 9.0\mic        & ... &  $<10   $ & SL &   ... & ...      \\
 $[$Ar\3$]$ 21.8\mic       & ... & $ <15     $ & LL  &    $     <10 $ & LH  \\
 $[$Fe\2$]$ 17.9\mic        & ... & $ <10     $ & LL  &  $    <4 $ & SH  \\
 $[$Fe\2$]$ 26.0\mic      & $34\pm3$ &  $60\pm15$ & LL   &  $ 15\pm5 $  & LH  \\
                      &             &  (+[O\4]) &              \\
 $[$Fe\3$]$ 23.0\mic       & ... & $ <15     $  & LL   &    $     <7 $  & LH  \\
 H$_2$ S(0) 28.2\mic           & ... & $ <10 $  & LL    &  $      <10 $  & LH   \\
 H$_2$ S(1) 17.0\mic             & ... &  $<10$    & LL    &   $     <10$  & SH \\
 H$_2$ S(2) 12.29\mic           & ... &  ...  &    SL  &  $    <10$  & SH  \\
\hline
\end{tabular}
\tablefoot{Fluxes are in units of $10^{-19}$\wm. Upper limits are $1\sigma$. Measurements by \citetalias{Wu07} were done with the high-resolution modules. For each measurement, we indicate the module used (SL, LL, SH, LH). }
\tablefoottext{a}{Assuming that the molecular hydrogen line flux is negligible.}
\tablefoottext{b}{Estimated using optimal extraction of AORkey 9008640 together with a correction factor based on [Ne\3] 15.5\mic. }
\end{table}

\section{X-ray source intrinsic spectrum}\label{secapp:xspec}

In this section, we explore the possible intrinsic X-ray spectra that are compatible with the XMM-\textit{Newton} observation, the observed optical and IR line fluxes (e.g., [Ne\5], [C\2], [O\1]), and the H$^0$ column density. All tests are performed with xspec \citep{Arnaud96}. We are mostly interested in photons with energies between $\sim0.1-2$\,keV because they are absorbed in the H\1\ region. Lower energy photons are absorbed in the H\2\ region and higher energy photons freely cross the H\2\ + H\1\ regions. For this reason, one must bear in mind that the fits to the XMM-\textit{Newton} observation are actually mostly constrained by the well detected hard X-ray component that is not absorbed. Recovering the soft-X-ray component heavily depends on the model chosen that  ties the hard- and soft-X-ray components in a physically motivated way. 

\begin{figure}
\centering
\includegraphics[width=9cm,clip,trim=0 0 0 0]{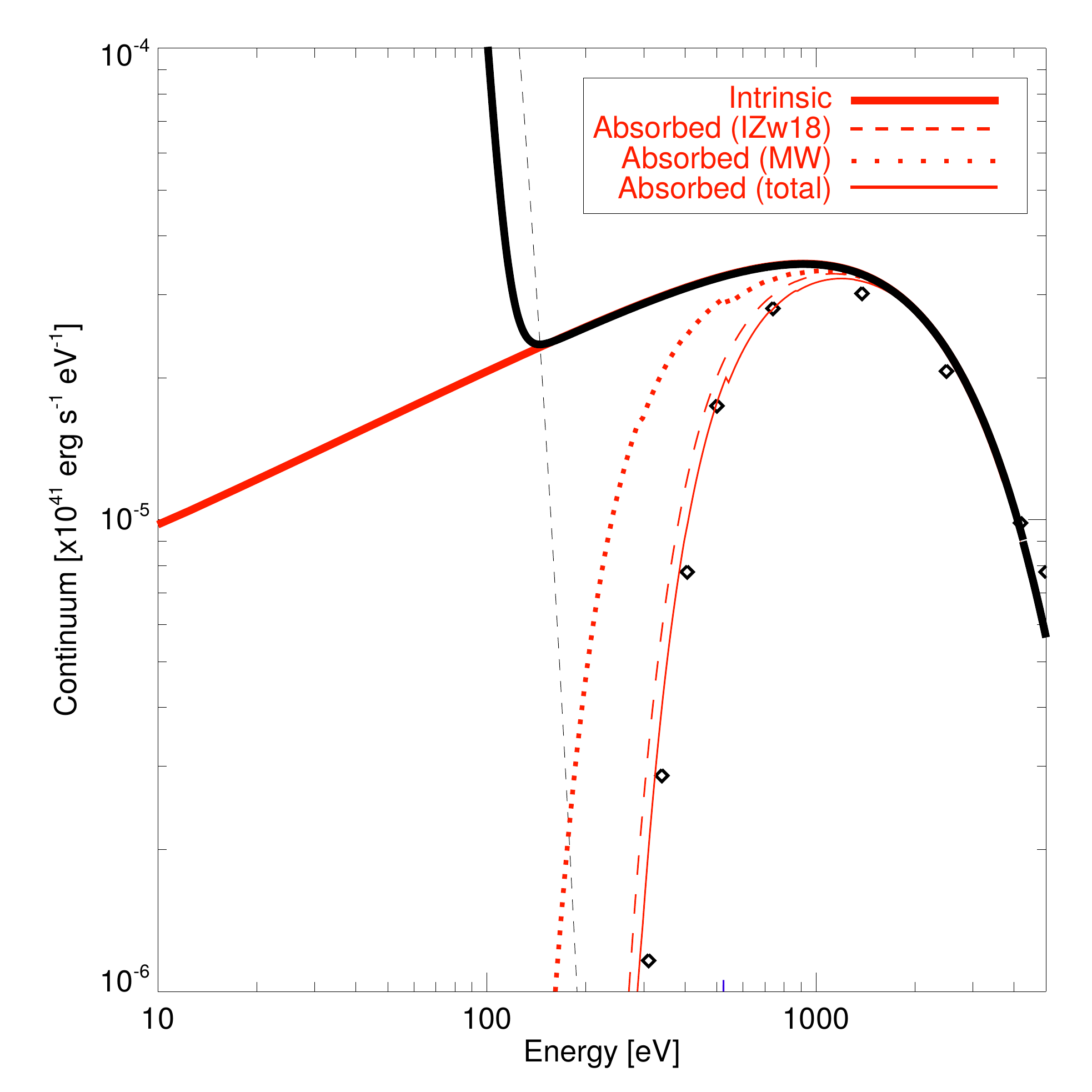}
\caption{X-ray spectrum of \izw. The X-ray emission is calculated with a \texttt{diskbb} model (distribution of blackbodies from an accretion disk). The diamonds show the unfolded XMM-\textit{Newton} spectrum. The dashed red line shows the spectrum absorbed by \izw\ alone while the dotted red line shows the spectrum absorbed by the Milky Way alone. 
The top red solid line shows the intrinsic (unabsorbed) X-ray spectrum. 
The thick black curve shows the sum of the radiation field components (two blackbodies in the optical and the X-ray source).  }
\label{fig:xspec_diskbb}
\end{figure}

We first consider a \texttt{diskbb} model. For the attenuation in the Milky Way, we used the foreground Galactic extinction $0.091$ \citep{Schlegel98,Schlafly11} and the $A_V$-to-column density conversion of \cite{Guver09}. We find $N_{\rm H}\approx2\times10^{20}$\,cm$^{-2}$, in good agreement with the value measured at $21$\,cm (e.g., \citealt{Stark92,Kalberla05}). The total absorbing hydrogen column density in the Milky Way (including molecular hydrogen) is $2.7\times10^{20}$\,cm$^{-2}$ \citep{Willingale13}. Figure\,\ref{fig:xspec_diskbb} shows the \texttt{diskbb} model and the absorbed spectrum. Most of the absorption occurs in \izw, with a column density of $10^{21}$\,cm$^{-2}$. The luminosity between $0.3-1$\,keV is $\approx10^{40}$\,erg\,s$^{-1}$ and the temperature at the inner edge of the accretion disk is $1.04$\,keV. These results are similar to those found by \cite{Kaaret13}. These values are unchanged when using the abundances measured in absorption by \textit{HST}/COS rather than abundances measured in emission in the H\2\ regions (Sect.\,\ref{sec:abundances}). The column density for the \texttt{diskbb} model is somewhat lower than expected based on H\1\ absorption measurements. The X-ray source is located within the stellar cluster (Sect.\,\ref{sec:xray}), toward which the H\1\ column density has been calculated to be $2.2\times10^{21}$\,cm$^{-2}$ \citep{Aloisi03,Lecavelier04,Lebouteiller13a}. The column density measured in absorption, however, represents an average value (toward the most FUV-bright stars within the $2\arcsec$ COS aperture), which may differ from that derived along the line of sight of the X-ray point source.

While the \texttt{diskbb} model was preferred in the present study, combinations of blackbodies were also considered. Such combinations allow us to fully explore the range of acceptable spectra without strong assumptions about the nature of the source. On the other hand, using blackbodies ensures that physically unrealistic sharp features and/or discontinuities are discarded from the analysis. 
The single $2\times10^6$\,K blackbody used in \citetalias{Pequignot08} neglected the hard X-ray tail ($\gtrsim2$\,keV) which barely impacts on the H\1\ region properties, yet it was a coarse first approximation, given the X-ray data now available. 

\begin{figure*}
\centering
\includegraphics[width=6.52cm,clip,trim=20 0 17 0]{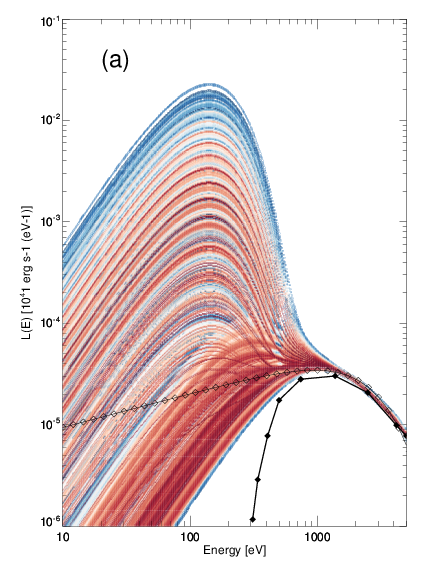}
\includegraphics[width=5.8cm,clip,trim=63 0 17 0]{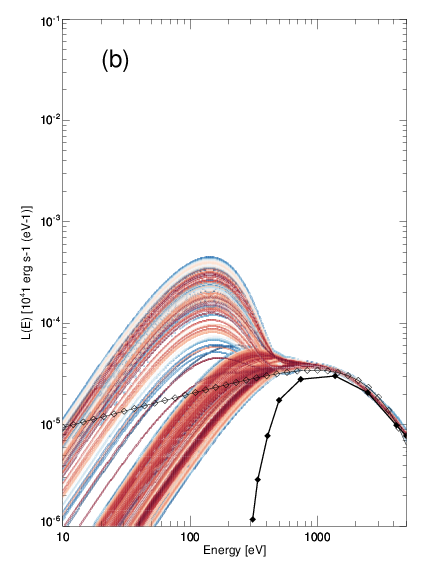}
\includegraphics[width=5.8cm,clip,trim=63 0 17 0]{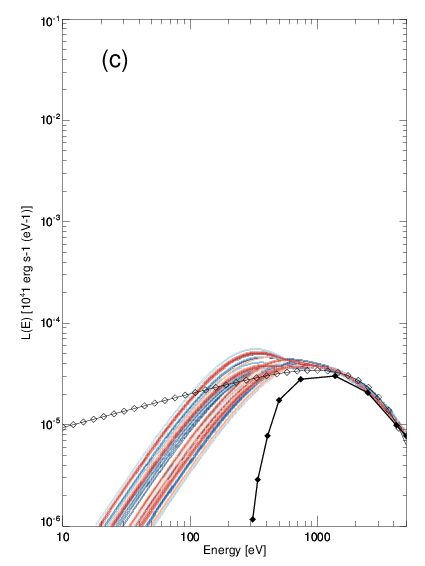}
\caption{Test of the $4$-blackbody combinations with xspec. The open diamonds show the \texttt{diskbb} model while the filled diamonds show the XMM-\textit{Newton} unfolded spectrum. The set of best blackbody combinations with column density $<10^{22}$\,cm$^{-2}$ are shown in panel (\textit{a}). A subset with column densities between $1-2.5\times10^{21}$\,cm$^{-2}$ is shown in panel (\textit{b}). Only in the combinations left in panel (\textit{c}) are the associated photoionization models fulfilling the observed upper limit on the [Ne\5] flux. For simplicity, we plot the X-ray spectra of panel (c) with the same luminosity as for the other panels, but in reality an increased luminosity is used, according to Sect.\,\ref{sec:const_xrays}. 
In each panel, the colors indicate the relative variation of the $\chi^2$ among the plotted combinations (lower $\chi^2$ in red, larger $\chi^2$ in blue). }
\label{fig:xspec_bbs}
\end{figure*}

Using a sum of four blackbodies with temperatures between $0.05-1.20$\,keV, and column densities between $0-1\times10^{22}$\,cm$^{-2}$ proved to provide sufficient flexibility (nine free parameters). Figure\,\ref{fig:xspec_bbs}a
shows a broad selection of combinations providing the best $\chi^2$. 
Unlike for the \texttt{diskbb} model, owing to the lack of physical ties between the hard and soft X-ray components, the soft X-rays are little constrained and a degeneracy
develops between the coldest blackbody fluxes and the (here poorly defined) column density. 
We constrain further the column density in the X-ray model assuming that the X-ray source lies within the NW stellar cluster, hence assuming a column density of $1-2.5\times10^{21}$\,cm$^{-2}$ (Fig.\,\ref{fig:xspec_bbs}b). 
Two different spectral shapes emerge, one with a soft X-ray excess and one without.  In Figure\,\ref{fig:xspec_bbs}b, the X-ray luminosity range of the models is $1.1-2.5\times10^{40}$\,erg\,s$^{-1}$. Below $\sim0.3$\,keV, the freedom left by the xspec approach with a blackbody combination is considerable.

Some of the degeneracy can be lifted using signatures of the X-ray emission on the surrounding ISM. 
Our standard photoionization model, in which the \texttt{diskbb} X-ray spectrum is adopted, predicts fluxes $3-5$ times weaker than the observational upper limits for the high-ionization lines [Ne\5] $14.3$\mic\ and [Ne\5] $24.3$\mic\ (Table\,\ref{tab:cloudycompir}). Also, no detection has been reported for [Ne\5] $3426$\AA\ in \izw. Producing [Ne\5] in the H\2\ region requires photon energies $\gtrsim100$\,eV. In Figure\,\ref{fig:xspec_bbs}b, the soft X-ray excess allowed by xspec around $100$\,eV can exceed the relatively flat \texttt{diskbb} model spectrum by more than one order of magnitude. Obviously, at least some of these blackbody combinations produce much stronger [Ne\5] lines than \texttt{diskbb}. Figure\,\ref{fig:xspec_bbs}c shows the combination spectra allowed when the additional constraint brought by the
upper limit on [Ne\5] is applied to photoionization model results\footnote{Since the additional constraint provided by [Ne\5] is related to the ionized gas irradiated by the X-ray source, we need to consider the X-ray luminosity around $4\times10^{40}$\,erg\,s$^{-1}$ used in our modeling (Sect.\,\ref{sec:const_xrays}).}. 

This illustrates how detailed photoionization modeling of the ISM around X-ray sources may provide powerful constraints on the low-energy tail of X-ray source spectra and therefore on the nature and properties of such sources. A preliminary inference from the present analysis is that the soft X-ray spectrum of the \izw-NW source cannot much exceed that corresponding to the \texttt{diskbb} model. Not only is the solution above $\sim0.4$\,keV well defined, but it turns out that, within a factor of $\sim2.5$, the set of blackbody combinations follows the diskbb spectrum down to less than $0.1$\,keV. The fall of the blackbody combinations toward low energies, essentially an artefact of the fitting assumptions, is of no consequence since the stellar continuum then dominates.

\end{document}